\documentclass[12pt]{iopart}

\usepackage{graphics}
\usepackage{amssymb}
\usepackage{section5}
\begin{document}


\title[The combined exact diagonalization- {\em ab initio} approach]{
The combined exact diagonalization- {\em ab initio} approach and its
application to correlated electronic states and Mott-Hubbard
localization in nanoscopic systems}

\author{Jozef Spa\l ek, Edward M. G\"orlich, Adam Rycerz, Roman Zahorbe\'nski}
\address{
Marian Smoluchowski Institute of Physics, Jagiellonian University,
ulica Reymonta 4, 30-059 Krak\'ow, Poland }

\date{ \today }

\begin{abstract}
 We overview the EDABI method developed recently and combining
the exact diagonalization and {\em ab initio} aspects of electron
states in correlated systems and apply it to nanoscopic systems.  In
particular, we discuss the localization-delocalization transition
for the electrons that corresponds to the Mott-Hubbard transition in
bulk systems. We show, that, the statistical distribution function
for electrons in a nanochain evolves from the Fermi-Dirac-like to
the Luttinger-liquid-like with the increasing interatomic distance.
The concept of Hubbard subbands is introduced to nanoclusters, and
corresponds to the HOMO-LUMO splitting in the molecular and organic
solid states. Also, the nanochains exhibit magnetic splitting
(Slater-like), even without the symmetry breaking, since the
spin-spin correlations extend over the whole system. Thus, the
correlated nanoscopic systems exhibit unique and universal features,
which differ from those of molecular and infinite systems. These
features define unique properties reflecting {\em "the Mott
physics"} on the nanoscale. We also employ the EDABI method to the
transport properties in nanoscopic systems. For example, we show
that the particle-hole symmetry is broken when the tunneling
conduction through $H_2$ molecule is calculated.
\end{abstract}

\pacs{61.64.+w, 36.40.-c, 73.63.-b}
\submitto{\JPCM}
\maketitle

\section{Introduction}
 \indent The studies of {\em nanosystems} are becoming increasingly important
 in view of their application in quantum nanoelectronics and
 related fields of research. Of particular importance are their
 quantum electronic properties, since they determine their behavior
 as concrete devices: quantum nanowire connectors and semiconducting
 elements, single-electron transistors, spin valves etc. Under these
 circumstances, solid-state and molecular {\em nanophysics} is
 developing very rapidly to provide proper quantitative (and
 qualitative)  characteristics of their static, transport, and
 optical properties \cite{ref1}. \\
 \indent Independently of the applications, the research in {\em
 nanophysics} is important for its own fundamental sake. Namely, the
 nanosystems are finite systems and therefore, most of the limiting
 situations considered in the condensed matter physics and involving the limit for the number of atoms $N
 \rightarrow \infty$, is simply inapplicable. Furthermore,
 the role of boundary conditions is very nontrivial, as  they should reflect the system actual configuration. Also, and probably most
 importantly, the question of electron-electron interactions,
 particulary for extended states, should be taken into account on
 equal footing with the single-electron aspects of their quantum
 states, since the screening processes are very often highly
 ineffective. Nevertheless, given the circumstances, the role of
 physics is to single out universal properties of the systems such
 as nanowires, clusters, quantum dots, etc. Such research involves
 also determining the conditions,  under which the bulk-solid concepts are
 applicable, since then the description can be simplified
 remarkably.\\
 \indent In this article we present our recent and earlier results
 concerning the studies of nanosystems containing as an intrinsic property the electron
 correlation effects induced by the Coulomb interactions, within the
 devised EDABI method combining both exact diagonalization and {\em
 ab initio} aspects of electronic states, as well as their transport
 properties \cite{SpalekPodsiadl00}. This approach allows for discussing the
 evolution of the physical properties of e.g. nanowires or clusters,
 both
 as a function of their interatomic distance, as well as to determine
 the values of microscopic single-particle and interaction
parameters. The visualization of the properies as a
 function of interatomic distance is particularly important for
 nanometer size systems, as they are studied customarily by placing
 them on a substrate with the lattice parameter, which differs from
 their equilibrium  interatomic distance (sometimes the substrate
 even stabilizes them).\\
 \indent Our most interesting results can be summarized as follows.
 First, we show how the system properties evolve from the
 Fermi-liquid-like to the atomic-like states for nanowires containing up to $N=16$
 simple atoms, passing through the Luttinger-liquid-like state, with the
 increasing interatomic distance $R$. This evolution is determined
 by calculating directly the system statistical distribution
 function $n_{k\sigma}$ of electrons and their dynamical spectral function.
 Finite-size scaling properties are introduced to determine the {\em
 Mott critical interatomic distance ( the Mott criterion)} for the
 transition from the delocalized to the localized states. Secondly, we
 show that the electronic states in nanosystems of $N \sim 10$
 atoms exhibit a magnetic splitting reminiscent of the {\em Slater
 splitting} in antiferromagnetic metallic systems \cite{Slater}. This
 type of symmetry change is associated with the breakdown of
the  discrete translational symmetry, as the
 antiferromagnetism sets in. In the case of nanosystems such symmetry
 breakdown is not required. It turns out, that the splitting appears
 if the spin-spin correlation length is of the system size. The
 splitting should be detected e.g. in nanowires containing strongly
 correlated electrons in a half-filled valence band configuration.
 Thirdly, the question arises whether with the increasing
 interatomic distance one should not observe the Hubbard split-band
 structure of nanowire and molecular (e.g. $H_N$ or $Li_N$)
 clusters? We show that nanocluster levels group into
 multiple Hubbard subbands (in bulk systems such a grouping is termed
 as HOMO and LUMO structures). Finally, in the case of a quantum dot
 composed of e.g. $H_2$ molecule attached to the semimacroscopic
 electrodes we show, that the electron-hole symmetry in the tunneling
 transport through the molecule is not preserved and is due to the
 difference in electronic binding energy of $H_2^+$ and $H_2^-$ states. Such
 result is not obtained if a parameterized model of a quantum dot
 properties is used \cite{RycerzSpalek06}.\\
 \indent The structure of this paper is as follows. In the next
 Section we overview the EDABI method \cite{Slater,RycerzSpalek06,ThesisAR,ThesisEG,ThesisRZ}, as well as
 discuss some of the many-electron general properties adopted to the
 analysis carried out in the next Sections. We also
 compare the method with the configuration interaction (CI) approach
 used in quantum chemistry. In Section 3 we discuss atomic systems and $H_N$ nanoclusters, whereas in Section 4
 the results for nanochains containing up to $N=16$ atoms are elaborated in detail. In Section 5 we
 employ the EDABI method to calculate the tunneling conductivity
 through the $H_2$ molecule and the Drude weight for the nanochains. We also discuss the role of boundary conditions there.\\
 \indent The present method and the research grew out of our earlier
 work on the thermodynamics of the Mott-Hubbard transition in correlated
 systems \cite{SpalekDattaHonig87}. There, the principal question was if the
 metallic and magnetic insulating states can be regarded as
 separates phases in the thermodynamic sense? The
 affirmative answer to the above question provided a partial answer to the
 Sir Nevill Mott question: {\em What is a metal?} In
 this article the corresponding principal question: {\em how small a
 piece of metal can be?} In other words, can a nanochain composed of e.g.
 $N=10$ Cu atoms be regarded already as a metallic system and in
 what sense? The answer is 'yes', but above a critical value of
 voltage applied to the chain and for not too large interatomic
 distance.

\section{Exact diagonalization combined with an {\em ab initio} approach}
\subsection{General features}
The second-quantization language is used when we have an interaction
between the quantum physical fields representing the classical
particles composing the system. The approach is usually formulated
in the occupation-number representation, expressing the possible
occupations of a given (\emph{complete} in quantum mechanical sense)
set of single-particle states, between which there are transitions
induced by their mutual interaction. Explicitly, the single-particle
basis $\{\Phi_i({\bf r})\}$ defines the field operator $\hat \Psi
({\bf r})$, in terms of which the many-particle Hamiltonian (or
Lagrangian) is defined. In the nonrelativistic case, the Hamiltonian
is defined as \cite{Schweber}
\[
\hat H = \int d^3{\bf r} \hat \Psi^\dagger ({\bf r}) H_1({\bf r})
\hat \Psi ({\bf r }) + \frac{1}{2} \int d^3r d^3r' \hat \Psi^\dagger
({\bf r}) \hat \Psi^\dagger ({\bf r'}) V({\bf r - r'}) \hat \Psi
({\bf r'})\hat \Psi ({\bf r})
\]
\begin{equation}
 \equiv \hat T + \hat V
 \label{Ham2Q}
\end{equation}
with
\begin{equation}
\hat \Psi ({\bf r}) = \sum_{i} \Phi_i ({\bf r}) a_i .
\label{Foper}
\end{equation}
$H_1({\bf r})$ represents Hamiltonian for a single particle in the
assembly of $N$ \emph{indistinguishable} particles, $V({\bf r -
r'})$ is the  interaction between a single pair of particles, $a_i$
is the annihilation operator of the particle in the single particle
state $\Phi_i({\bf r})$. One should underline that the basis
$\{\Phi_i({\bf r})\}$ should be {\em complete}, but otherwise
arbitrary (it does
not have to be orthogonal \cite{Feynman}).\\
\indent The complete set of $\{ \Phi_i ({\bf r}) \}$ in
(\ref{Foper}) is in practical calculations not infinite, so we have
to resort to a finite subset of $[ \Phi_i ({\bf r}) ]$ when
performing explicit computations, i.e. constructing
\emph{many-particle models}. In this manner, we make the basis {\em
incomplete} in the quantum-mechanical sense, even though the model
may be well justified on physical grounds as it may contain
principal dynamic processes involved. In the first step, we
diagonalize $H$ in the Fock space for given trial basis $[ \Phi_i
({\bf r}) ]$. Then, as a next step, one can optimize the finite
(\emph{incomplete}) subset taken by e.g. minimizing the ground state
energy $E_G \equiv \langle \Phi_0 \vert \hat H \vert \Phi_0 \rangle$
(where $\vert \Phi_0 \rangle$ is the ground state wave function in
the Fock-space representation), with respect to the selected subset
$[ \Phi_i ({\bf r}) ]$. Such procedure is highly nontrivial, since
the determination of $E_G=E_G\{\Phi_i ({\bf r}),\nabla \Phi_i ({\bf
r})\}$ requires first the diagonalization of the parameterized
Hamiltonian (\ref{Ham2Q}) in the Fock space \cite{Noone} (with the
microscopic parameters containing both $[ \Phi_i ({\bf r}) ]$ and $[
\nabla \Phi_i ({\bf r}) ]$) and, only  after that, setting up an
{\em effective wave equation} for each $ \Phi_i ({\bf r})$, with
$E_G$ treated as a functional of the trial basis $[\Phi_i ({\bf
r})]$. The resultant renormalized or {\em self-adjusted wave
equation} (SWE) should have a universal meaning to the same degree,
as has the starting Hamiltonian (\ref{Ham2Q}) in that {\em
incomplete} basis. We call this method of approach EDABI ( a
combined {\bf E}xact {\bf D}iagonalization - {\em {\bf AB} {\bf
I}nitio} approach ). In this manner, only then  the approach to the
interacting system can be regarded as \emph{completed}, particularly
in the situation when the interparticle interaction cannot be
regarded as weak, e.g. for {\em correlated systems}. Obviously, when
the correlations are weak, the approach reduces to the Hartree-Fock
approximation. This approach has been implemented so far to some
exactly soluble models of correlated electrons  and to nanochains
\cite{RycerzSpalek01,SpalekRycerz01,SpalekWojcik92} containing up to
$N=16$ atoms using as a trial basis, with adjustable Slater or
Gaussian orbitals \cite{RycerzSpalek06}. Here we present  a rather
extensive analysis of our method, applicable to both fermion and
boson systems, as well as construct explicitly the multiparticle
wave function is the simplest atomic situations. The last aspect of
the work may be applied to both atomic and molecular systems
providing e.g. a systematic approximation scheme in the
quantum-chemical calculations. In particular, the explicit
many-particle wave-function construction allows for a comparison
with the method of multiconfigurational interaction (MCI) approach
\cite{Shavitt} utilized in quantum chemistry. Simply put, we develop
a relatively straightforward (but not simple!) workable scheme,
which is \emph{applicable to both many-electron
atoms, molecules and molecular ions, as well as to clusters and nanoscopic systems.}\\
\indent One can summarize important features of our approach as
follows. First, the interaction and the single particle terms in the
many-particle Hamiltonian are treated on an equal footing. Second,
the single-particle wave function appears in a nonexplicit form in
the expression for the microscopic parameters and is determined
explicitly from the variational principle for the ground-state
energy $E_G = \langle \hat T \rangle + \langle \hat V \rangle \equiv
E_G \{ \Phi_i ({\bf r}) \}$, which leads to the self-adjusted wave
equation \cite{Schroed}. Third, the explicit construction of the
multiparticle wave function $\Psi ({\bf r_1, ... ,r_N})$ in terms of
the field operators \cite{Robertson} provides a way of systematizing
the approximation schemes. Fourth, one does not count twice the
interaction, which should be taken with care in e.g. in LDA+U
method. All of these features are particularly important for the
systems, for which the interaction cannot be treated as a
perturbation. Among such systems are cluster and various correlated
Fermi and Bose systems, particularly of low dimensionality. Probably
the most important formal feature of the present approach is that
that by reversing the order of solving the problem (treating first
the interaction in the Fock space and only then determining the
single particle wave function in the Hilbert space), we
\emph{complete} the treatment of strongly correlated systems.
Unfortunately, our approach is executable so far only in limited
number of situations. As a side result we obtain also the values of
\emph{microscopic parameters} for the parametrized models
\cite{Mattis}. In effect, the physical properties can be analyzed as
a function of interatomic distance, not only as a function of model
parameters, and thus provide
us with the global minimum for system at hand and for given interatomic distance. \\
\indent The {\em self-adjusted wave equation} (SWE), as we shall
see, is of nonlocal and nonlinear nature. Hence, it is very
difficult to solve it directly. Nonetheless, the main purpose of the
present paper is to present the solution in the closed variational
form and illustrate its character in simple situations, ranging from
the
atomic physics to nanophysics.\\
\subsection{Renormalized (self-adjusted) wave equation (SWE)}
\indent  We start with Hamiltonian (\ref{Ham2Q}) and write down in
the explicit-spin basis, in which the spin is regarded as an
additional coordinate, i.e. in the form
\[
\hat H = \sum_\sigma \int d^3{\bf r} \hat \Psi^\dagger_\sigma
({\bf r}) H_1({\bf r}) \hat \Psi_\sigma ({\bf r })
\]
\begin{equation}
+ \frac{1}{2}
\sum_{\sigma_1 \sigma_2} \int \int d^3{\bf r_1} d^3{\bf r_2} \hat
\Psi^\dagger_{\sigma_1} ({\bf r_1}) \hat \Psi^\dagger_{\sigma_2}
({\bf r_2}) V({\bf r_1 - r_2}) \hat \Psi_{\sigma_2} ({\bf
r_2})\hat \Psi_{\sigma_1} ({\bf r_1}).
\label{HamSpin}
\end{equation}
We define the spin-dependent field operator as
\begin{equation}
\hat \Psi_\sigma ({\bf r}) = \sum_{i=1}^{M} w_i ({\bf r})
\chi_\sigma a_{i \sigma}, \label{FoperSpin}
\end{equation}
where $\{ w_i({\bf r})\}$ is a {\em complete} single-particle basis
with the set of quantum numbers denoted by $i$. Note that we regard
the Hamiltonians $H_1({\bf r})$ and $V({\bf r_1 - r_2})$ as spin
independent (it is straightforward to generalize the formalism to
the case with spin-dependent Hamiltonian, i.e. when magnetic field
or spin-orbit interaction are included). Substituting
(\ref{FoperSpin}) into (\ref{HamSpin}) we obtain the usual form of
the Hamiltonian
\begin{equation}
H = \sum_{i j \sigma} t_{i j} a^\dagger_{i \sigma} a_{j \sigma} +
\frac{1}{2} \sum_{i j k l \sigma_1 \sigma_2} V_{i j k l}
a^\dagger_{i \sigma_1} a^\dagger_{j \sigma_2} a_{l \sigma_2} a_{k
\sigma_1}, \label{HamParam}
\end{equation}
with the microscopic parameters defined by
\begin{equation}
t_{i j} \equiv \langle w_i \vert H_1 \vert w_j \rangle \equiv \int
d^3{\bf r} w^\star_i({\bf r}) H_1({\bf r}) w_j({\bf r}),
\label{Tint}
\end{equation}
and
\begin{equation}
V_{i j k l} \equiv \langle w_i w_j \vert V \vert w_k w_l \rangle
\end{equation}
\[
\equiv
\int d^3{\bf r_1} d^2d^3{\bf r_2} w^\star_i({\bf r_1})
w^\star_j({\bf r_2}) V({\bf r_1 - r_2}) w_k({\bf r_1}) w_l({\bf
r_2}) \label{Vint}
\]
In the standard form (\ref{HamParam}) of the many-particle
Hamiltonian the single- and many- particle aspects of the problem
are separated in the sense that the calculation of the hopping
parameters$t_{i j}$ and  their interaction correspondants $V_{i j k
l}$, both containing the single-particle wave-functions $\{ w_i({\bf
r})\}$, is separated from the diagonalization procedure of the
Hamiltonian in the Fock space \cite{Noone2}. The latter procedure is
dependent only on the nature of the commutation relation between the
annihilation ($a_{i \sigma}$) and creation ($a^\dagger_{j \sigma}$)
operators. Thus this two-step procedure can be seen explicitly when
we calculate the system ground-state energy
\begin{equation}
E_G \equiv \langle H \rangle = \sum_{i j \sigma} t_{i j} \langle
a^\dagger_{i \sigma} a_{j \sigma} \rangle + \frac{1}{2} \sum_{i j
k l \sigma_1 \sigma_2} V_{i j k l} \langle a^\dagger_{i \sigma_1}
a^\dagger_{j \sigma_2} a_{l \sigma_2} a_{k \sigma_1} \rangle,
\label{EG}
\end{equation}
where the averaging $\langle \dots \rangle$ takes place over all
accessible occupancies of given single particle states $\vert i
\sigma_1 \rangle$, $\vert j \sigma_2 \rangle$, $\vert k \sigma_1
\rangle$, and $\vert l \sigma_2 \rangle$. Obviously, if we want to
consider all occupancies (a grand canonical ensemble), we
diagonalize $H - \mu N$, where $N$ is the total number of
particles and only a posteriori impose the conditions that $N =
\sum_{i \sigma} \langle
n_{i \sigma} \rangle$, with $n_{i \sigma} \equiv a^\dagger_{i \sigma} a_{i \sigma}$ \\
\indent  So far, the approach is standard \cite{Schweber,Feynman}.
We have proposed
\cite{SpalekPodsiadl00,RycerzSpalek01,SpalekRycerz01,SpalekWojcik92}
to close the solution (i.e. the complete calculation of e.g. $E_G$)
with the determination of the single-particle basis $\{ w_i({\bf
r})\}$ by treating the expression (\ref{EG}) as a functional of the
set of functions $\{ w_i({\bf r})\}$ and their gradients. In such
situation the renormalized (self-adjusted) wave equation is
determined from the Euler equation for the functional
\begin{equation}
E \{ w_i({\bf r})\} \equiv E_G \{ w_i({\bf r})\} - \mu N - \sum_{i
\geq j} \lambda_{i j} \left( \int d^3{\bf r} w^\star_i({\bf r})
w_j({\bf r}) - \delta_{i j} \right), \label{EFunc}
\end{equation}
where
\[
N=\sum_\sigma \int d^3({\bf r}) \langle \hat
\Psi^\dagger_\sigma({\bf r}) \hat \Psi_\sigma({\bf r}) \rangle =
\sum_{i j \sigma} \int d^3{\bf r} w^\star_i({\bf r}) w_j({\bf r})
\langle a^\dagger_{i \sigma} a_{j \sigma} \rangle,
\]
$N$ is the number of particles in the system, and $\lambda_{i j}$
are the Lagrange multipliers, when the single-particle basis is
nonorthonormal.\\
\indent The general form of this equation in the stationary case
and in the grand canonical-ensemble formalism is
\begin{equation}
\frac{\delta (E_G - \mu N)}{\delta w^\star_i({\bf r})} - \nabla
\frac{\delta (E_G - \mu N)}{\delta (\nabla w^\star_i({\bf r}))} -
\sum_{i \geq j} \lambda_{i j} w_j({\bf r}) = 0. \label{GenForm}
\end{equation}
We will make a {\em fundamental} postulate concerning this equation:
{\em As this equation does not contain explicitly the
(anti)commutation relations between the creation and annihilation
operators, it is equally valid for both fermions and bosons and
determines a rigorous, within the class of states included in the
definition of $\hat{\Psi} ({\bf r})$, wave equation for a single
particle wave function in the ground state, in the millieu of
remaining $(N-1)$ particles.} Additionally, as is implicit in the
treatment above, we have defined one global spin-quantization axis
for all single particle states used to define $\hat \Psi_\sigma
({\bf r})$. In some spin noncolinear systems this is insufficient
and will require a more refined treatment. Also, if we use the
particle conserving approach to calculate $E_G$, then we put $\mu
\equiv 0$ in (\ref{GenForm}). Likewise, for the orthonormal basis
$\{ w_i({\bf r})\}$ used to define $\hat \Psi_\sigma ({\bf r})$ in
(\ref{FoperSpin}), one puts $\lambda_{ij} \equiv 0$. In the latter
case the system (\ref{GenForm}) represents a set of Euler equations
for renormalized Wannier functions. In what follows we discuss
examples of application of this equation to fermion systems (it can
be applied to Bose systems in the same manner). But first, we define
the renormalized many particle wave function as complementing the
above renormalized single particle states $\{ w_i({\bf r})\}$ and
then discuss the differences with the MCI approach.
\subsection{Multiparticle wave function from the
second-quantization approach} 
\begin{figure}
\resizebox{0.75\columnwidth}{!}{%
  \includegraphics{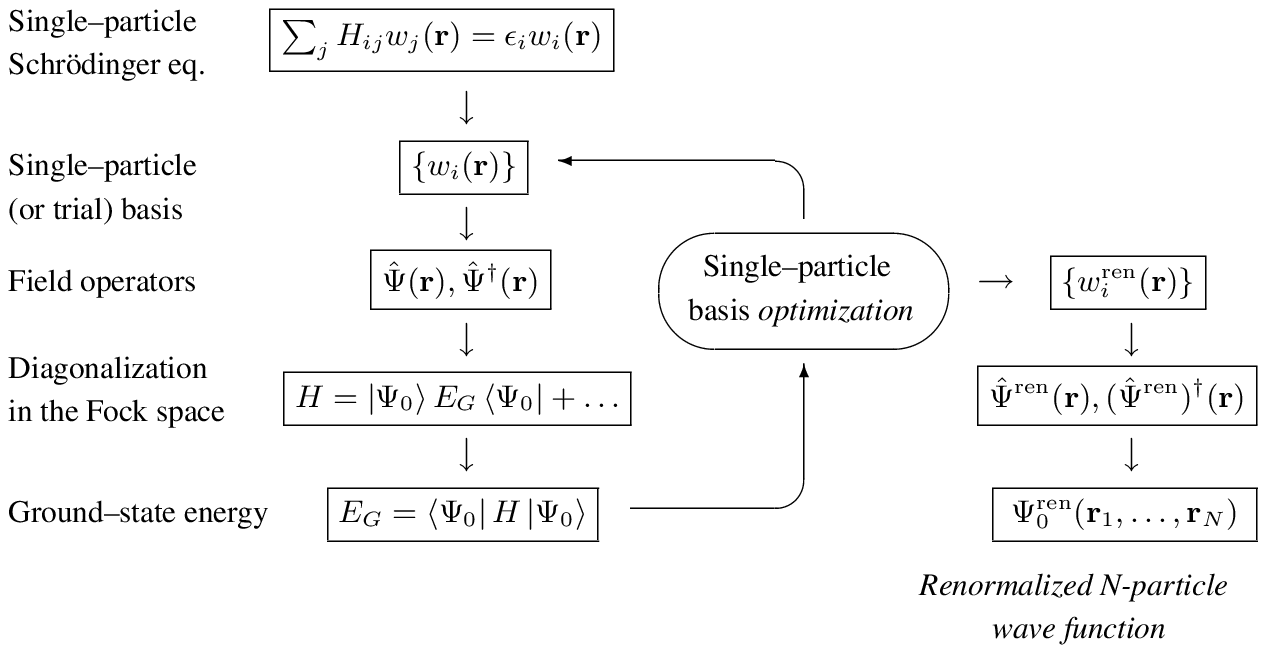}
  }
 \caption{Flow-chart describing the scheme of the EDABI method.
For details see main text. When selecting the Gaussian starting
single-particle set, the topmost block should be disregarded.}
\label{Fig1}
\end{figure}

\indent The general N-particle state
$\vert \Phi_N \rangle$ in the Fock space can be defined through
the corresponding $N$-particle wave function $\Psi_\alpha ({\bf
r_1, \dots , r_N})$ in the Hilbert space in the following manner
\cite{Schweber,Robertson}
\begin{equation}
\vert \Phi_N \rangle = \frac{1}{\sqrt{N!}} \int d^3{\bf r_1}
\dots {\bf r_N} \Psi_N ({\bf r_1, \dots , r_N}) \hat
\Psi^\dagger ({\bf r_1}) \dots \hat \Psi^\dagger ({\bf r_N}) \vert
0 \rangle, \label{NPartState}
\end{equation}
where $\vert 0 \rangle$ is the vacuum state. One can reverse this
relation and a simple algebra yields the following expression for
the wave function $\Psi_\alpha ({\bf r_1, \dots , r_N})$ in terms
of $\vert \Phi_N \rangle$
\begin{equation}
\Psi_\alpha ({\bf r_1, \dots , r_N}) = \frac{1}{\sqrt{N!}} \langle
0 \vert \hat \Psi ({\bf r_1}) \dots \hat \Psi ({\bf r_N}) \vert
\Phi_N \rangle. \label{NPartWave}
\end{equation}
In other words, to obtain the wave function in the coordinate
representation, we not only annihilate N-particles from the state
$\vert \Phi_N \rangle$, but also project out thus obtained result
onto the Fock vacuum state and normalize it by the factor
$(N!)^{-1/2}$. Usually, the formula (\ref{NPartWave}) is not used as
we proceed from first to second quantization. Now, {\em the crucial
point} is based on the observation that if we substitute in the
field operator $\hat \Psi ({\bf r})$ the renormalized wave functions
obtained from Eq.(\ref{GenForm}), then we should obtain the
renormalized field operator and as a consequence, the renormalized
multiparticle wave function $\Psi_\alpha ({\bf r_1, \dots ,r_N})$
from relation (\ref{NPartWave}). This last step of inserting
renormalized field operator completes the procedure of a formal
treatment of many-particle system, which avoids writing down
explicitly the N-particle Schr\"odinger equation. The whole approach
is schematically represented in Fig. \ref{Fig1}. This scheme
provides an exact renormalized single-particle wave function from
Eq. (\ref{GenForm}) and the exact $N$-particle wave function
provided we can diagonalize the second-quantized  model Hamiltonian
(\ref{HamParam}) for the problem at hand. We shall see next, that we
can approach the true solution also step by step.
\subsection{Finite-basis approximation for the field operator:
Difference with the multiconfiguration interaction (MCI) approach}
\indent The field operator $\hat \Psi ({\bf r})$ defined in terms of
the sum over a complete basis $\{ w_i({\bf r}) \}$ contains an
infinite number of single-particle states. We assume that, in
general, we represent the field operator by $M$ wave functions $\{
w_i({\bf r}) \}$. Explicitly,
\begin{equation}
\hat \Psi ({\bf r}) \equiv \sum_{i=1}^\infty w_i({\bf r}) a_i
\simeq \sum_{i=1}^M w_i({\bf r}) a_i, \label{ApproxFoper}
\end{equation}
with $i$ representing a complete set of quantum numbers and $M$
being a finite number. This approximation represents one of the most
\emph{fundamental} features of constructing theoretical models. The
neglected states usually represent highly excited (and thus
negligible) states of the system. We can then write the approximate
N-particle wave function ($ N \leq M$) in the following manner
\begin{equation}
\Psi_\alpha ({\bf r_1, \dots r_n}) = \frac{1}{\sqrt{N!}}
\sum_{i_1, \dots,i_N = 1}^M \langle 0 \vert a_{i_N} \dots a_{i_1}
\vert \Phi_N \rangle w_{i_1}({\bf r_1}) \dots w_{i_N}({\bf
r_N}).
\label{ApproxWave}
\end{equation}
Recognizing that within the occupation-number space spanned on the
states $\{ \vert i_k \rangle \}_{k=1\dots M}$, we have the N-particle state in the Fock space of
the form
\begin{equation}
\vert \Phi_N \rangle = \frac{1}{\sqrt{N!}} \sum_{j_1, \dots,
j_N = 1}^{M} C_{j_1 \dots j_N} a^\dagger_{j_1} \dots
a^\dagger_{j_N} \vert 0 \rangle,
\label{ApproxState}
\end{equation}
where $C_{j_1 \dots j_N}$ represents the coefficients of the
expansion to be determined from a diagonalization procedure.
Substituting (\ref{ApproxState}) to (\ref{ApproxWave}) we obtain
\begin{equation}
\Psi_\alpha({\bf r_1, \dots,r_N}) = \label{ApproxWave1}
\end{equation}
\[
\frac{1}{N!}\sum_{i_1, \dots, i_N = 1}^{M} \sum_{j_1, \dots, j_N =
1}^{M} \langle 0 \vert a_{i_1} \dots a_{i_N} a^\dagger_{j_1} \dots
a^\dagger_{j_N}\vert 0 \rangle C_{j_1 \dots j_N} w_{i_1}({\bf r_1})
\dots w_{i_N}({\bf r_N}).
\]
The expression provides $N!$ nonzero
terms each equal to $(-1)^P$, where P represents the sign of the
permutation of quantum numbers $(j_1 \dots j_N)$ with respect to
selected collection $(i_1 \dots i_N)$. In other words, we can write
that
\begin{equation}
\Psi_\alpha({\bf r_1, \dots,r_N}) = \label{ApproxWave2}
 \frac{1}{N!}\sum_{i_1, \dots,i_N = 1}^{M} C_{i_1 \dots i_N} (A,S)[
w_{i_1}({\bf r_1}) \dots w_{i_N}({\bf r_N}) ].
\end{equation}
We have the same expansion coefficients for both wave function in
the fock space $\vert \Phi_N \rangle$ and that in Hilbert
space$\Psi_\alpha({\bf r_1, \dots,r_N})$! Therefore, the above
expression represents the multiconfigurational-interaction wave
function of $N$ particles distributed among $M$ states with the
corresponding weights $C_{i_1 \dots i_N}$ for each configuration,
and $(A,S)$ represents respectively the antisymmetrization (Slater
determinant) or the symmetrization (simple product $w_{i_1}({\bf
r_1}) \dots w_{i_N}({\bf r_N})$) for the fermions and bosons,
respectively. Whereas MCI used in quantum chemistry \cite{Szabo}
bases on variational optimizations of both the coefficients $C_{i_1
\dots i_N}$ and the basis $\{ w_i({\bf r}) \}$, here the
coefficients $C$ are determined from diagonalization in the Fock
space, spanned on $M$ states in the Hilbert space and determined
from (\ref{GenForm}). The presence of SWE (\ref{GenForm}) thus
supplements the usual MCI approach. \\
\indent Summarizing, the differences between the EDABI and the MCI
methods, both of which belong to the class of multi-determinant
expansion of N-particle wave function, are threefold:
\renewcommand{\labelenumi}{\em \roman{enumi})}
\begin{enumerate}
\item {\bf Historical.} MCI evolved from variational methods of
quantum physics and chemistry to include the electronic
correlations and hence, to obtain lower value of $E_G$ by starting
from many-particle Schr\"odinger equation. EDABI represents a
procedure of calculating single-particle wave function starting
from  parameterized models of strongly correlated electrons. \item
{\bf Technical.} In MCI, we optimize simultaneously the coefficient
expressing the weights of different determinants (representing
different micro-configurations), as well as the parameters of the
trial single-particle basis. In EDABI, we diagonalize the
Hamiltonian expressed in the Fock space (with the help of either
analytic or numerical methods), combined with a simultaneous
optimization of the orbital size in the resultant ground state.
\item {\bf Essential.} In the case of analytically soluble models,
EDABI leads formally to the explicit form of the renormalized wave
equation, which represents a \emph{ nonlinear Schr\"odinger
equation of nonlocal type}. This circumstance opens up a new
direction of studies in \emph{ mathematical quantum physics}.
Additionally, it allows for a direct determination of dynamical
correlation functions, transport properties, etc. in the
convenient, second-quantization, language.
\end{enumerate}

\subsection{An interpretation of the approach: rationale behind the self-adjusted wave function in the many-body system}
Usually,  the choice of starting single-particle basis $\{ w_i({\bf
r}) \}$ is dictated by the physics of the system at hand. Since the
creation and annihilation processes are characterized only by the
quantum numbers $i$ of those starting single-particle states, one
can say that they represent a {\em particle language} characterizing
transitions between those states. The ground state energy obtained
from the diagonalization in the Fock space defines resultant
single-particle states, which can be called the {\em self-adjusted}
states, after the optimization of the single-particle wave function
has been carried out via solving the {\em self-adjusted wave
equation} (SWE)\cite{Noone3,Noone3a} or its variational version. In
other words, we allow the initial particle wave function $w_i ({\bf
r})$ to \emph{adjust} to the correlated state. In such scheme the
particle and the wave aspects of the single-particle states are
intertwinned formally, illustrating among others the particle-wave
complementarity, this time in a formal manner. For example, the Born
probability density of finding a particle is taken here as
$\sum_\sigma \langle \hat{\Psi}^\dagger_\sigma ({\bf r})
\hat{\Psi}_\sigma ({\bf r}) \rangle / N$. In what follows
we essentially illustrate the method on concrete examples and apply it to nano systems.\\
\indent One may ask {\em the basic question}: Why to revert the
usual sequence of solving the single-particle wave equation first,
and only then constructing the field operators in the
second-quantization Hamiltonian (or Lagrangian), by solving the
many-particle Hamiltonian first and only then readjusting the
orbitals? The reason for this is as follows. As said above, in many
cases (e.g. for correlated and/or low-dimensional fermionic systems)
we encounter the situation when the interaction cannot be regarded
as a perturbation and therefore should be treated \emph{at least} on
equal footing with the single-particle aspect of the problem. This
is because, in general, the interaction may change the class of the
stationary-state wave function. Such a situation is beautifully
illustrated on example of metal-insulator phase transition of the
Mott-Hubbard type, at which the metallic state represented by the
Bloch-type wave functions switches to the localized (Wannier-type)
states even though those representations are regarded as equivalent
from a single-particle point of view. In other words, the
interaction determines the particle wave function. Furthermore, the
electron as a separate entity (lepton) is preserved even in the
highly correlated milieu of other particles and therefore,
constructing the self-consistent wave equation (SWE) has a sense as
it provides its wave function adjusted to
the environment.\\
\subsection{Single-particle basis selection and the
particle-density space profiles} As we have already mentioned, the
selected single-particle basis \( w_i(\mathbf{r}) \equiv
\{\Phi_{i}(\mathbf{r})\} \) is determined from the variational
principle for \( E_{G}\{\Phi_{i}(\mathbf{r}), \nabla
\Phi_{i}(\mathbf{r})\} \), and it should satisfy \textit{the
self-adjusted wave equation}. Here we solve this equation
variationally and take a trial basis \(
\{\Phi_{i}(\alpha;\mathbf{r})\} \) dependent on a finite number of
parameters \( \alpha \equiv \{\alpha_{p}\}_{p=1,...,K} \). Moreover,
if the basis is orthonormal, then the equation can be simplified
then to the form
\begin{equation}\label{eqromek1}
\frac{\delta(E_{G}-\mu N)}{\delta \alpha}=0.
\end{equation}
In most applications we select adjustable atomic orbitals as the
basic functions \( \{\Phi_{i} (\alpha;\mathbf{r})\} \), which need
to be orthonormalized with a special kind of orthonormalization
procedure. Additionally, if the number of particles \( N \) is
conserved by the Hamiltonian \( H \) then the term \( \mu N \) in
(\ref{eqromek1}) is absent, and the equation reduces to the
ordinary energy minimization \( E_{G}\equiv\langle H \rangle \)
with respect to the variation parameters \( \{ \alpha_p \}\).

\subsection{A remainder: Wannier basis for an extended periodic system}

Let us consider the multi-particle systems for a single-band case
for atoms located at positions $\{ {\bf R}_i \}$ and the basic wave
functions \( \{\Phi_{i}(\alpha;\mathbf{r})\} \) of the form \(
\{\Phi(\alpha;\mathbf{r}-\mathbf{R} _{i})\} \). The
orthonormalization procedure for periodic structures can be obtained
by starting with the expression for the \textit{Bloch} function

\begin{equation}
\Phi_{\mathbf{q}}(\alpha;\mathbf{r})=N_{\mathbf{q}}\sum^{N}_{j=1}e^{i\mathbf{q}\cdot\mathbf{R}_{j}}
\Phi_{j}(\alpha;\mathbf{r}),
\end{equation}
where \( N_{\mathbf{q}} \) is a normalization coefficient. The normalization condition then takes form
\begin{equation}
\begin{array}{c}
\langle\Phi_{\mathbf{q}}(\alpha;\mathbf{r})|\Phi_{\mathbf{q'}}(\alpha;\mathbf{r})\rangle=
N^{*}_{\mathbf{q}}N_{\mathbf{q'}}\sum_{jl}e^{-i\mathbf{q}\cdot\mathbf{R}_{j}+i\mathbf{q'}\cdot\mathbf{R}_{l}}
\langle\Phi_{j}(\alpha;\mathbf{r})|\Phi_{l}(\alpha;\mathbf{r})\rangle \\
=N^{*}_{\mathbf{q}}N_{\mathbf{q'}}\sum_{j}e^{i(\mathbf{q'}-\mathbf{q})\cdot\mathbf{R}_{j}}
\sum_{l}e^{i\mathbf{q'}\cdot(\mathbf{R}_{l}-\mathbf{R}_{j})}S_{lj}
\equiv\delta_{\mathbf{q}\mathbf{q'}},
\end{array}
\end{equation}
where \( S_{lj} \) is an overlap integral. The last sum does not
depend on the relative distance, hence
\begin{equation}
\langle\Phi_{\mathbf{q}}(\alpha;\mathbf{r})|\Phi_{\mathbf{q'}}(\alpha;\mathbf{r})\rangle=
N^{*}_{\mathbf{q}}N_{\mathbf{q'}}N\delta_{\mathbf{q}\mathbf{q'}}
\sum_{l}e^{i\mathbf{q'}\cdot(\mathbf{R}_{l}-\mathbf{R}_{j})}S_{lj}
\equiv\delta_{\mathbf{q}\mathbf{q'}},
\end{equation}
and thus
\begin{equation}
|N_{\mathbf{q}}|=(N\sum_{l}e^{i\mathbf{q}\cdot(\mathbf{R}_{l}-\mathbf{R}_{j})}S_{lj})^{-1/2}.
\end{equation}

As we can see, the above wave functions \(
\{\Phi_{\mathbf{q}}(\alpha;\mathbf{r})\} \) are the \textit{Bloch}
functions, from which we can construct the \textit{Wannier}
functions \( \{w_{l}(\alpha;\mathbf{r})\} \) in the usual manner,
i.e.

\begin{equation}
\begin{array}{c}
w_{l}(\alpha;\mathbf{r})=\frac{1}{\sqrt{N}}\sum_{\mathbf{q}}e^{-i\mathbf{q}\cdot\mathbf{R}_{l}}
\Phi_{\mathbf{q}}(\alpha;\mathbf{r}) \\
=\frac{1}{\sqrt{N}}\sum_{\mathbf{q}}e^{-i\mathbf{q}\cdot\mathbf{R}_{l}}
N_{\mathbf{q}}\sum_{j}e^{i\mathbf{q}\cdot\mathbf{R}_{j}}\Phi_{j}(\alpha;\mathbf{r}) \\
=\sum_{j}\Phi_{j}(\alpha;\mathbf{r})(\frac{1}{\sqrt{N}}\sum_{\mathbf{q}}N_{\mathbf{q}}
e^{-i\mathbf{q}\cdot(\mathbf{R}_{l}-\mathbf{R}_{j})}) \\
\equiv\sum_{j}\beta_{lj}\Phi_{j}(\alpha;\mathbf{r}),
\end{array}
\end{equation}
where
\begin{equation}
\beta_{lj}=\frac{1}{N}\sum_{\mathbf{q}}\frac{e^{-i\mathbf{q}\cdot(\mathbf{R}_{l}-\mathbf{R}_{j})}}
{\sqrt{\sum_{m}e^{-i\mathbf{q}\cdot(\mathbf{R}_{n}-\mathbf{R}_{m})}S_{mn}}}.
\end{equation}

The orthonormal Wannier set \( \{w_{l}(\alpha;\mathbf{r})\} \)
obtained with the help of the above procedure is thus a set of
linear combinations of the originally non-orthonormal atomic
orbitals. Moreover, the coefficients \( \{\beta_{lj}\} \) depend
only on \( \mathbf{R}_{l}-\mathbf{R}_{j} \) and satisfy the relation
\( \beta_{lj}=\beta^{*}_{jl} \).

\subsection{Wannier functions for finite systems}

The above procedure can be applied with modifications to finite
cluster systems. We start from the decomposition

\begin{equation}\label{eqromek2}
w_{l}(\alpha;\mathbf{r})=\sum^{M}_{j=1}\beta_{lj}\Phi_{j}(\alpha;\mathbf{r}),
\end{equation}
with the normalization condition in the form
\begin{equation}
\begin{array}{c}
\langle w_{l}(\alpha;\mathbf{r})|w_{l'}(\alpha;\mathbf{r}) \rangle=
\sum_{jk}\beta^{*}_{lj}\beta_{l'k}\langle\Phi_{j}(\alpha;\mathbf{r})|\Phi_{k}(\alpha;\mathbf{r})\rangle \\
=\sum_{jk}\beta^{*}_{lj}\beta_{l'k}S_{kj}\equiv\delta_{ll'}.
\end{array}
\end{equation}
In the matrix language, this condition can be rewritten as
\begin{equation}
\beta S \beta^{+}=1,
\end{equation}
or, equivalently as
\begin{equation}
\beta^{+}\beta=S^{-1}.
\end{equation}
We choose the \( \beta \) matrix in the form \( \beta=\beta^{+} \);
in effect this choice leads to the relation
\begin{equation}
\beta=S^{-1/2}.
\end{equation}

The above method is known as the L\"owdin method of determining
the orthonormal basis, and for the overlap integral matrix becomes
unity if elements of the system are separated from each other at
large distances. So, the \( \beta \) matrix then can be written
down as
\begin{equation}
\beta=1+\sum^{\infty}_{n=1}\frac{(-1)^{n}\triangle^{n}}{2^{n}n!}\prod^{n}_{k=1}(2k-1)
=1+\sum^{\infty}_{n=1}\triangle^{n}\prod^{n}_{k=1}(\frac{1}{2k}-1),
\end{equation}
where \( \triangle=S-1 \). However, the above series does not
converge in the tight binding approach and it has to be modified to
the form
\begin{equation}
\beta=\left( \frac{(1+C) S}{1+C} \right)^{-1/2},
\end{equation}
and thus
\begin{equation}
\beta=(1+C)^{-1/2}[1+\sum^{\infty}_{n=1}\triangle^{n}_{C}\prod^{n}_{k=1}(\frac{1}{2k}-1)],
\end{equation}
where \( \triangle_{C} \equiv S/(1+C)-1 \). The parameter $C$ allows
us to manipulate convergence of the series, for it appears in the
convergence condition
\begin{equation}
\|\triangle_{C}\|<1,
\end{equation}
that needs to be satisfied for any kind of the \( \|\cdot\| \) metrics. Therefore, we choose
the metrics
\begin{equation}
\|\triangle_{C}\|_{\infty} \equiv \max_{ij}\{|\frac{S_{ij}}{1+C}-\delta_{ij}|\}
=\max\{\frac{C}{1+C};\frac{\max_{i \neq j}|S_{ij}|}{1+C}\}.
\end{equation}
As we can see, the metrics satisfies the condition when the parameter
$C$ is equal to
\begin{equation}
C=\max_{i \neq j}|S_{ij}|.
\end{equation}
The method, we have just shown, can be applied to both single-band
and to multiple-band systems. The method can also be applied to
disordered systems. In what follows we will select as a starting
atomic basis $\{ \Phi_i ({\bf r}) \}$ either Slater or Gaussian
(STO-3G) basis when constructing the Wannier function, and
subsequently, optimize their size in the correlated state.

\subsection{Particle density profiles in space}

In the previous Section we dealt with the N-particle wave function
\( \Psi(\mathbf{r}_{1},..., \mathbf{r}_{N}) \). Here we show how
it can be applied to the evaluation of particle density \(
n(\mathbf{r}) \). We start with the usual definition of the
probability density for a single particle:

\begin{equation}
\rho(\mathbf{r}_{N})=\int d^{3}\mathbf{r}_{1}...\mathbf{r}_{N-1}
|\Psi(\mathbf{r}_{1},...,\mathbf{r}_{N})|^{2}.
\end{equation}
We utilize now the expression (4) for the wave function we
obtained in Section II. In effect,

\begin{equation}
|\Psi(\mathbf{r}_{1},...,\mathbf{r}_{N})|^{2}=\frac{1}{N!}
\langle \Phi_{N}|\widehat{\Psi}^{+}(\mathbf{r}_{N})...\widehat{\Psi}^{+}(\mathbf{r}_{1})|0 \rangle
\langle 0|\widehat{\Psi}(\mathbf{r}_{1})...\widehat{\Psi}(\mathbf{r}_{N})|\Phi_{N} \rangle.
\end{equation}
We have the particle-number conservation and hence we can rewrite (37) as
\begin{equation}
|\Psi(\mathbf{r}_{1},...,\mathbf{r}_{N})|^{2}=\frac{1}{N!}
\langle \Phi_{N}|\widehat{\Psi}^{+}(\mathbf{r}_{N})...\widehat{\Psi}^{+}(\mathbf{r}_{1})
\widehat{\Psi}(\mathbf{r}_{1})...\widehat{\Psi}(\mathbf{r}_{N})|\Phi_{N} \rangle.
\end{equation}
This is because we can insert \( \sum_{N}|N \rangle\langle N| \) instead of \( |0 \rangle\langle 0|
\). The expression for the field operator \( \widehat{\Psi}(\mathbf{r}) \), defined with the help
of the orthonormal basis \( \{w_{i}(\mathbf{r})\} \) in (4), leads to the
following relation

\begin{equation}
\widehat{\Psi}^{+}(\mathbf{r})\widehat{\Psi}(\mathbf{r})=
\sum_{i_{1}\sigma_{1}i_{2}\sigma_{2}}w^{*}_{i_{1}}(\mathbf{r})\chi^{*}_{\sigma_{1}}
w_{i_{2}}(\mathbf{r})\chi_{\sigma_{2}}a^{+}_{i_{1}\sigma_{1}}a_{i_{2}\sigma_{2}},
\end{equation}
and thus
\begin{equation}
\widehat{\Psi}^{+}(\mathbf{r})\widehat{\Psi}(\mathbf{r})=
\sum_{i_{1}i_{2}\sigma}w^{*}_{i_{1}}(\mathbf{r})w_{i_{2}}(\mathbf{r})
a^{+}_{i_{1}\sigma}a_{i_{2}\sigma}.
\end{equation}
Therefore, we can apply the above relation to the expression for \( \rho(\mathbf{r}_{N}) \).
Namely, by noting that

\begin{equation}
\rho(\mathbf{r}_{N})=\int d^{3}\mathbf{r}_{2}...\mathbf{r}_{N-1}
(\int d^{3}\mathbf{r}_{1}|\Psi(\mathbf{r}_{1},...,\mathbf{r}_{N})|^{2}),
\end{equation}
we can obtain the following series of helpful identities:
\begin{equation}
\begin{array}{c}
\int d^{3}\mathbf{r}_{1}|\Psi(\mathbf{r}_{1},...,\mathbf{r}_{N})|^{2} \\
=\frac{1}{N!}\int d^{3}\mathbf{r}_{1}
\langle \Phi_{N}|\widehat{\Psi}^{+}(\mathbf{r}_{N})...\widehat{\Psi}^{+}(\mathbf{r}_{1})
\widehat{\Psi}(\mathbf{r}_{1})...\widehat{\Psi}(\mathbf{r}_{N})|\Phi_{N} \rangle \\
=\frac{1}{N!}\sum_{i_{1}i_{2}\sigma}(\int d^{3}\mathbf{r}_{1}
w^{*}_{i_{1}}(\mathbf{r}_{1})w_{i_{2}}(\mathbf{r}_{1}))
\langle \Phi_{N}|\widehat{\Psi}^{+}(\mathbf{r}_{N})...a^{+}_{i_{1}\sigma}
a_{i_{2}\sigma}...\widehat{\Psi}(\mathbf{r}_{N})|\Phi_{N} \rangle \\
=\frac{1}{N!}
\langle \Phi_{N}|\widehat{\Psi}^{+}(\mathbf{r}_{N})...\widehat{\Psi}^{+}(\mathbf{r}_{2})
\sum_{i\sigma}n_{i\sigma}
\widehat{\Psi}(\mathbf{r}_{2})...\widehat{\Psi}(\mathbf{r}_{N})|\Phi_{N} \rangle \\
=\frac{1}{N!}
\langle \Phi_{N}|\widehat{\Psi}^{+}(\mathbf{r}_{N})...\widehat{\Psi}^{+}(\mathbf{r}_{2})
\widehat{\Psi}(\mathbf{r}_{2})...\widehat{\Psi}(\mathbf{r}_{N})|\Phi_{N} \rangle.
\end{array}
\end{equation}
The expectation value of the particle-number operator $\sum_{i\sigma}n_{ i \sigma }$ is equal to unity for
the multi-particle state \( \hat{\Psi}(\mathbf{r}_{2})...\hat{\Psi}(\mathbf{r}_{N})
|\Phi_{N}\rangle\sim|\Phi_{N=1}\rangle \). Obviously, the subsequent procedure applied \( N-2 \)
times leads to the net result
\begin{equation}
\rho(\mathbf{r}_{N})=\frac{1}{N}
\langle \Phi_{N}|\widehat{\Psi}^{+}(\mathbf{r}_{N})
\widehat{\Psi}(\mathbf{r}_{N})|\Phi_{N} \rangle.
\end{equation}
This result can be understood if we introduce explicitly the
particle-density operator
\begin{equation}
\widehat{n}(\mathbf{r})\equiv\widehat{\Psi}^{+}(\mathbf{r})\widehat{\Psi}(\mathbf{r}),
\end{equation}
and thus the particle density is
\begin{equation}
n(\mathbf{r})\equiv N\rho(\mathbf{r})=\langle \Phi_{N}|
\widehat{\Psi}^{+}(\mathbf{r})\widehat{\Psi}(\mathbf{r})|\Phi_{N} \rangle.
\end{equation}

Hence, we obtain the explicit expression for the density of particles using Eq.(4), i.e.
\begin{equation}
n(\mathbf{r})=\sum_{i_{1}i_{2}\sigma}w^{*}_{i_{1}}(\mathbf{r})w_{i_{2}}(\mathbf{r})
\langle \Phi_{N}|a^{+}_{i_{1}\sigma}a_{i_{2}\sigma}|\Phi_{N} \rangle,
\end{equation}
or, equivalently
\begin{equation}
\label{nodr47}
n(\mathbf{r})=\sum_{i}|w_{i}(\mathbf{r})|^{2}\langle n_{i} \rangle+
\sum_{i_{1} \neq i_{2}}w^{*}_{i_{1}}(\mathbf{r})w_{i_{2}}(\mathbf{r})
\sum_{\sigma}\langle a^{+}_{i_{1}\sigma}a_{i_{2}\sigma} \rangle,
\end{equation}
where averages \( \langle ... \rangle \) are taken with the N-particle ground state \( |\Phi_{N}
\rangle \).

The first of two terms represents the contribution of the particle-number operator to the total
density of particles and appears in the \textit{Hartree-Fock} approach as the only term. The second
term can provide a significant contribution when the averages \( \langle a^{+}_{i_{1}\sigma}
a_{i_{2}\sigma}\rangle \) is of the same magnitude as the first
contribution. Explicitly, if the N-particle function has the form
of a simple determinant
\begin{equation}
\Psi_{i_1 \dots i_N} \left( {\bf r_1}, \dots , {\bf r_N} \right) =
\frac{1}{\sqrt{N!}} \left| \begin{array}{ccc}
w_{i_1} ({\bf r_1}) & \ldots & w_{i_N} ({\bf r_1}) \\
\vdots & \ddots & \vdots \\
w_{i_1} ({\bf r_N}) & \ldots & w_{i_N} ({\bf r_N})
\end{array} \right|,
\label{Determ}
\end{equation}
then the occupancies $\langle n_i \rangle \equiv 1$ in
(\ref{nodr47}) an $\langle a^\dagger_{i_1 \sigma} a_{i_2 \sigma}
\rangle \equiv 0$ for $i_1 \neq i_2$. This is not the case when the
multi-configurational form (\ref{NPartWave}) of $\Psi ({\bf r_1},
\dots , {\bf r_N})$ is taken, as we shall see in the following. In
the remaining part of this paper we discuss the results obtained
within this method of approach for various nanosystems.

\section{Atomic systems and nanoclusters}
In this Section we start with the simplest examples of lightest
atoms and ions to illustrate the specific features of the EDABI
method, as well as to provide an elementary example of the
renormalized wave equation. In particular, we systematically enrich
up the trial basis to build up the Fock space.
\subsection{A didactic example: He atom}
We start by selecting as $\{ w_i({\bf r})\}$ just two 1s-like states
for the He atom $\Phi_\sigma({\bf r}) = (\alpha^3/\pi)^{1/2}
\exp(-\alpha r) \chi_\sigma$, where $\alpha$ is the effective
inverse radius of the states. In other words, the simplest trial
field operator is of the form
\begin{equation}
\hat{\Phi}({\bf r}) = \Phi_\uparrow ({\bf r}) a_\uparrow +
\Phi_\downarrow({\bf r}) a_\downarrow , \label{Psi1s}
\end{equation}
where $a_\sigma$ is the annihilation operator of particle in the
state $\Phi_\sigma({\bf r})$. The Hamiltonian in the second
quantization for this two-element basis has then the form
\begin{equation}
H = \epsilon_a ( n_\uparrow +n_\downarrow ) + U n_\uparrow
n_\downarrow ,
\label{HHe1s}
\end{equation}
where $n_\uparrow = a^\dagger_\uparrow a_\uparrow$, whereas
\begin{equation}
\epsilon_a = \langle \Phi_\sigma \vert H_1  \vert \Phi_\sigma
\rangle ,
\end{equation}
and
\begin{equation}
U = \langle \Phi_\sigma \Phi_{\overline{\sigma}} \vert V  \vert
\Phi_{\overline{\sigma}} \Phi_\sigma \rangle
\end{equation}
are the matrix elements of the single-particle part defined as
\begin{equation}
H_1 = -\frac{\hbar^2}{2 m} \nabla_1^2 -\frac{\hbar^2}{2 m}
\nabla_2^2 - \frac{2 e^2}{\kappa_0 r_1} - \frac{2 e^2}{\kappa_0
r_2} \stackrel{a.u.}{\equiv} -\nabla_1^2 - \nabla_2^2 -
\frac{4}{r_1} - \frac{4}{r_2}
\end{equation}
and of the Coulomb interaction
\begin{equation}
V = \frac{e^2}{\kappa_0 |{\bf r_1} - {\bf r_2}|}
\stackrel{a.u.}{\equiv} \frac{2}{|{\bf r_1} - {\bf r_2}|} ,
\end{equation}
with the corresponding definitions in atomic units after the second
equality sign. The only eigenvalue of (\ref{HHe1s}) is obtained for
the state $a^\dagger_\uparrow a^\dagger_\downarrow |0>$ and is $E =
2 \epsilon_a + U$. This total energy is then minimized with respect
to $\alpha$ to obtain the well-known variational estimate of both
$\alpha$ and the ground state energy $E_G$, as discussed before
\cite{SpalekPodsiadl00}. However, we may look at the problem
differently. As the approximate  field operator can be defined in an
arbitrary basis, we may regard the eigenvalue $E$ as a functional of
$\Phi_\sigma({\bf r})$, since the functions are under the integral
expressions. Therefore, the true wave function is obtained from the
Euler equation for the functional under the proviso that the wave
function is normalized. This means that we minimize the functional
\begin{equation}
E\{ \Phi_\sigma({\bf r}) \} = \sum_\sigma \int d^3r
\Phi^*_\sigma({\bf r}) H_1({\bf r}) \Phi_\sigma({\bf r})
\end{equation}
\[
+ \frac{1}{2} \sum_\sigma \int d^3r d^3r' |\Phi_\sigma({\bf r})|^2
V_{12}({\bf r} - {\bf r'}) |\Phi_{\overline{\sigma}}({\bf r})|^2 .
\]
In effect, the Euler equation take the form of he unrestricted
Hartree-Fock equations for $\Phi_\sigma ({\bf r})$
\begin{equation}
\left( \nabla^2 - \frac{2 e^2}{\kappa_0 r} \right)\Phi_\sigma({\bf
r})+ \Phi_\sigma({\bf r}) \int d^2r' \frac{e^2}{\kappa_0 |{\bf r}
-{\bf r'}|} |\Phi_{\overline{\sigma}}({\bf r'})|^2 = \lambda
\Phi_\sigma({\bf r}) . \label{HFeqn}
\end{equation}
Thus we can see that taking in the simplest case just two spin
orbitals we obtain either well-known variational estimate \cite{Bethe}
for
$\alpha$ and $E_G$  for He atom: $\alpha = 27/(16 a_0)$ and $E_G =
-5.695 Ry$, where $a_0 \simeq 0.53$\AA \ is the 1s Bohr orbit
radius.\\
\indent Obviously, the proposed expression (\ref{Psi1s}) of the
field operator is the simplest one, though it leads to nontrivial
results even though  the trial atomic basis $\{ \Phi_\sigma({\bf r})
\}$ is far from being complete in the quantum-mechanical sense.
However, we can improve systematically on the basis by selecting a
reacher basis than that in (\ref{Psi1s}). The further step in this
direction is discussed next.
\subsection{The basis enrichment for the lightest atoms and ions: He, Li}
We can expand the field operator in the basis involving the higher
order irreducible representations of the rotation group with
$n=2$, which in the variational scheme involve including, apart
from the $\Psi_{1s}({\bf r})$ orbital, also the higher
$\Psi_{2s}({\bf r})$ and $\Psi_{2pm}({\bf r})$, with $m = \pm 1 ,
0$ (i.e. the next shell); all of them involving the adjustment of
the corresponding orbital characteristics $\alpha_i$, $i = 1s, 2s$
and $2pm$. The field operator is then
\[
\hat{\Psi}({\bf r}) = \sum_\sigma \left[ w_{1s} ({\bf r}) \chi_{1
\sigma} a_{1 \sigma} + w_{2s}({\bf r}) \chi_{2 \sigma} a_{2 \sigma}
+ \sum_{m = -1}^{+1} w_{2pm}({\bf r}) \chi_{m \sigma} a_{2pm \sigma}
\right]
\]
\begin{equation}
\equiv \sum_{i \sigma} w_i({\bf r}) \chi_{i
\sigma} a_{i \sigma} , \label{Psi1s2s2pm}
\end{equation}
where $w_i({\bf r})$ are orthogonalized orbitals obtained from the
nonorthogonal atomic \cite{Noone4} basis $\{ \Psi_i({\bf r}) \}$
in a standard manner. The Fock space spanned on $2 + 2 + 6 = 10$
trial spin orbitals contains $D = { 2 M \choose N_e }$ dimensions,
where $M = 5$ now and $N_e = 2, 3$ is the number of electrons for
He and Li, respectively. This means that $D = 45$ and $120$ in
those two cases and we have to diagonalize the Hamiltonian
matrices of that size to determine the ground and the lowest
excited states.
\begin{table}
\caption{Optimized Bohr-orbit radii $a_i = \alpha_i^{-1}$ of 1s, 2s,
and 2p orbits (in units of $a_0$), the overlap $S$ between
renormalized 1s and 2s states, and the ground state energy for the
lightest atoms and ions (five Slater orbitals taken).}
\label{Toptradii} \begin{center}
\begin{tabular}{cccccc}
\hline \hline
 & $a_{1s}$ & $a_{2s}$ & $a_{2p}$ & $S$ & $E_G$ (Ry) \\
\hline $H$ & 1 & 2 & 2 & 0 & -1 \\
 $H^-$ & 0.9696 & 1.6485 & 1.017 & -0.1 & -1.0487 \\
 $He$  & 0.4274 & 0.5731 & 0.4068 & -0.272 & -5.79404 \\
 $He^-$ & 1.831 & 1.1416 & 0.4354 & -0.781 & -5.10058 \\
 $Li$ & 0.3725 & 1.066 & 0.2521 & 0.15 & -14.8334 \\
 $Be^+$ & 0.2708 & 0.683 & 0.1829 & 0.109 & -28.5286 \\
\hline \hline
\end{tabular}
\end{center}
\end{table}
One should note that we construct and subsequently diagonalize the
$<i|H|j>$ matrix in the Fock space for (fixed) parameters
$\epsilon_a, t_{ij}$, and $V_{kl}$. After the diagonalization has
been carried out, we readjust the wave function and start the
whole procedure again until the absolute minimum is reached (cf. Fig. \ref{Fig1}).\\
\indent By diagonalizing the corresponding Hamiltonian matrices and
subsequently, minimizing the lowest eigenvalue with respect to the
parameters $\alpha_i$ - the inverse radial extensions of the
corresponding wave functions, we obtain the results presented in
Table \ref{Toptradii} (the values $a_{2pm}$ are all equal within the
numerical accuracy $\sim10^{-6})$. For example, the ground state
energy of He is $E_G = -5.794$ Ry, which is close to the accepted
"exact" value \cite{Bethe} $-5.8074$, given  the simplicity of our
approach. Further improvement is feasible by either including the
$n=3$ states or by resorting to the Gaussian basis; these are not
performed in this Part (see next Sections).
Instead, we elaborate on two additional features.\\
\indent First, we can represent the ground-state two-particle
spin-singlet wave function for He atom taking $\hat{\Psi}({\bf r|})$
in the form (\ref{Psi1s2s2pm}), which has the following form
\cite{ThesisEG}
\begin{eqnarray}
|\Psi^{He}_0> = ( -0.799211 a^+_{1s\downarrow} a^+_{1s\uparrow} +
0.411751 a^+_{1s\downarrow} a^+_{2s\uparrow} - 0.411751
a^+_{1s\uparrow} a^+_{2s\downarrow}
\\ \nonumber
- 0.135451a^+_{2s\downarrow} a^+_{2s\uparrow} + 0.0357708
a^+_{2p0\downarrow} a^+_{2p0\uparrow} + 0.0357641
a^+_{2p1\downarrow} a^+_{2p-1\uparrow}
\\ \nonumber
- 0.0357641 a^+_{2p1\uparrow}
a^+_{2p-1\downarrow} ) |0>,
\end{eqnarray}
Similarly, the $S^z = + 1/2$ state for Li atom is of the form
\begin{eqnarray}
|\Psi^{Li}_0> = ( 0.997499 a^+_{1s\downarrow} a^+_{1s\uparrow}
a^+_{2s\uparrow} -0.0570249 a^+_{1s\uparrow} a^+_{2s\downarrow}
a^+_{2s\uparrow}
\\ \nonumber
+ 0.0039591 a^+_{1s\uparrow} a^+_{2p0\downarrow} a^+_{2p0\uparrow} +
0.00395902 a^+_{1s\uparrow} a^+_{2p1\downarrow} a^+_{2p-1\uparrow}
\\ \nonumber
-0.00395894 a^+_{1s\uparrow} a^+_{2p1\uparrow} a^+_{2p-1\downarrow}
- 0.023783 a^+_{2s\uparrow} a^+_{2p0\downarrow} a^+_{2p0\uparrow}
\\ \nonumber
-0.0237806 a^+_{2s\uparrow} a^+_{2p1\downarrow} a^+_{2p-1\uparrow}
+0.0237806 a^+_{2s\uparrow} a^+_{2p1\uparrow} a^+_{2p-1\downarrow}
) |0> .
\end{eqnarray}
We see that the probability of encountering the configuration $1s^2$
in He is less than $2/3$, whereas the corresponding configuration
$1s^2 2s$ for Li almost coincides with that for the hydrogenic-like
picture. The reason for the difference is that the overlap integral
between $1s$ and $2s$ states $S = <1s|2s>$ in the former case is
large and the virtual transitions $1s \rightleftharpoons 2s$ do not
involve a substantial change in of the Coulomb energy. Those wave
functions can be used to evaluate any ground-state characteristic by
calculating $<\Psi_G|\hat{O}|\Psi_G>$ for $\hat{O}$ represented in
the 2nd quantized form. For example, the atom dipole moment operator
is $\hat{\bf d} = e \int d^3r
\hat{\Psi}^\dagger({\bf r}) {\bf x} \hat{\Psi}({\bf r})$, etc.\\
\indent The second feature is connected with determination of the
microscopic parameters $V_{ijkl}$ in our Hamiltonian, since their
knowledge is crucial for atomic cluster calculations, as well as
the determination of physical properties of extended systems as a
function of the lattice parameter.
 Namely, we
can rewrite the Hamiltonian (\ref{HamParam}) for the case of single
atom within the basis (\ref{Psi1s2s2pm}) in the form
\[
H = \sum_{i \sigma} \epsilon_i n_{i \sigma} + t \sum_{\sigma}
\left( a^\dagger_{2 \sigma} a_{1 \sigma} + a^\dagger_{1 \sigma}
a_{2 \sigma} \right) + \sum_{i = 1}^{5} U_i
   n_{i \uparrow} n_{i \downarrow} +
\frac{1}{2} \sum_{i \neq j} K_{i j} n_i n_j
\]
\begin{equation}
 -\frac{1}{2}\sum_{ i
\neq j} J_{i j} \left( {\bf S_i \cdot S_j} - \frac{1}{2} n_i n_j
\right) + \sum_{i \neq j} J_{ij} a^\dagger_{i \uparrow} a^\dagger_{i \downarrow} a_{j \downarrow} a_{j \uparrow}
+ \sum_{i \neq j \sigma} V_{ij} n_{i \overline{\sigma}} a^\dagger_{i \sigma}  a_{j \sigma}
.
\label{ParHam}
\end{equation}
t is the hopping integral between 1s and 2s states, $U_i$ are the
intraorbital Coulomb interactions, $K_{i j}$ are their
interorbital correspondants, $V_{i j}$ is the so-called correlated
hopping integral, and $J_{i j}$ is the direct exchange integral,
for states $i$ and $j = 1,\ldots ,5$. The principal parameters for
the atoms and selected ions are provided in Table \ref{Tmicropar}.
We can draw the following interpretation from this analysis. The
calculated energy difference $\Delta E$ for $He$ between the
ground state singlet and the first excited triplet is $-2.3707 -
(-5.794) \simeq 3.423 Ry$ ( the singlet $1s\uparrow 2s\downarrow$
is still 1 Ry higher). The corresponding energy of the Coulomb
interaction in $1s^2$ configuration is $U_1 = 3.278$, the value
comparable to $\Delta E$. Additionally, the Coulomb interaction in
$1s\uparrow 2s\downarrow$ state is $\approx 1.5 Ry$, a
substantially lower value. The relative energetics tells us why we
have a substantial admixture of the excited $1s\uparrow
2s\downarrow$ state to the singlet $1s^2$. In other words, a
substantial Coulomb interaction ruins hydrogenic-like scheme,
although the actual values could be improved further by enriching
still the trial basis.
\begin{table}
\caption{Microscopic parameters (in Ry) of the selected atoms and
ions all quantities are calculated for the orthogonalized atomic
states. t is the 1s-2s hopping magnitude, $U_i$ is the intraorbital
 Coulomb interaction (i=1s(1), 2s(2), m=0(3), and m=$\pm1$(p)), whereas $K_{ij}$ and $J_{ij}$ are the interorbital Coulomb and exchange interaction parameters. }
\label{Tmicropar}
\begin{tabular}{ c  c  c c c c  c c c c c c}
\hline \hline
 & $t$ & $U_1$ & $U_2$ & $U_3$ & $U_p$ & $K_{12}$ & $K_{13}$ & $K_{23}$ & $J_{12}$ & $J_{13}$ & $J_{23}$  \\
\hline $H^-$ & 0.057 & 1.333 & 0.369 & 0.77 & 0.728 & 0.519 & 0.878 & 0.457 & 0.061 & 0.138 & 0.035 \\
 $He$ & 1.186 & 3.278 & 1.086 & 1.924 & 1.821 & 1.527 & 2.192 & 1.289 & 0.212 & 0.348 & 0.115 \\
 $He^-$ & -1.1414 & 1.232 & 0.764 & 1.798 & 1.701 & 0.929 & 1.421 & 1.041 & 0.269 & 0.28 & 0.102 \\
 $Li$ & -0.654 & 3.267 & 0.533 & 3.105 & 2.938 & 0.749 & 3.021 & 0.743 & 0.06 & 0.606 & 0.014 \\
 $Be^+$ & -0.929 & 4.509 & 0.869 & 4.279 & 4.049 & 1.191 & 4.168 & 1.175 & 0.105 & 0.837 & 0.025 \\
\hline \hline
\end{tabular}
\end{table}
\\
\indent One may ask how the renormalized wave equation would look
in the present situation? The answer to this question is not brief
already for the basis containing $M=5$ starting states $\{
w_i({\bf r}) \}$; we return to this question in a slightly simpler
case of molecular states we consider next.

\begin{figure}
\resizebox{0.75\columnwidth}{!}{%
  \includegraphics{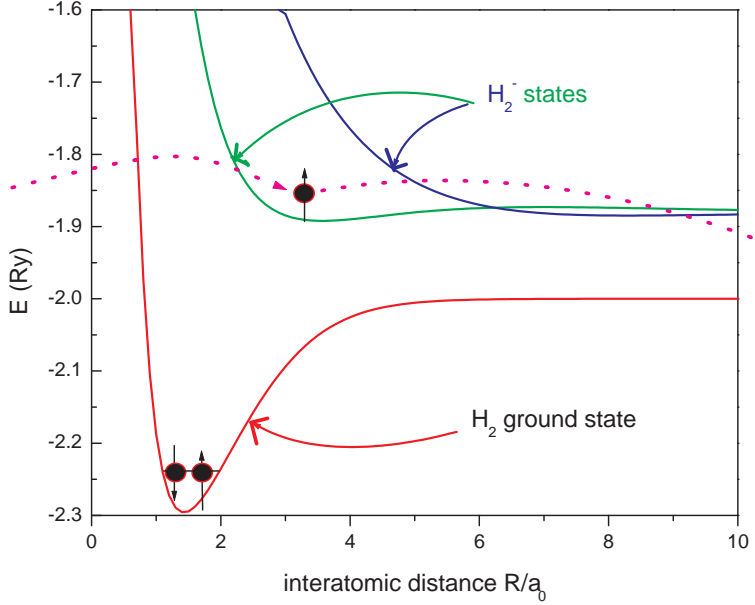}
}
 \caption{The level scheme of the $H_2$ ground state and the lowest
$H_2^-$ states as a function of the interatomic distance $R$. The
hopping electron illustrates the relevance of $H_2^-$ ionic
configuration when measuring tunneling conductivity of $H_2$ system.
} \label{Fig2}
\end{figure}

\subsection{$H_2$ molecule and the  $H_2^-$ and $H_2^+$ ions}
\begin{figure}
\resizebox{0.85\columnwidth}{!}{%
  \includegraphics{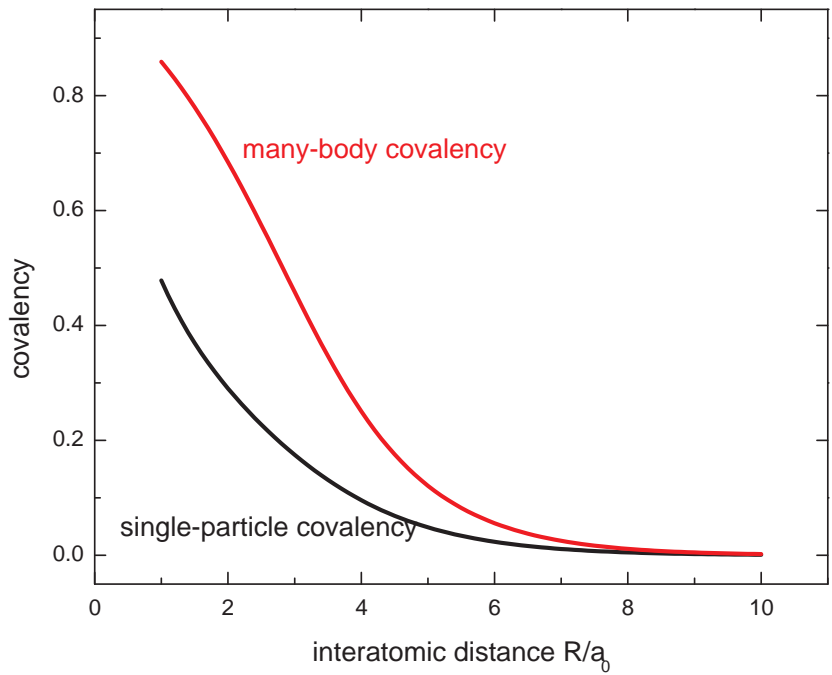}
} \caption{The single-particle ($\gamma$) and many-body
($\gamma_{mb}$) covalency factors for the $H_2$ wave functions. For
details see main text.} \label{Fig3}
\end{figure}

\begin{figure}
\resizebox{0.85\columnwidth}{!}{%
  \includegraphics{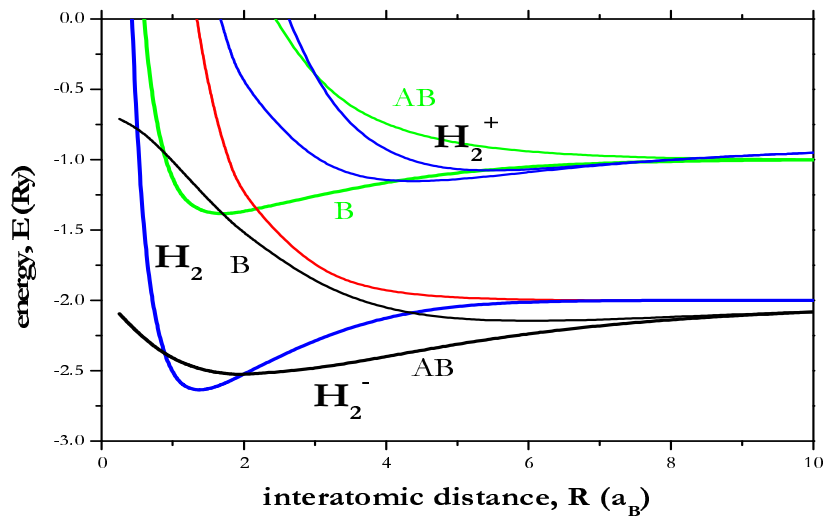}
  }
 \caption{Position of $H_2^+$ bonding(B) and antibonding (AB) energy levels with respect to those of $H_2$ and $H_2^-$, drawn as a function of intermolecular distance.}
 \label{Fig:H2BAB}
\end{figure}

In this Subsection we consider $H_2$ and $Li_2$ molecules, and
$H_2^-$ ion molecule. For the illustration of the method we have
plotted in Fig. \ref{Fig2} the level scheme for the $H_2$ and
$H_2^-$ systems. We consider first the situation with only one
1s-like orbital per atom. For $H_2$ we have $ { 4 \choose 2 } = 6$
two particle states \cite{SpalekPodsiadl00}. For that purpose, we
start with the parameterized Hamiltonian (\ref{ParHam}), where
subscripts "i" and "j" label now the two atomic sites and hence $U_1
= U_2 = U$, $ K_{12} = K$, $J_{12} = J$, $V_{12} = V$, and
$\epsilon_1 = \epsilon_2 = \epsilon_a$. The lowest eigenstate for
$H_2$ is
\begin{equation}
E_G \equiv \lambda_5 = 2 \epsilon_a + \frac{1}{2} ( U + K ) + J -
\frac{1}{2} \left[ ( U - K )^2 + 16 ( t + V )^2 \right]^{1/2} ,
\label{lambda5}
\end{equation}
and the corresponding singlet ground state in the Fock space has
the form
\[
| G > = \frac{1}{ \sqrt{2 D (D - U + K)} } \times
\]
\begin{equation}
\left\{  \frac{4 ( t + V )}{ \sqrt{2} } ( a^\dagger_{1 \uparrow}
a^\dagger_{2 \downarrow}- a^\dagger_{1 \downarrow} a^\dagger_{2
\uparrow} ) -  \frac{( D - U + K)}{\sqrt{2}}( a^\dagger_{1 \uparrow}
a^\dagger_{2 \downarrow} + a^\dagger_{1 \downarrow} a^\dagger_{2
\uparrow} ) \right\} | 0 > , \label{Gvect}
\end{equation}
where
\[
D \equiv \left[ (U - K)^2 + 16 (t + V)^2 \right] ^ {1/2}.
\]
The lowest spin-singlet  eigenstate has an admixture of symmetric ionic state $\frac{1}{\sqrt{2}}( a^\dagger_{1 \uparrow} a^\dagger_{2
\downarrow} + a^\dagger_{1 \downarrow} a^\dagger_{2 \uparrow} )$.
Therefore, to see the difference with either the Hartree-Fock or
Heitler-London approach to $H_2$ we construct the two-particle
wave function for the ground state according to the prescription
\begin{equation}
\Phi_0 ({\bf r}_1,{\bf r}_2) \equiv \frac{1}{\sqrt{2}} <0| \hat
\Psi ({\bf r}_1) \hat \Psi ({\bf r}_2) | G >.
\label{GSprescript}
\end{equation}
Taking $\hat \Psi ({\bf r}) = \sum_{i = 1}^2 \sum_{\sigma =
\uparrow}^\downarrow \Phi_i({\bf r}) \chi_\sigma ({\bf r})$, we
obtain that
\begin{equation}
\Phi_0 ({\bf r}_1,{\bf r}_2) = \frac{2 ( t + V )}{\sqrt{2 D (D - U
+ K)}} \Phi_c ({\bf r}_1,{\bf r}_2) - \frac{1}{2} \sqrt{\frac{D - U
+ K}{2 D}} \Phi_i ({\bf r}_1,{\bf r}_2),
\end{equation}
where the covalent part is
\begin{equation}
\Phi_c ({\bf r}_1,{\bf r}_2) =
\end{equation}
\[
 \left[ w_1({\bf r}_1) w_2({\bf
r}_2) + w_1({\bf r}_2) w_2({\bf
r}_1) \right] \left[ \chi_\uparrow({\bf r}_1) \chi_\downarrow({\bf r}_2) - \chi_\downarrow({\bf r}_1) \chi_\uparrow({\bf
r}_2) \right],
\]
whereas the ionic part takes the form
\begin{equation}
\Phi_i ({\bf r}_1,{\bf r}_2) =
\end{equation}
\[
\left[ w_1({\bf r}_1) w_1({\bf
r}_2) + w_2({\bf r}_1) w_2({\bf r}_2) \right] \left[
\chi_\uparrow({\bf r}_1) \chi_\downarrow({\bf r}_2) -
\chi_\downarrow({\bf r}_1) \chi_\uparrow({\bf r}_2) \right].
\]
The  ratio of the coefficients  before $\Phi_c ({\bf r}_1,{\bf
r}_2)$ and $\Phi_i ({\bf r}_1,{\bf r}_2)$ can be regarded as the
\emph{many-body covalency} $\gamma_{mb}$. This value should be
distinguished from the \emph{single-particle covalency} $\gamma$
appearing in the definition of the orthogonalized atomic orbital
$w_i({\bf r})$:
\begin{equation}
w_i({\bf r}) = \beta \left[ \Phi_i({\bf r}) - \gamma \Phi_j({\bf
r})\right],
\end{equation}
with $ j \neq i $. The two quantities are drawn in Fig.
\ref{Fig3}. The many-body covalency $\gamma_{mb}$ represents a
true degree of multiparticle configurational mixing.
\\
\indent In Table \ref{TH2micro} we list the energies and the values
of the microscopic parameters for $H_2$ system with optimized
orbitals, whereas in Table \ref{TH2mmicro} the same is provided for
$H_2^-$ molecular ion. One should notice a drastic difference for
the so-called {\em correlated hopping} matrix element V in the two
cases. The same holds true for the direct exchange integral J
(ferromagnetic). This exchange integral is always decisively smaller
than that for the antiferromagnetic kinetic exchange, $J_{kex} = 4
(t+V)^2/(U - K)$. The virtual interatomic hopping processes leading
to the strong kinetic exchange are the source of the singlet nature
of the $H_2$ ground state. The $H_2^-$ ground state is unstable with
respect to the dissociation into $H_2$ and $e^-$, contrary to the
$H^-$ case. However, the energetics of such state is important when
calculating e.g. the metallization of molecular hydrogen or
determining the tunnelling conductivity through $H_2$ molecule, as
shown schematically in Fig. \ref{Fig2}. This last Figure illustrates
the method of determining the energetics of excited states of $H_2$
by measuring e.g. the tunnelling conductivity (i.e. via $H_2^-$
intermediate state, see below).\\
\indent For the sake of completeness, we have provided in Fig.
\ref{Fig:H2BAB} the energy levels vs. $R$ for $H_2$,$H_2^-$, and
$H_2^+$ ions and have labeled both the bonding (B) and antibonding
(AB) level positions. We see that the distance between $H_2$ and
$H_2^-$ levels is much smaller then the distance between $H_2$ and
$H_2^+$ states. This will lead to the asymmetry in the tunneling
conductivity when reversing the bias voltage sign, as discussed in
detail in Section 5.

\begin{table}
\caption{Ground-state energy and microscopic parameters (in Ry) for
$H_2$ molecule. The last column represents
 the kinetic exchange integral characterizing intersite antiferromagnetic exchange}
\label{TH2micro}
\begin{tabular}{rcccccccc} \hline \hline $R/a$ & $E_G/N$ & $\epsilon_a$ & $t$ & $U$ & $K$ & $V$ [mRy] & $J$ [mRy] &
$\frac{4(t+V)^2}{U-K}$[mRy]
 \\  \hline 1.0 & -1.0937 & -1.6555 & -1.1719 & 1.8582 & 1.1334 & -13.5502 &
   26.2545 & 7755.52 \cr 1.5 & -1.1472 & -1.7528 & -0.6784 & 1.6265 & 0.9331 &
    -11.6875 & 21.2529 & 2747.41 \cr 2.0 & -1.1177 & -1.722 & -0.4274 & 1.4747 &
   0.7925 & -11.5774 & 16.9218 & 1130.19 \cr 2.5 & -1.0787 & -1.6598 & -0.2833 &
   1.3769 & 0.6887 & -12.0544 & 13.1498 & 507.209 \cr 3.0 & -1.0469 & -1.5947 &
    -0.1932 & 1.3171 & 0.6077 & -12.594 & 9.8153 & 238.939 \cr 3.5 & -1.0254 &
    -1.5347 & -0.1333 & 1.2835 & 0.5414 & -12.8122 & 6.9224 & 115.143 \cr 4.0 &
    -1.0127 & -1.4816 & -0.0919 & 1.2663 & 0.4854 & -12.441 & 4.5736 &
   55.8193 \cr 4.5 & -1.006 & -1.4355 & -0.0629 & 1.2579 & 0.4377 & -11.4414 &
   2.8367 & 26.9722 \cr 5.0 & -1.0028 & -1.3957 & -0.0426 & 1.2539 & 0.3970 &
    -9.9894 & 1.6652 & 12.9352 \cr 5.5 & -1.0012 & -1.3616 & -0.0286 & 1.2519 &
   0.3623 & -8.3378 & 0.9334 & 6.1455 \cr 6.0 & -1.0005 & -1.3324 &
    -0.01905 & 1.251 & 0.3327 & -6.7029 & 0.5033 & 2.8902 \cr 6.5 & -1.00024 &
    -1.3073 & -0.0126 & 1.2505 & 0.3075 & -5.2242 & 0.2626 & 1.3452 \cr 7.0 &
    -1.0001 & -1.2855 & -0.0083 & 1.2503 & 0.2856 & -3.9685 & 0.1333 &
   0.6197 \cr 7.5 & -1.00004 & -1.2666 & -0.0054 & 1.2501 & 0.2666 &
    -2.9509 & 0.066 & 0.2826 \cr 8.0 & -1.00002 & -1.25 & -0.0035 &
   1.25006 & 0.25 & -2.1551 & 0.032 & 0.1277 \cr 8.5 & -1.00001 & -1.2353&
    -0.0023 & 1.25003 & 0.2353 & -1.5501 & 0.01523 & 0.0572 \cr 9.0 & -1. &
    -1.2222 & -0.0015 & 1.25001 & 0.2222 & -1.1005 & 0.0071 &
   0.0254 \cr 9.5 & -1. & -1.2105 & -0.0009 & 1.25001 & 0.2105 & -0.7725 &
   0.0033 & 0.0112 \cr 10.0 & -1. & -1.2 & -0.0006 & 1.25 & 0.2 & -0.5371 &
   0.0015 & 0.0049 \cr   \hline \hline
\end{tabular}
\end{table}

\begin{table}
\caption{Same as in Table \ref{TH2micro} for $H_2^-$ ion}
\label{TH2mmicro}
\begin{tabular}{rcccccccc} \hline \hline $R/a$ & $E_G/N$ &
$\epsilon_a$ & $t$ & $U$ & $K$ & $V$ & $J$ &
$\frac{4(t+V)^2}{U-K}$
\\ \hline 1.0 & -0.4591 & -1.6607 & -0.5869 & 1.1414 & 0.7360 & -0.0105 &
   0.0163 & 3.5220 \cr 1.5 & -0.7659 & -1.6647 & -0.4285 & 1.1279 & 0.6983 &
    -0.0085 & 0.0161 & 1.7782 \cr 2.0 & -0.8813 & -1.6259 & -0.3083
& 1.0979 &
   0.6474 & -0.0078 & 0.0150 & 0.8871 \cr 2.5 & -0.9264 & -1.5737 & -0.2221 &
   1.0692 & 0.5961 & -0.0079 & 0.0133 & 0.4476 \cr 3.0 & -0.9423 & -1.5204 &
    -0.1603 & 1.0466 & 0.5476 & -0.0086 & 0.0113 & 0.2286 \cr 3.5 & -0.9460 &
    -1.4704 & -0.1154 & 1.0305 & 0.5025 & -0.0093 & 0.0091 & 0.1179 \cr 4.0 &
    -0.9450 & -1.4252 & -0.0826 & 1.0196 & 0.4608 & -0.0099 & 0.0071 & 0.0612 \cr
   4.5 & -0.9426 & -1.3848 & -0.0585 & 1.0126 & 0.4226 & -0.0101 & 0.0052 &
   0.0319 \cr 5.0 & -0.9402 & -1.3491 & -0.0410 & 1.0080 & 0.3881 & -0.0099 &
  0.0037 & 0.0167 \cr 5.5 & -0.9384 & -1.3176 & -0.0284 & 1.0051 & 0.3573 &
    -0.0093 & 0.0025 & 0.0088 \cr 6.0 & -0.9373 & -1.2901 & -0.0194 & 1.0032 &
   0.3300 & -0.0085 & 0.0017 & 0.0046 \cr 6.5 & -0.9365 & -1.2621 & -0.0130 &
   0.9905 & 0.3058 & -0.0075 & 0.0011 & 0.0025 \cr 7.0 & -0.9363 & -1.2402 &
    -0.0086 & 0.9876 & 0.2847 & -0.0065 & 0.0007 & 0.0013 \cr 7.5 & -0.9365 &
    -1.2211 & -0.0056 & 0.9856 & 0.2662 & -0.0055 & 0.0004 & 0.0007 \cr 8.0 &
    -0.9367 & -1.2044 & -0.0036 & 0.9844 & 0.2498 & -0.0046 & 0.0003 & 0.0004 \cr
   8.5 & -0.9372 & -1.1897 & -0.0022 & 0.9839 & 0.2352 & -0.0037 & 0.0002 &
   0.0002 \cr 9.0 & -0.9376 & -1.1768 & -0.0013 & 0.9839 & 0.2222 & -0.0030 &
   0.00009 & 0.00010 \cr 9.5 & -0.9380 & -1.1653 & -0.0008 & 0.9842 & 0.2105 &
    -0.0024 & 0.00005 & 0.00005 \cr 10.0 & -0.9384 & -1.1549 & -0.0004 & 0.9848 &
   0.2000 & -0.0018 & 0.00003 & 0.00003 \cr \hline \hline
\end{tabular}
\end{table}

\subsection{Hydrogen clusters, $H_N$}
\begin{figure}
\resizebox{0.75\columnwidth}{!}{%
  \includegraphics{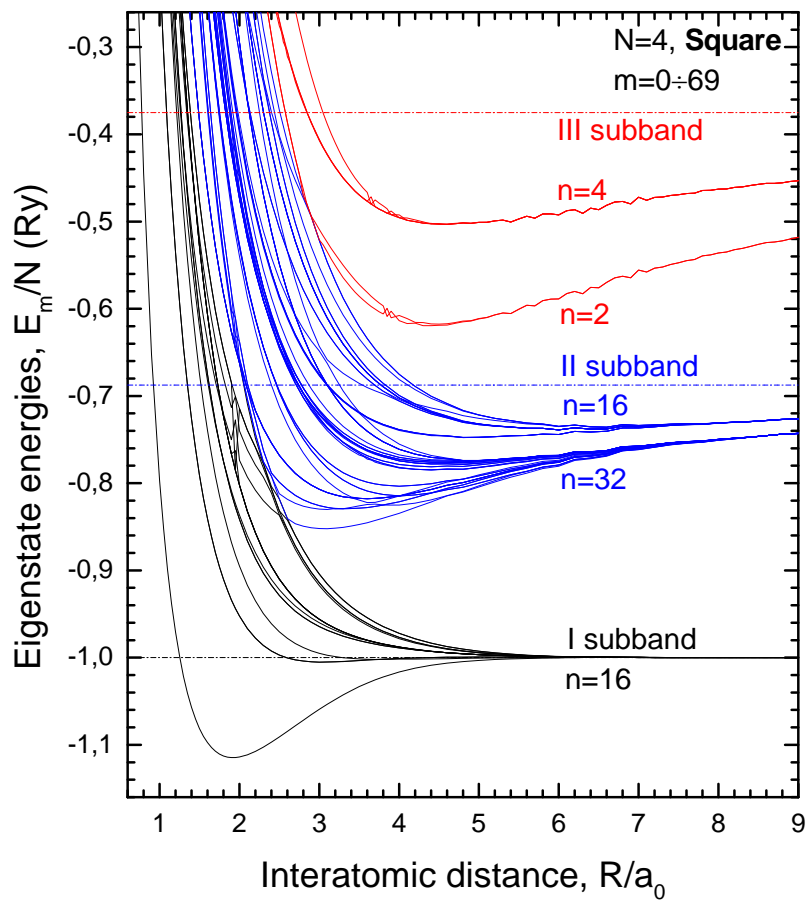}
} \caption{Ground- and excited states energies for the $H_4$ square
configuration as a function of the interatomic distance.}
\label{Fig4}
\end{figure}

\begin{figure}
\resizebox{0.75\columnwidth}{!}{%
  \includegraphics{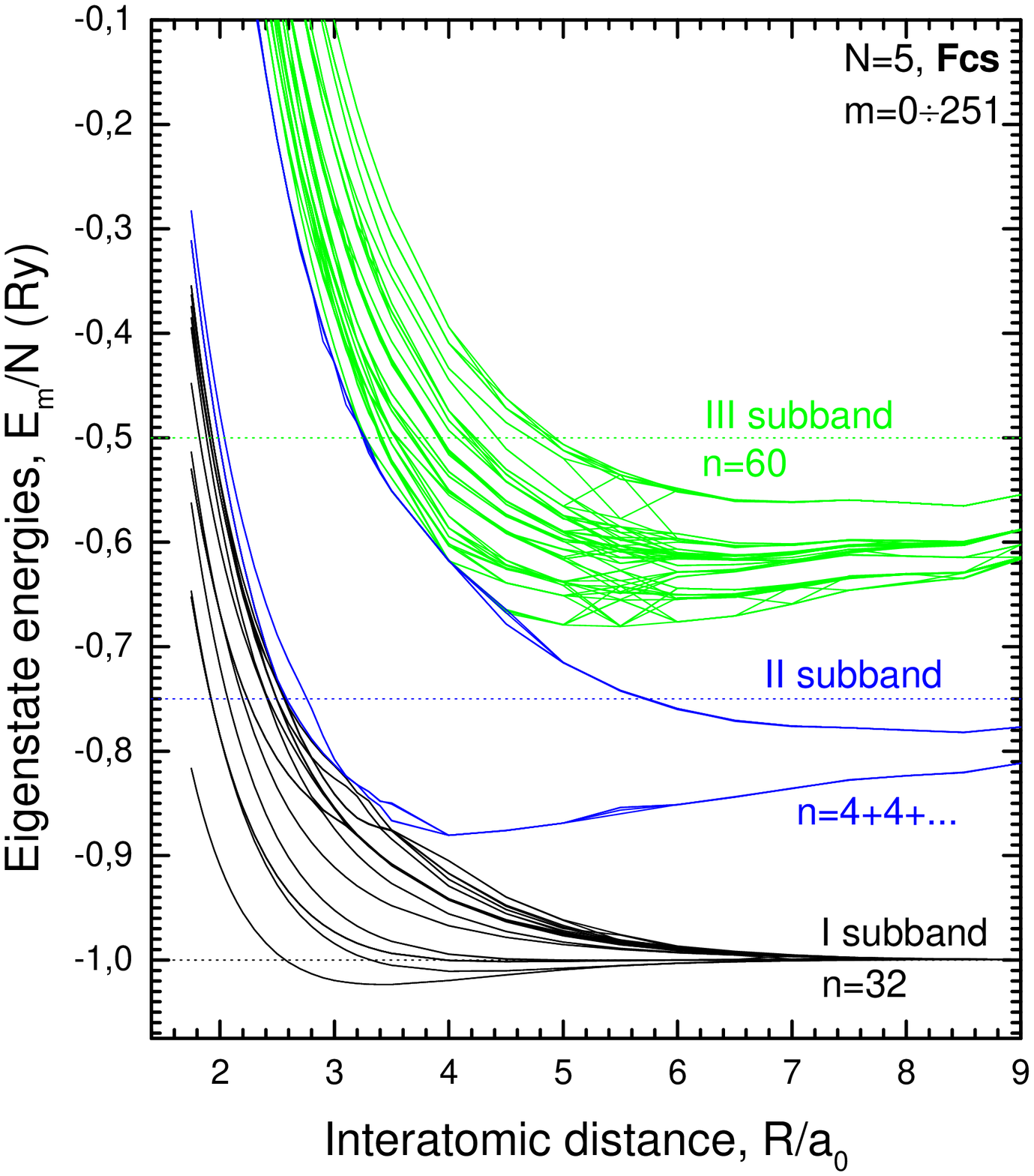}
} \caption{Ground- and excited states energies for the $H_5$
face-centered-square (fcs) configuration as a function of the
interatomic distance.} \label{Fig5}
\end{figure}

\begin{figure}
\resizebox{0.75\columnwidth}{!}{%
  \includegraphics{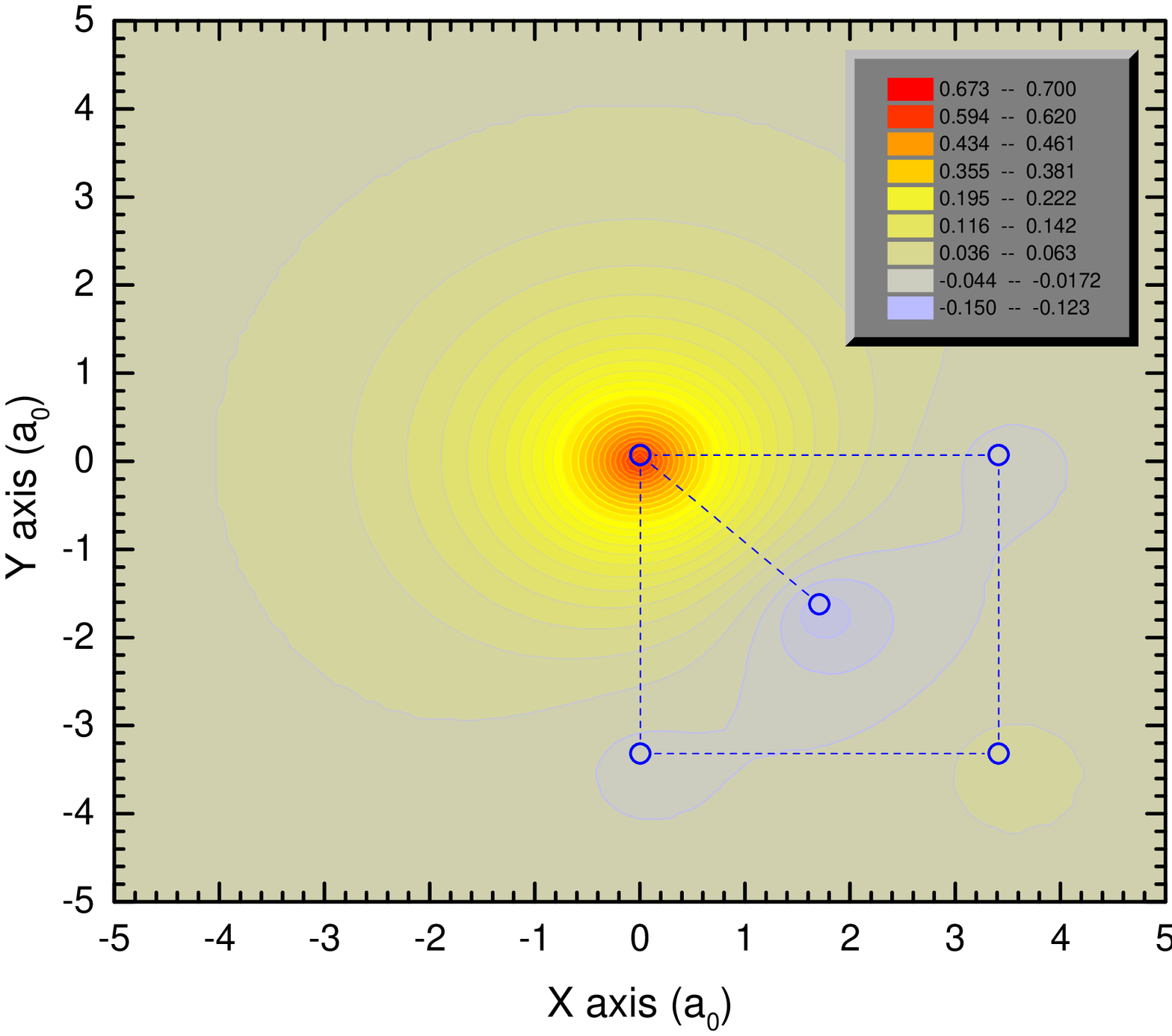}
} \caption{The renormalized Wannier function profile for for the
central atom in the fcs structure.} \label{Fig6}
\end{figure}

\begin{figure}
\resizebox{0.75\columnwidth}{!}{%
  \includegraphics{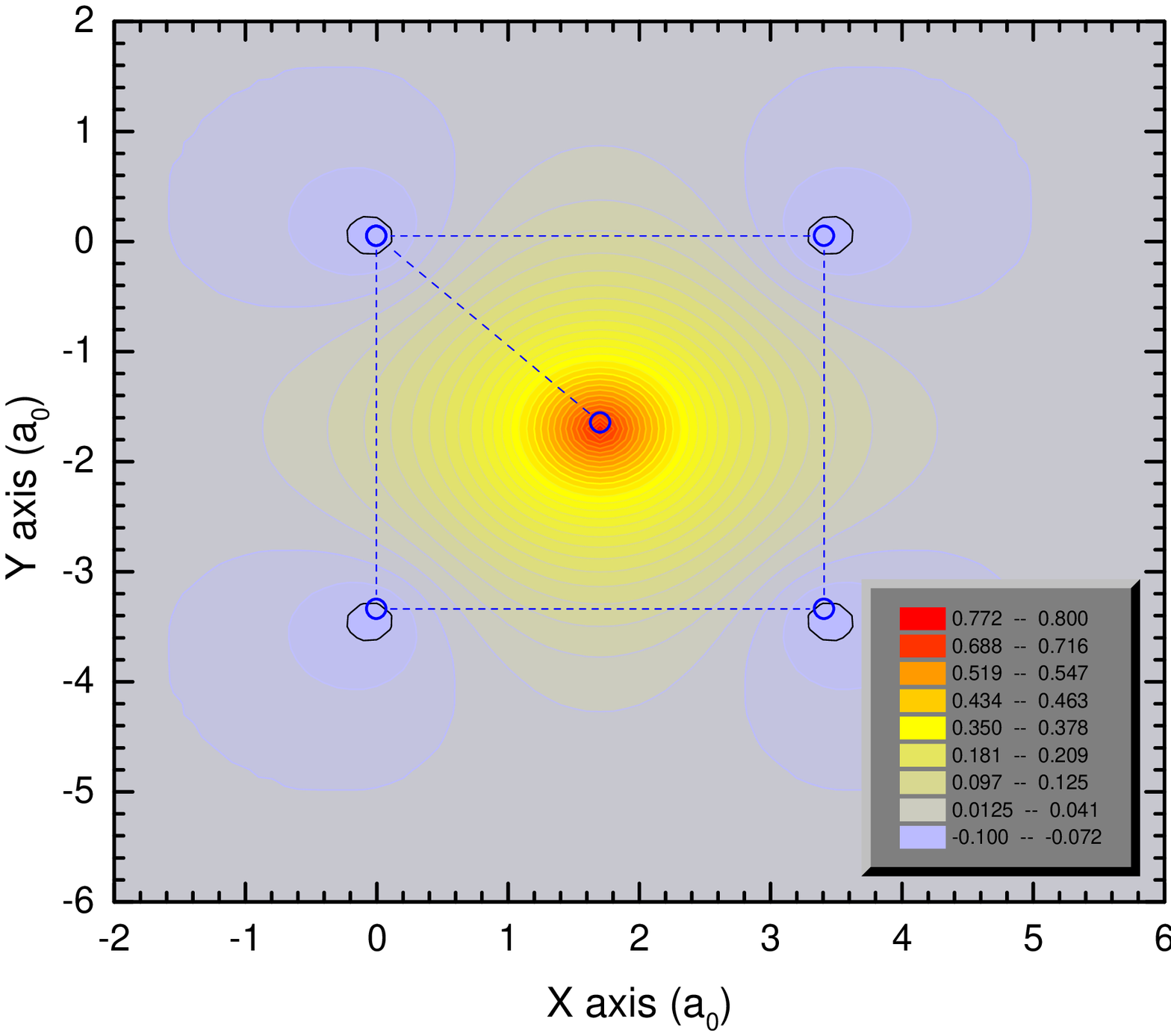}
} \caption{The renormalized Wannier function profile for for the
corner atom in the fcs structure.} \label{Fig7}
\end{figure}

\begin{figure}
\resizebox{0.75\columnwidth}{!}{%
  \includegraphics{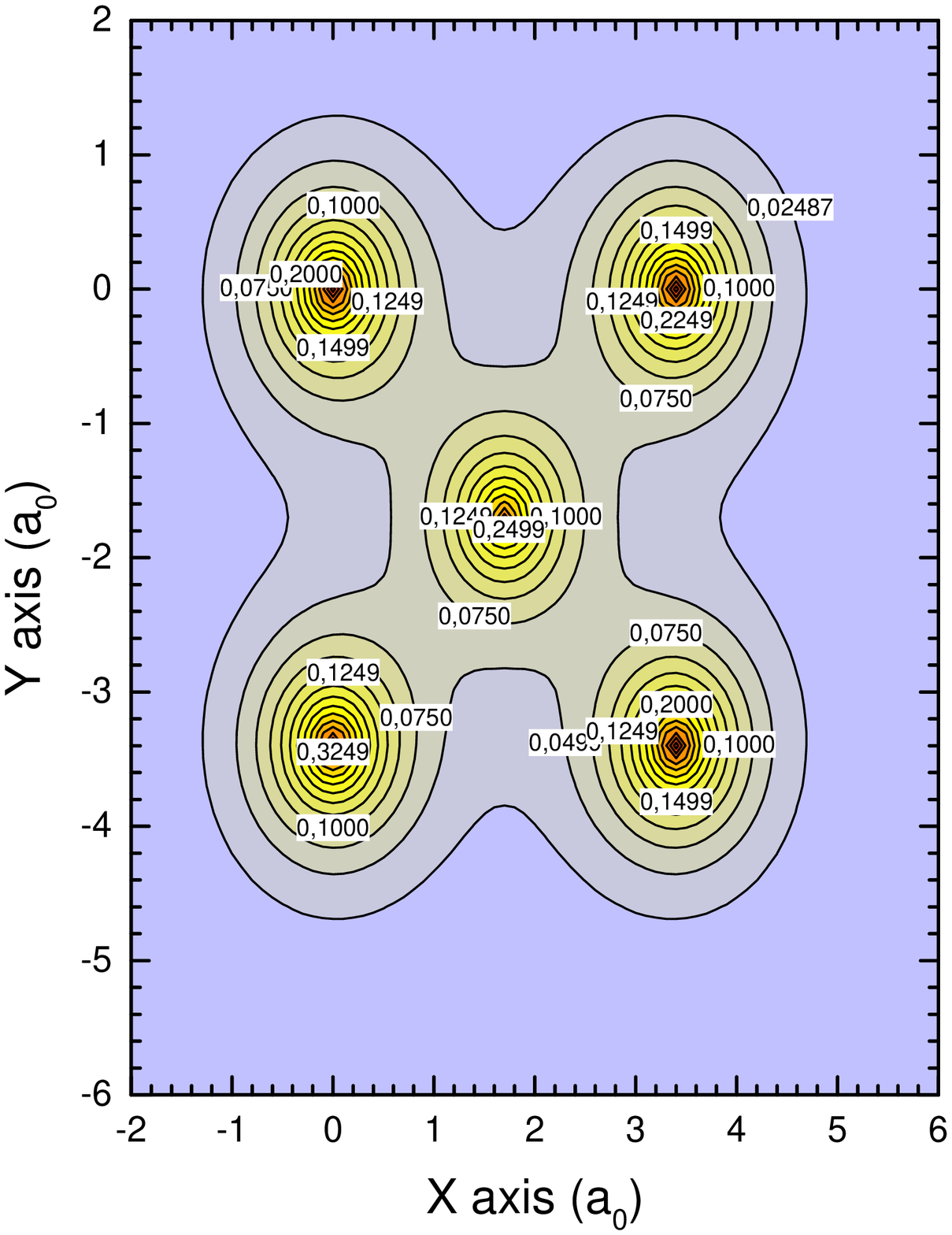}
} \caption{Electron density profile for the $H_5$
face-centered-square (fcs) structure.} \label{Fig8}
\end{figure}

\begin{figure}
\resizebox{0.75\columnwidth}{!}{%
  \includegraphics{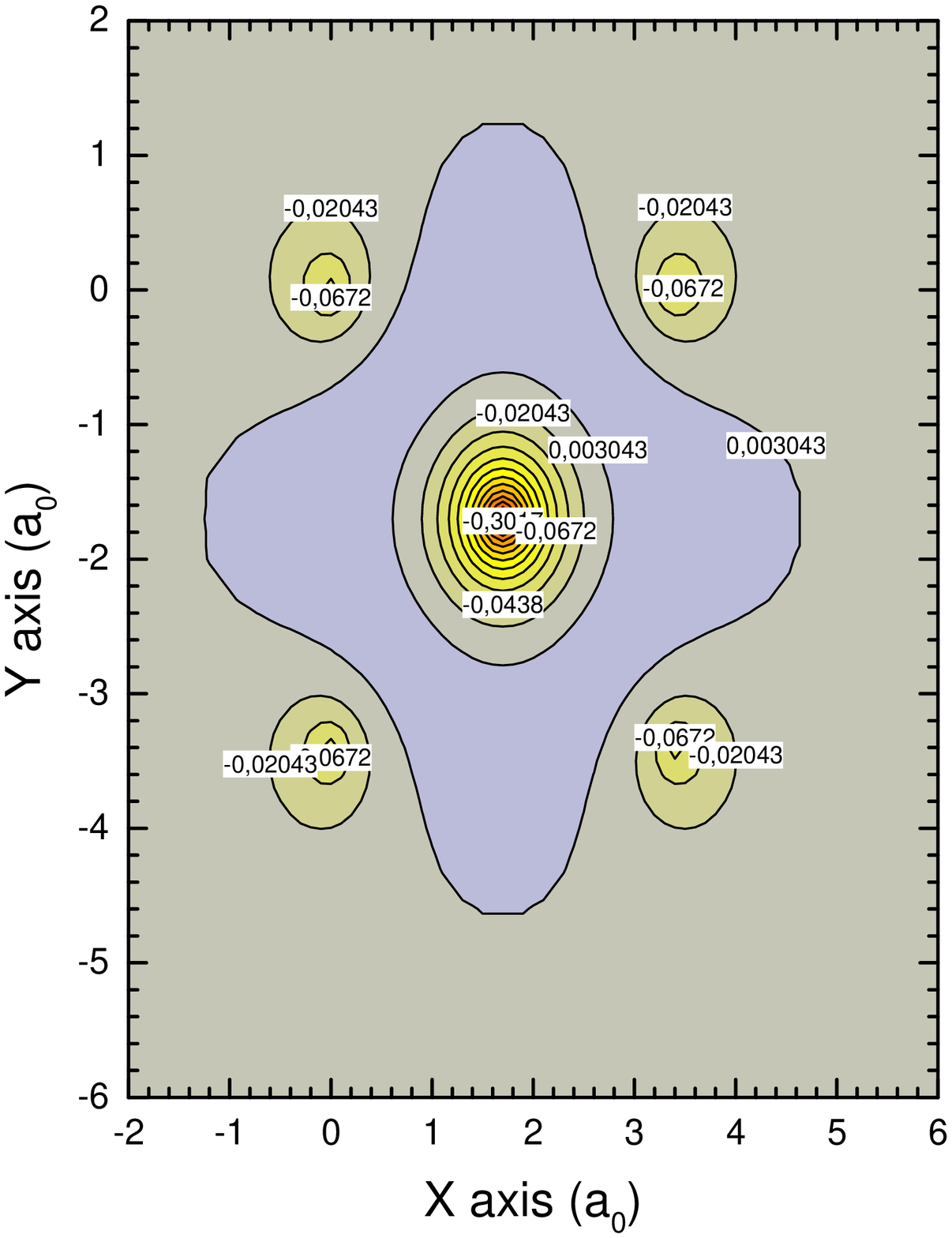}
} \caption{Difference between the calculated density of states and
the Hartree-Fock results for the $H_5$ fcs structure.} \label{Fig9}
\end{figure}

As next application we consider hydrogen-cluster $H_N$ systems, with
$N \leq 6$ atoms. We take the atomic-like 1s orbitals $\{\Phi_i
({\bf r} )\}$ of an adjustable size $a \equiv \alpha^{-1}$,
composing the orthogonalized  atomic (Wannier) functions $\{w_i
({\bf r} )\}_{i=1,\ldots ,N}$. The cluster of $N$ atoms with $N$
electrons contains ${ 2 N \choose N }$ states and the
second-quantized Hamiltonian is of the form (\ref{ParHam}), with 3-
and 4- site terms added. The 3- and 4- site interaction terms are
difficult to calculate in the Slater basis (see below). Therefore,
we have made an ansatz \cite{ThesisRZ} namely, we impose the
condition on the trial Wannier function that the 3- and 4-site
matrix elements $V_{ijkl}$ vanish. This allows for an explicit
expression of the 3- and 4- site matrix elements $V'_{ijkl}$ in the
atomic representation via the corresponding one- and two- site
elements. In Figs. \ref{Fig4} and \ref{Fig5} we present the results
for the ground- and excited- states energies for the square and
face-centered-square (fcs) configurations, $N = 4$ and $5$,
respectively.
The states are grouped into manifolds, which are characterized by
the number of double occupancies of a single state $w_i({\bf r})$,
appearing in the system. The horizontal lines mark the ground state,
states with one and two double electron occupancies in the atomic
limit (i.e. for large interatomic distance). The manifolds thus
correspond to the {\em Hubbard subbands} introduced for strongly
correlated solids \cite{Hubbard}. As far as we are aware of, our
results are the first manifestation of the energy manifold evolution
into well separated  subbands with the interatomic distance
increase. The first two subbands correspond to HOMO and LUMO levels
determined in quantum-chemical calculation \cite{Szabo}. In. Fig.
\ref{Fig6} and \ref{Fig7} we represent the renormalized Wannier
function profiles for the face centered square configuration of
$N=5$ atoms, for the central and the corner positions, respectively.
Note the small negative values on the nearest-neighbor sites to
assure the orthogonality of the functions centered on different
sites. Obviously, the atom in the center of the square is
inequivalent to the remaining four corner atoms, as can be see
explicitly in Fig. \ref{Fig8}, where the density profile  $n({\bf
r})$ in the cluster plane, according to formula (\ref{nodr47}), has
been drawn.
The electron density near the central atom is decisively smaller, a
clear sign of electron-correlation effects induced by the repulsive
Coulomb interaction. This assessment is corroborated in Fig.
\ref{Fig9}, where the difference to the Hartree-Fock part of the
density profile has been presented. These densities should be
possible to be determined with the help of scanning tunneling
microscopy (STM).
\subsection{Energetically stable $H_4$ clusters and fermionic nanoladders}
\begin{figure}
\resizebox{0.75\columnwidth}{!}{%
  \includegraphics{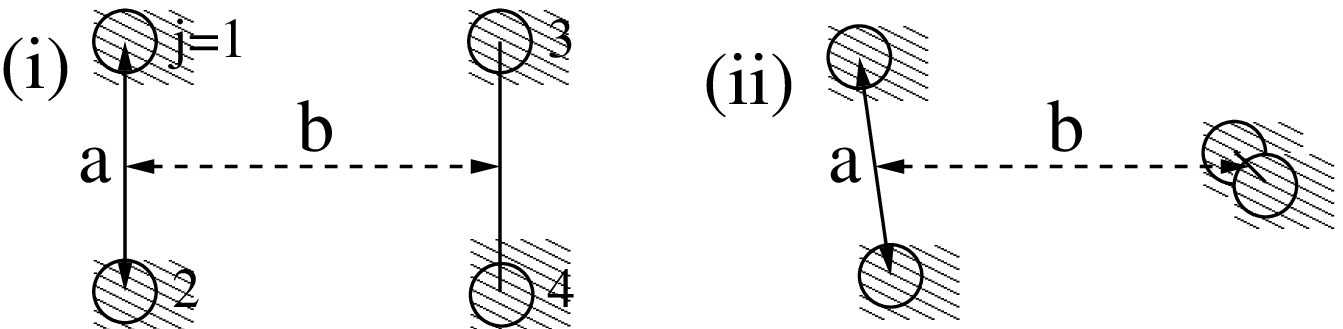}
} \caption{Schematic representation of the $H_4$ cluster geometries:
(i) parallel and (ii) perpendicular orientation of $H_2$ molecules.
Geometrical parameters of the cluster are: the bond length $a$ and
the intermolecular distance $b$. The numbering order of the lattice
sites $j$ is also provided.} \label{Fig10}
\end{figure}

The $H_N$ clusters of regular-polygon shape are not stable
energetically, as the ground state energies in Figs. \ref{Fig4}
and \ref{Fig5} at the minimum (per atom) is above that for the
$H_2$ case. This is because a stable, say, $H_4$ cluster will
reflect strong molecular bonding of each pair of $H_2$ molecules
to saturate the covalent character of the bonding. To prove that
this is indeed the case we have considered an example of a
rectangular cluster differentiating between the lateral distance
(say, along the bond of the length a) and the horizontal
intermolecular distance b. We consider both  the planar and the
twisted by $90^\circ$ configurations, as drawn in Fig.
\ref{Fig10}.
The Gaussian STO-3G basis of an adjustable size $\alpha^{-1}$ has
been used in this case, so the 3- and 4- site interaction terms in
the atomic basis are included explicitly. Additionally, the
Hamiltonian in the second quantization, only  containing the
principal one- and two- site interactions, is taken in  the form
\cite{ThesisAR}
\begin{equation}
H = \epsilon^{eff}_a \sum_{i \sigma} n_{i \sigma} + \sum_{i \neq
j} t_{i j} a^{\dagger}_{i \sigma} a_{j \sigma} + U \sum_i n_{i
\uparrow} n_{i \downarrow} +\frac{1}{2} \sum_{i \neq j} K_{i j}
\delta n_i \delta n_j ,
\label{HamAdam}
\end{equation}
where $\delta n_i = 1 - n_i$, $n_i = \sum_\sigma n_{i \sigma}$ and
\begin{equation}
\epsilon^{eff}_a = \epsilon_a + \frac{1}{2 N} \sum_{i \neq j}
\left( K_{i j} + \frac{2}{R_{i j}} \right),
\end{equation}
with the last term in the parenthesis representing the Coulomb
repulsion (in atomic units) between the protons placed at the
distance $R_{i j}$, together with the electron-electron intersite
repulsion $K_{i j}$ (such a redefinition is necessary to achieve the
proper atomic limit, when the nearest-neighbor distance $R
\rightarrow \infty$. Also, we have to take into account three
different hopping integrals: between the nearest neighbors $t_1 =
t_{1 2} = t_{3 4}$, between the second neighbors $t_2 = t_{1 3} =
t_{2 4}$, and the third neighbors $t_3 = t_{1 4} = t_{2 3}$.
Likewise, we have three intersite Coulomb interactions $\{ K_n \}_{n
= 1, 2,
3}$.\\
\indent The basic ground-state characteristics for the two spatial
arrangements is provided in Tables \ref{TabAdam1} and
\ref{TabAdam2}.
\begin{table}
\caption{The optimal bond length $a_{\min}$, inverse orbital size
$\alpha_{\min}$, and the ground--state energy $E_G$ per atom for the
planar H$_4$ cluster $(i)$. The corresponding energy of the
molecular dimer $E_G(2\mathrm{H}_2)$ is also provided.}
\label{TabAdam1}
\begin{center}
\begin{tabular}{rllll}
\hline\hline
$b/a_0$ & $a_{\rm min}/a_0$ & $\alpha_{\rm min}a_0$
 & $E_G/N$ & $E_G(2\mathrm{H}_2)/N$ \\ \hline
1.7 &  1.3627 & 1.2231 &  -0.928424 &  -0.922411 \\
2.0 &  1.3291 & 1.2395 &  -1.019152 &  -1.016365 \\
2.5 &  1.3518 & 1.2304 &  -1.098551 &  -1.097314 \\
3.0 &  1.3829 & 1.2157 &  -1.131770 &  -1.131167 \\
3.5 &  1.4075 & 1.2041 &  -1.144613 &  -1.144317 \\
4.0 &  1.4238 & 1.1969 &  -1.148598 &  -1.148454 \\
5.0 &  1.4373 & 1.1911 &  -1.148093 &  -1.148056 \\
6.0 &  1.4390 & 1.1908 &  -1.145975 &  -1.145964 \\
8.0 &  1.4375 & 1.1924 &  -1.143651 &  -1.143649 \\
10.0 & 1.4366 & 1.1929 &  -1.142756 &  -1.142755 \\
20.0 & 1.4357 & 1.1940 &  -1.141908 &  -1.141908 \\ \hline
$\infty$ &  1.4356 & 1.1943 & -1.141783 & \\
\hline\hline
\end{tabular}
\end{center}
\end{table}

\begin{table}[!p]
\caption{The optimal bond length $a_{\min}$, inverse orbital size
$\alpha_{\min}$, and the ground--state energy $E_G$ per atom for the
cluster geometry $(ii)$. The corresponding energy of the molecular
dimer $E_G(2\mathrm{H}_2)$ is also provided.} \label{TabAdam2}
\begin{center}
\begin{tabular}{rllll}
\hline\hline
$b/a_0$ & $a_{\rm min}/a_0$ & $\alpha_{\rm min}a_0$
 & $E_G/N$ & $E_G(2\mathrm{H}_2)/N$ \\ \hline
1.6 &  1.5796 & 1.1568 &  -0.928235 &  -0.923390 \\
2.0 &  1.3759 & 1.2206 &  -1.038891 &  -1.037968 \\
2.5 &  1.3725 & 1.2193 &  -1.108116 &  -1.107803 \\
3.0 &  1.3947 & 1.2091 &  -1.136931 &  -1.136809 \\
3.5 &  1.4153 & 1.2000 &  -1.147566 &  -1.147519 \\
4.0 &  1.4292 & 1.1937 &  -1.150310 &  -1.150293 \\
5.0 &  1.4397 & 1.1894 &  -1.148674 &  -1.148672 \\
6.0 &  1.4400 & 1.1897 &  -1.146200 &  -1.146200 \\
8.0 &  1.4377 & 1.1926 &  -1.143705 &  -1.143705 \\
10.0 & 1.4367 & 1.1929 &  -1.142774 &  -1.142774 \\
20.0 & 1.4357 & 1.1940 &  -1.141908 &  -1.141908 \\ \hline
$\infty$ &  1.4356 & 1.1943 & -1.141783 & \\
\hline\hline
\end{tabular}
\end{center}
\end{table}
The results are listed as a function of intermolecular distance
and include: the bond length $a_{min}$, the inverse
Gaussian-function size $\alpha$, the energy of $H_4$ and the
energy of two $H_2$ molecules (per atom). The principal feature of
these results is that the system is energetically stable against
the dissociation into two $H_2$ molecules (cf. the last row in the
two situations). The global minima are:
\begin{enumerate}
\item for the planar geometry: $b_{min} = 4.32 a_0$, $a_{min} =
1.4303 a_0$, $\alpha_{min} = 1.1937 a_0^{-1}$, and $E_G^{min}/N =
-1.149061 Ry$; and
\item for the twisted $90^o$ geometry: $b_{min} = 4.13 a_0$, $a_{min} =
1.4318 a_0$, $\alpha_{min} = 1.11927 a_0^{-1}$, and $E_G^{min}/N =
-1.150396 Ry$.
\end{enumerate}
So, the geometry (ii) is the most stable in vacuum and should
constitute a building block for $H_2$ molecular crystal. Note we
have not included the zero point motion of the nuclei
\cite{SpalekPodsiadl00}. The ratio $b/a$ is roughly four, so substantial molecular identity of $H_2$ pairs survives.\\
\indent As in all cases before, the knowledge of microscopic
parameters is crucial for the larger-cluster or solid-state
configurations ( Hamiltonian (\ref{HamAdam}) provides whole dynamics
within the subspace with one orbital per atom). Therefore, in
\ref{TabAdam3} and \ref{TabAdam4} we list them for the two
configurations considered, as a function of intermolecular distance
$b$ ( the last row represents the corresponding values
for $H_2$ in the ground state).\\
\indent The stability of the $H_4$ cluster raises a very interesting
question of stability of $H_{2 N}$ ladders, which can be regarded as
the simplest fermionic ladders with the frustration of electron
spins in the twisted configuration.
\begin{table}[!p]
\caption{Microscopic parameters (in Ry) for the H$_4$ cluster
configuration $(i)$ calculated in the Gaussian STO--3G basis.
Corresponding values of the inverse orbital size $\alpha_{\min}$ and
the bond length $a_{\min}$ are provided in Table \ref{TabAdam1}.}
\label{TabAdam3}
\begin{center}
\begin{tabular}{rllllllll}
\hline\hline
$b/a_0$ & $\epsilon_a^{eff}$ & $t_1$ & $t_2$ & $t_3$ &
  $U$ & $K_1$ & $K_2$ & $K_3$ \\
\hline
1.7 & -0.2354 & -0.8610 & -0.6622 &  0.0822
 &  1.811 &  1.027 &  0.947 &  0.728 \\
2.0 & -0.3088 & -0.8791 & -0.5137 &  0.0617
 &  1.802 &  1.032 &  0.872 &  0.692 \\
2.5 & -0.4233 & -0.8390 & -0.3268 &  0.0352
 &  1.748 &  1.007 &  0.750 &  0.619 \\
3.0 & -0.4925 & -0.7983 & -0.2067 &  0.0220
 &  1.702 &  0.984 &  0.649 &  0.553 \\
3.5 & -0.5319 & -0.7685 & -0.1290 &  0.0150
 &  1.671 &  0.967 &  0.567 &  0.497 \\
4.0 & -0.5533 & -0.7492 & -0.0785 &  0.0103
 &  1.653 &  0.957 &  0.500 &  0.449 \\
5.0 & -0.5689 & -0.7316 & -0.0267 &  0.0037
 &  1.639 &  0.949 &  0.401 &  0.373 \\
6.0 & -0.5707 & -0.7275 & -0.0084 &  0.0007
 &  1.638 &  0.948 &  0.334 &  0.318 \\
8.0 & -0.5692 & -0.7269 & -0.0006 & -0.0000
 &  1.640 &  0.949 &  0.250 &  0.243 \\
10.0 & -0.5684 & -0.7268 & -0.0000 & -0.0000
&  1.641 &  0.950 &  0.200 &  0.197 \\
20.0 & -0.5673 & -0.7272 & -0.0000 & -0.0000
 &  1.642 &  0.951 &  0.100 &  0.100 \\ \hline
 $\infty$ & -0.5671 & -0.7273 & 0 & 0 & 1.642 & 0.951 & 0 & 0 \\
\hline\hline
\end{tabular}
\end{center}
\end{table}

\begin{table}[!p]
\caption{Microscopic parameters (in Ry) for the H$_4$ cluster
configuration $(ii)$ calculated in the Gaussian STO--3G basis.
Corresponding values of the inverse orbital size $\alpha_{\min}$ and
the bond length $a_{\min}$ are provided in Table \ref{TabAdam2}.}
\label{TabAdam4}
\begin{center}
\begin{tabular}{rllllllll}
\hline\hline
$b/a_0$ & $\epsilon_a^{eff}$ & $t_1$ & $t_2$ & $t_3$ &
  $U$ & $K_1$ & $K_2$ & $K_3$ \\
\hline
1.6 & -0.5352 & -0.5423 & -0.2986 & -0.2986
 &  1.650 &  0.911 &  0.816 &  0.816 \\
2.0 & -0.4511 & -0.7496 & -0.2230 & -0.2230
 &  1.732 &  0.991 &  0.772 &  0.772 \\
2.5 & -0.4863 & -0.7820 & -0.1451 & -0.1451
 &  1.711 &  0.986 &  0.680 &  0.680 \\
3.0 & -0.5221 & -0.7721 & -0.0923 & -0.0923
 &  1.683 &  0.972 &  0.598 &  0.598 \\
3.5 & -0.5467 & -0.7558 & -0.0571 & -0.0571
 &  1.661 &  0.961 &  0.530 &  0.530 \\
4.0 & -0.5613 & -0.7423 & -0.0342 & -0.0342
 &  1.646 &  0.953 &  0.473 &  0.473 \\
5.0 & -0.5715 & -0.7293 & -0.0115 & -0.0115
 &  1.637 &  0.947 &  0.387 &  0.387 \\
6.0 & -0.5719 & -0.7264 & -0.0038 & -0.0038
 &  1.636 &  0.947 &  0.326 &  0.326 \\
8.0 & -0.5692 & -0.7269 & -0.0003 & -0.0003
 &  1.640 &  0.949 &  0.247 &  0.247 \\
10.0 & -0.5685 & -0.7268 & -0.0000 & -0.0000
 &  1.641 &  0.950 &  0.198 &  0.198 \\
20.0 & -0.5673 & -0.7272 & -0.0000 & -0.0000
 &  1.642 &  0.951 &  0.100 &  0.100 \\ \hline
 $\infty$ & -0.5671 & -0.7273 & 0 & 0 & 1.642 & 0.951 & 0 & 0 \\
\hline\hline
\end{tabular}
\end{center}
\end{table}
\begin{figure}
\resizebox{1.2\columnwidth}{!}{%
  \includegraphics{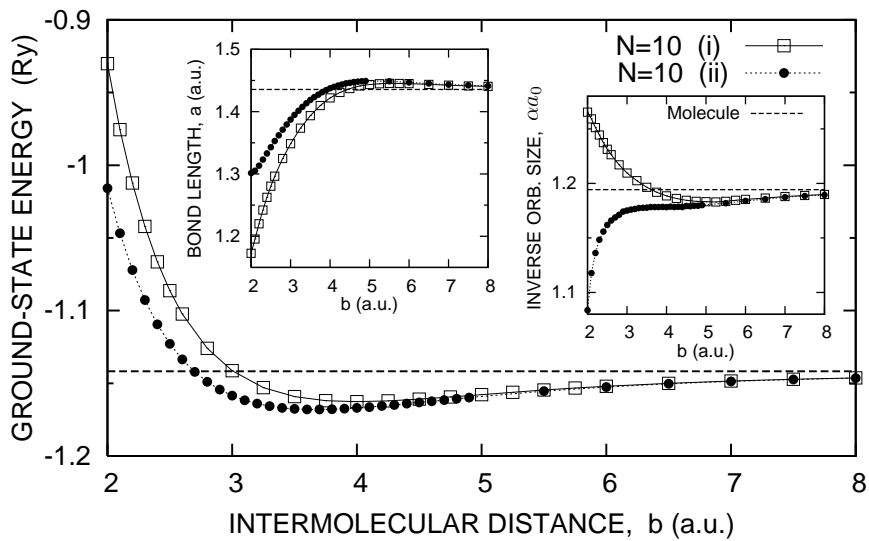}
}\caption{Ground--state energy per atom for the planar ladder of
$N\!=\!12$ atoms $(a)$ with a fixed intermolecular distance $b$
(specified in the units of Bohr radius $a_0$), and $(b)$ the
corresponding values of the optimal inverse orbital size
$\alpha_{\min}$. Periodic boundary conditions are used. The results
for the atomic chain (\emph{dashed} lines) with a lattice parameter
$a\lesssim b_\mathrm{crit}$ are shown for the comparison.}
\label{Fig:eglad}
\end{figure}
\begin{figure}[!t]
\resizebox{\columnwidth}{!}{%
  \includegraphics{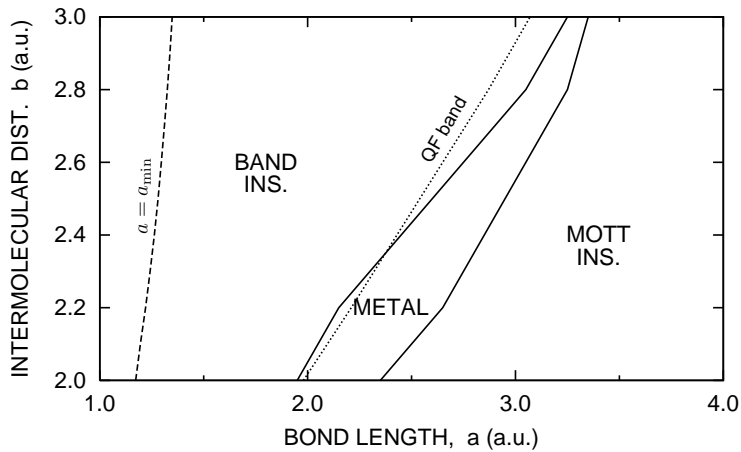}
}\caption{Phase diagram of the planar fermionic ladder.
The optimal value of the bond length $a_{\min}$ is drawn as a \emph{dashed}
line. \emph{Dotted} line marks the effective quarter--filling obtained from
the noninteracting band model.}
\label{phdiag}
\end{figure}
For that purpose we have considered planar (i) and $90^o$ twisted
(ii) ladder composed of $8 \div 12$ atoms. In Fig. \ref{Fig:eglad}
we display the ground state energy (per atom) for the nanoladder
containing $N=10$ atoms with periodic boundary conditions. The
insets provide the values of the optimal bond length $a_{min}$
(left) and the inverse atomic (Gaussian) orbital size $\alpha_{min}$
(right). The Gaussian orbitals (STO-3G basis) have been used to
define the single-particle basis. The horizontal dashed line marks
$H_2$ ground state energy. One should also note that due to the
closed-shell molecular-crystal configuration ($b/a \sim 4$) of the
ground state, the data almost do not depend on the system size (e.g.
analogical results for $N=8$ fit onto those shown in Fig.
\ref{Fig:eglad} up to the pixel size). The characteristics of the
global minima are:
\begin{enumerate}
\item for the planar geometry: $b_{min} = 4.00 a_0$, $a_{min} =
1.422 a_0$, $\alpha_{min} = 1.189 a_0^{-1}$, and $E_G^{min}/N =
-1.1626 Ry$; and
\item for the twisted $90^o$ geometry: $b_{min} = 3.67 a_0$, $a_{min} =
1.426 a_0$, $\alpha_{min} = 1.178 a_0^{-1}$, and $E_G^{min}/N =
-1.1680 Ry$.
\end{enumerate}
These values lead to the binding energies of the molecules in the
nanoladders: $\Delta E_G^{(i)}/N = 20.8 mRy$ and $\Delta
E_G^{(ii)}/N = 26.2 mRy$, respectively, i.e. about three times
larger than for the $H_4$ clusters.\\
\indent One can also draw some conclusions about the nature of
electronic states in those nanoladders. Namely, by calculating the
so called charge stiffness (Drude weight) and the charge gap (cf.
Ref. \cite{ThesisAR}) one can draw a phase diagram shown in Fig. 13
for a planar nanoladder on the plane $a-b$. For either $a$ or $b$
large we expect the Mott insulating state of spins $S=1/2$. However,
it is interesting to note that for bond lengths in the range $2 \div
3$ and an appropriate intermolecular distance $(b \sim a)$ a
quasimetalic state is possible and is followed by a band insulator
for small $b < a$. The metal phase sandwiching the two, band and
Mott insulating phases appears well beyond the optimal configuration
for the nanoladder (see the dashed line representing $a=a_{min}$ as
a function of $b$). So it can appear only when nanoladder is
artificially made on a supporting substrate. The dotted line marks
the model of qurted-filled  (QF) band system discussed in Ref.
\cite{ThesisAR}.

\section{Monoatomic nanowires: from nanometal to Mott insulator}
\subsection{Ground state properties}
\begin{figure}
\resizebox{0.75\columnwidth}{!}{%
  \includegraphics{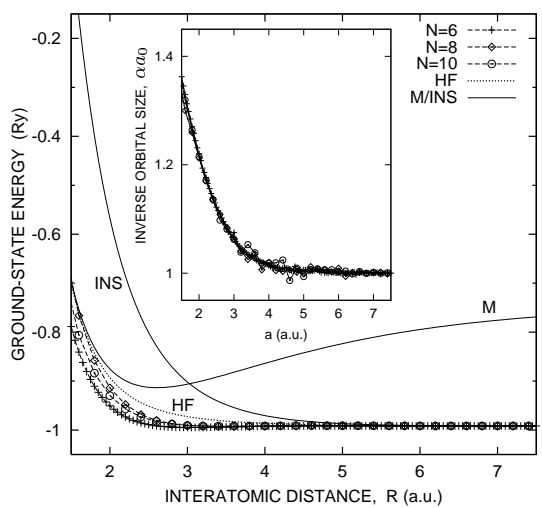}
} \caption{Ground state energy per atom vs. R for the linear chain
with $N = 6 \div 10$ atoms with periodic boundary conditions. The
STO-3G Gaussian basis for representation of atomic orbitals forming
the Wannier function has been used. The inset provides a universal
behavior of the inverse size $\alpha$ of the orbitals. For details
see main text.} \label{Fig11}
\end{figure}

The EDABI method was applied first to nanochains starting with
Slater basis of atomic states
\cite{ThesisAR,SpalekRycerz01,RycerzSpalek01}. In this Section we
present the new results starting from STO-3G "atomic" basis of
adjustable size. The selection of this basis makes the inclusion of
3- and 4-site interactions possible. We limit here to the situation
with one electron per atom. Strictly speaking, we study a nanochain
of hydrogen atoms, which can also model the behavior of single
valence electron nanowires.  \\
\indent We start again with Hamiltonian (\ref{HamAdam}) describing
the so-called extended Hubbard model for a system with one orbital
per atom (here the orbitals are taken as Gaussians of an adjustable
size, out of which we compose the Wannier functions). The
diagonalization in the Fock space is performed with the help of
Lanczos method described in detail elsewhere
\cite{ThesisAR,ThesisEG}. In Fig. \ref{Fig11} we plot the ground
state energy of chains containing $N = 6 - 10$ hydrogen atoms, as a
function of interatomic distance $R$ (in units of $a_0$). The
Wannier functions are calculated in the tight-binding approximation,
with six attractive atomic Coulomb wells representing the periodic
potential. In the inset we display the inverse size $\alpha$ (in
units of inverse $a_0$) of the Gaussian functions (note that the
actual size of atoms is reduces in the correlates state). The
continuous $E_G$ lines INS and M represent respectively to the Mott
insulating state represented by $E_G^{INS} = \epsilon_a^{eff}$, and
to the ideal metallic state, for which
\begin{equation}
E_G^M = \epsilon_a^{eff} - \frac{4 |t|}{\pi} + \frac{1}{2 N}
\sum_{i \neq j} K_{ij} \langle \delta n_i \delta n_j \rangle,
\label{E70}
\end{equation}
where the correlation is taken for the one-dimesional ideal gas
\begin{equation}
\langle \delta n_i \delta n_j \rangle = 2 \frac{ \sin^2 ( \pi
|i-j| /2 )}{( \pi |i-j| )^2}. \label{E71}
\end{equation}
\indent For the sake of completeness, we have also included the
Hartree-Fock (HF) result for $E_G$ (the dotted line) calculated for
the antiferromagnetic (Slater) state; it provides a better estimate
of $E_G$ than either M or INS curves. For discussing HF solution we
take the Hamiltonian (\ref{HamAdam}) in the form
\[
H^{HF} = \epsilon_a^{eff} \sum_{i \sigma} n_{i \sigma} + t \sum_{i
\sigma} \left( a^\dagger_{i \sigma} a_{i+1 \sigma} + h. c. \right)+
\]
\[
+ U \sum_i \left( \langle n_{i \uparrow} \rangle n_{i \downarrow} +
\langle n_{i \downarrow} \rangle n_{i \uparrow} - \langle n_{i
\uparrow} \rangle \langle n_{i \downarrow} \rangle \right) +
\]
\[
\sum_{i < j} K_{ij} \left( \langle \delta n_i \rangle \delta n_j
+\delta n_i \langle \delta n_j \rangle - \langle \delta n_i
\rangle \langle \delta n_j \rangle  \right) -
\]
\begin{equation}
\sum_{i<j \sigma}
K_{ij} \left( \langle a^\dagger_{i \sigma} a_{j \sigma} \rangle
a^\dagger_{j \sigma} a_{i \sigma} + \langle a^\dagger_{j \sigma}
a_{i \sigma} \rangle a^\dagger_{i \sigma} a_{j \sigma} - \left|
\langle a^\dagger_{i \sigma} a_{j \sigma} \rangle \right|^2
\right).
\end{equation}
The HF solution involves calculations of the ground-state energy
with a simultaneous self-consistent determination of the sublattice
magnetization $m = \langle n_{i \uparrow} - n_{i \downarrow}
\rangle$ and of the hopping correlation function $\langle
a^\dagger_{i \sigma} a_{j \sigma} \rangle$,
which will not be discussed in detail here \cite{ThesisAR} (obviously, $\langle \delta n_i \rangle = 0$).\\
\begin{figure}
\resizebox{1.0\columnwidth}{!}{%
  \includegraphics{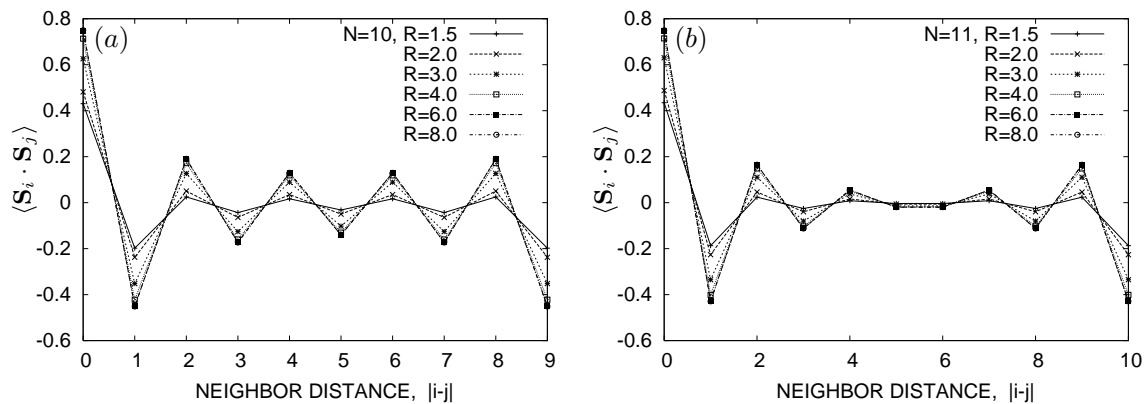}
}\caption{Parity effect on spin ordering: spin--spin correlations
  for nanochains of $N=10$ $(a)$ and $N=11$ $(b)$ atoms. The values
  of the interatomic distance $R$ are specified in the atomic units
  ($a_0=0.529\mbox{ \AA}$).}
\label{ssfig}
\end{figure}
\indent The most spectacular are the spin-spin correlations in the
collective spin-singlet ($\sum_{i=1}^N {\bf S}_i \equiv 0$) state.
In Fig. \ref{ssfig}a we display the  corresponding spin-spin
correlation function $\langle {\bf S}_i \cdot {\bf S}_j \rangle$ as
a function of the distance $|i-j|$ for $N=10$ and several values of
$R$.  We observe N\'{e}el-like state correlations in this spin
singlet state as it oscillates through the whole system length. For
$N=11$ these quasi oscillatory correlations are strongly suppressed
due to the spin-frustration effects, as shown explicitly in Fig.
\ref{ssfig}b. Changing the lattice constant $R$ does not alter the
picture qualitatively. This long-range feature of spin-spin
correlations will have profound consequences on the single-particle
spectrum, as we discuss below. Here the boundary conditions
discussed below (cf. Sec. 4.5) have been considered, although the
difference between the cases a and b survives even for the periodic
boundary conditions. Additional features of the $N$ dependence of
$<{\bf S}_i \cdot {\bf S}_j>$ have been discussed in
\cite{RycerzSpalek06a}
\begin{figure}
\resizebox{0.75\columnwidth}{!}{%
  \includegraphics{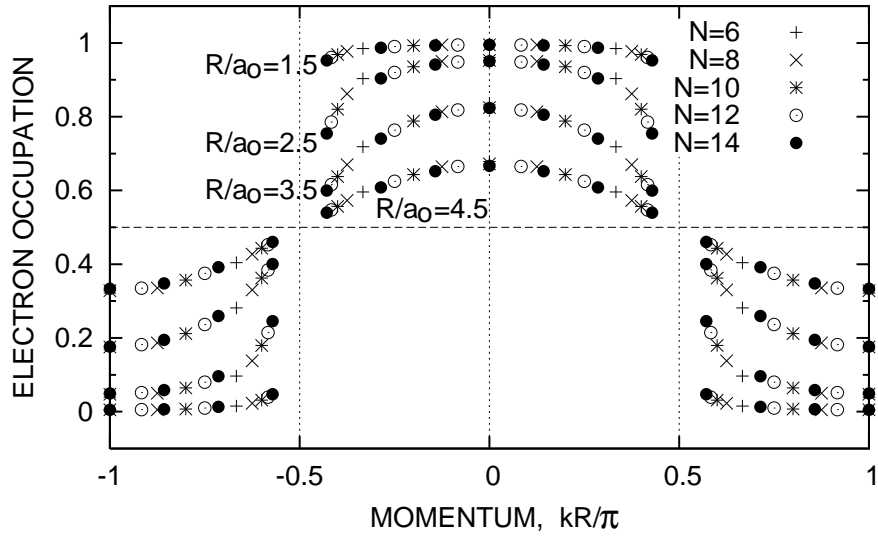}
}\caption{Momentum distribution $\overline{n}_{{\bf k} \sigma}$ for
the number of atoms $N = 6 \div 14$ as it evolves with the
increasing lattice constant$R$.} \label{Fig:MomentumDistribution}
\end{figure}
\begin{figure}
\resizebox{0.75\columnwidth}{!}{%
  \includegraphics{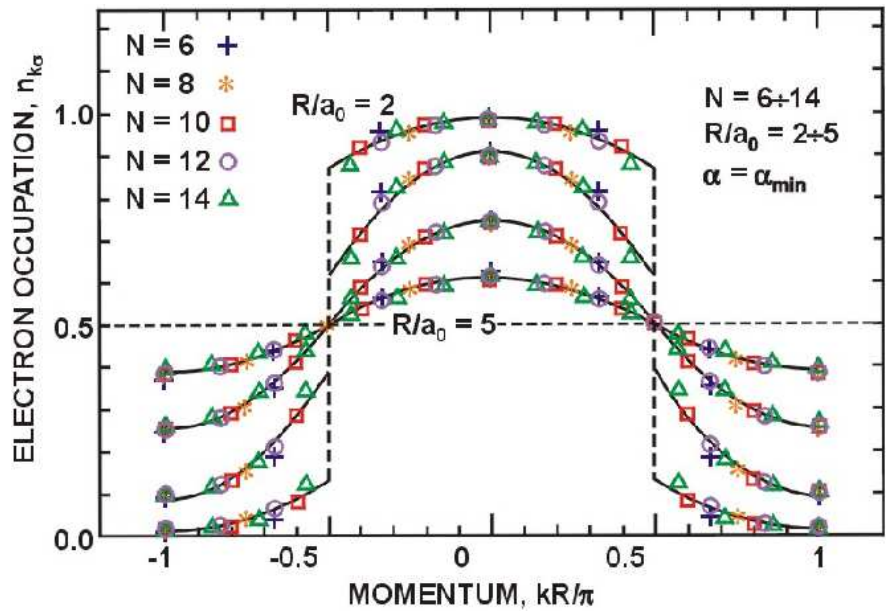}
} \caption{Statistical distribution $n_{\mathbf{k}\sigma}$ for
electrons in a chain of $N=6\div 14$ atoms with periodic boundary
conditions. The interatomic distance $R$ is specified in units of
Bohr radius $a_0$. The continuous line represents the parabolic
interpolation, which is of the same type for both $k>k_F$ and
$k<k_F$.} \label{nkscol}
\end{figure}
\begin{figure}
\resizebox{0.75\columnwidth}{!}{%
  \includegraphics{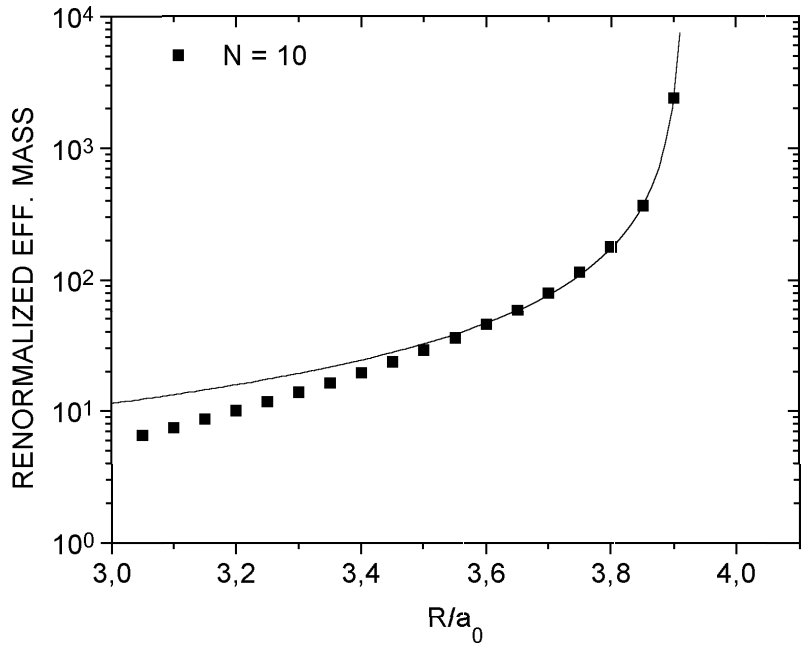}
} \caption{The critical behavior of the quasiparticle mass at the
Fermi level; for details see main text.} \label{meffint}
\end{figure}
\begin{figure}
\resizebox{0.75\columnwidth}{!}{%
  \includegraphics{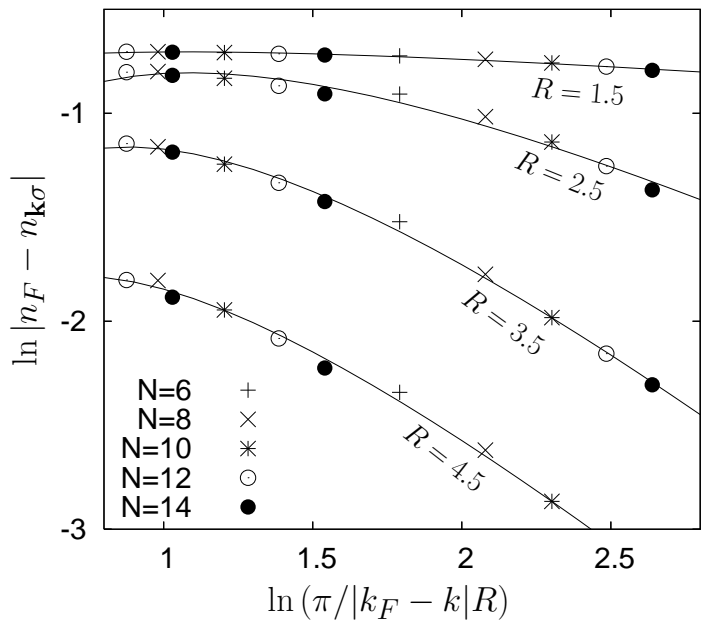}
} \caption{Luttinger--liquid scaling for a half--filled
one--dimensional chain of $N=6\div 14$ atoms. Interatomic distance
$R$ is specified in the units of $a_0$.} \label{llnnpabc}
\end{figure}
\begin{table}[!p]
\caption{Optimized inverse orbital size, microscopic parameters and
the ground--state energy for $N=10$ atoms calculated in Slater--type
basis, as a function of interatomic distance. Intersite Coulomb
repulsion $K_1$ is included on the mean--field level in
$\epsilon^{eff}_a$, Hubbard $U$ term is treated exactly.
Single--particle potential contains \emph{six} Coulomb wells.}
\label{Tab9}
\begin{center}
\begin{tabular}{rclllll}
\hline\hline $R/a_0$ & $\alpha_{\min}a_0$
 & $\epsilon^{eff}_a$ & $t$ & $U$ & $K$ & $E_G/N$ \\
\hline 1.5 & 1.806 &
  0.9103 &  -1.0405 & 2.399 & 1.695 &  0.0665 \\
2.0 & 1.491 &
  -0.1901 & -0.5339 & 1.985 & 1.172 & -0.5179 \\
2.5 & 1.303 &
  -0.6242 & -0.3076 & 1.722 & 0.889 & -0.7627 \\
3.0 & 1.189 &
  -0.8180 & -0.1904 & 1.553 & 0.713 & -0.8800 \\
3.5 & 1.116 &
  -0.9104 & -0.1230 & 1.440 & 0.596 & -0.9391 \\
4.0 & 1.069 &
  -0.9559 & -0.0815 & 1.365 & 0.513 & -0.9693 \\
4.5 & 1.039 &
  -0.9784 & -0.0546 & 1.317 & 0.451 & -0.9848 \\
5.0 & 1.022 &
  -0.9896 & -0.0370 & 1.288 & 0.403 & -0.9926 \\
6.0 & 1.013 &
  -0.9977 & -0.0165 & 1.269 & 0.334 & -0.9982 \\
7.0 & 1.001 &
  -0.9995 & -0.0072 & 1.252 & 0.286 & -0.9996 \\
8.0 & 1.001 &
  -0.9999 & -0.0031 & 1.251 & 0.250 & -0.9999 \\
10.0 & 1.000 &
  -1.0000 &  0.0003 & 1.250 & 0.200 & -1.0000 \\
\hline\hline
\end{tabular}
\end{center}
\end{table}
\begin{table}[!b]
\caption{Same as in Table \ref{Tab9}, calculated in Gaussian--type
STO--3G basis.} \label{Tab10}
\begin{center}
\begin{tabular}{rclllll}
\hline\hline $R/a_0$ & $\alpha_{\rm min}a_0$
 & $\epsilon^{eff}_a$ & $t$ & $U$ & $K$ & $E_G/N$ \\
\hline 1.5 & 1.309 &
  0.1311  & -0.8643 & 2.002 & 1.154 & -0.5684 \\
2.0 & 1.205 &
  -0.5342 & -0.4595 & 1.718 & 0.908 & -0.8154 \\
2.5 & 1.120 &
  -0.7893 & -0.2750 & 1.530 & 0.750 & -0.9139 \\
3.0 & 1.067 &
  -0.8975 & -0.1776 & 1.412 & 0.639 & -0.9567 \\
3.5 & 1.038 &
  -0.9465 & -0.1197 & 1.342 & 0.558 & -0.9756 \\
4.0 & 1.020 &
  -0.9698 & -0.0820 & 1.299 & 0.494 & -0.9841 \\
4.5 & 1.013 &
  -0.9812 & -0.0562 & 1.276 & 0.442 & -0.9881 \\
5.0 & 1.005 &
  -0.9868 & -0.0382 & 1.260 & 0.399 & -0.9901 \\
6.0 & 1.003 &
  -0.9908 & -0.0170 & 1.251 & 0.333 & -0.9914 \\
7.0 & 1.000 &
  -0.9915 & -0.0072 & 1.247 & 0.286 & -0.9917 \\
8.0 & 1.000 &
  -0.9917 & -0.0027 & 1.246 & 0.250 & -0.9917 \\
10.0 & 1.000 &
  -0.9917 & -0.0003 & 1.246 & 0.200 & -0.9917 \\
\hline\hline
\end{tabular}
\end{center}
\end{table}
\begin{table}[!p]
\caption{Same as in Table \ref{Tab9}, calculated in Gaussian--type
STO--3G basis, with the inclusion of the \emph{long--range} Coulomb
interactions.} \label{Tab11}
\begin{center}
\begin{tabular}{rclllll}
\hline\hline $a/a_0$ & $\alpha_{\rm min}a_0$
 & $\epsilon^{eff}_af$ & $t$ & $U$ & $K$ & $E_G/N$ \\
\hline 1.5 & 1.322 &
  0.1340  & -0.8684 & 2.014 & 1.156 & -0.7691 \\
2.0 & 1.208 &
  -0.5338 & -0.4603 & 1.721 & 0.909 & -0.9377 \\
2.5 & 1.119 &
  -0.7894 & -0.2748 & 1.528 & 0.749 & -0.9824 \\
3.0 & 1.063 &
  -0.8977 & -0.1770 & 1.407 & 0.639 & -0.9924 \\
3.5 & 1.030 &
  -0.9466 & -0.1192 & 1.334 & 0.557 & -0.9932 \\
4.0 & 1.011 &
  -0.9697 & -0.0817 & 1.288 & 0.493 & -0.9922 \\
4.5 & 1.006 &
  -0.9812 & -0.0562 & 1.269 & 0.442 & -0.9917 \\
5.0 & 1.006 &
  -0.9868 & -0.0382 & 1.260 & 0.399 & -0.9915 \\
6.0 & 1.002 &
  -0.9908 & -0.0170 & 1.250 & 0.333 & -0.9917 \\
7.0 & 1.000 &
  -0.9915 & -0.0072 & 1.247 & 0.286 & -0.9917 \\
8.0 & 1.000 &
  -0.9917 & -0.0027 & 1.246 & 0.250 & -0.9917 \\
10.0 & 1.000 &
  -0.9917 & -0.0003 & 1.246 & 0.200 & -0.9917 \\
\hline\hline
\end{tabular}
\end{center}
\end{table}
\begin{figure}
\resizebox{0.75\columnwidth}{!}{%
  \includegraphics{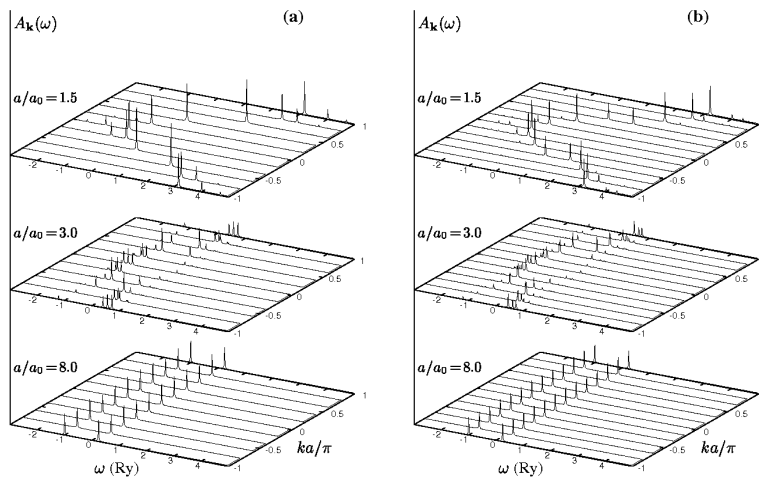}
}\caption{Spectral functions $A( {\bf k}, \omega )$ for $N=10$ (a)
and $N=12$ atoms, with periodic boundary conditions. A clear
dispersion of the states is observed for $R = 1.5 a_0$, which
transform into the atomic peaks with the increasing $R$, through an
incoherent regime for $R/a_0 \approx 3$}
\label{Fig:SpectralFunction}
\end{figure}
\begin{figure}
\resizebox{1\columnwidth}{!}{%
  \includegraphics{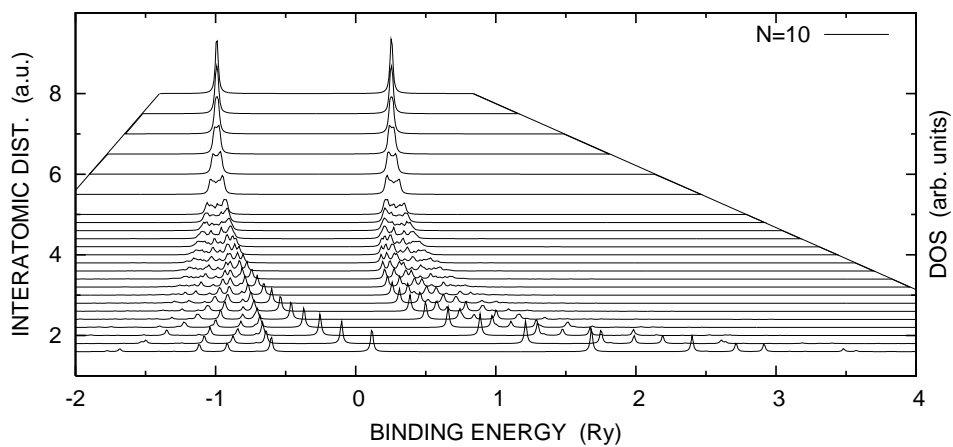}
}\caption{Density of single-particle states for $N=10$ atoms. Note a
pronounced incoherent spectrum for $R$ in the range $3 \div 5 a_0$.}
\label{Fig16}
\end{figure}
\begin{figure}
\resizebox{\columnwidth}{!}{%
  \includegraphics{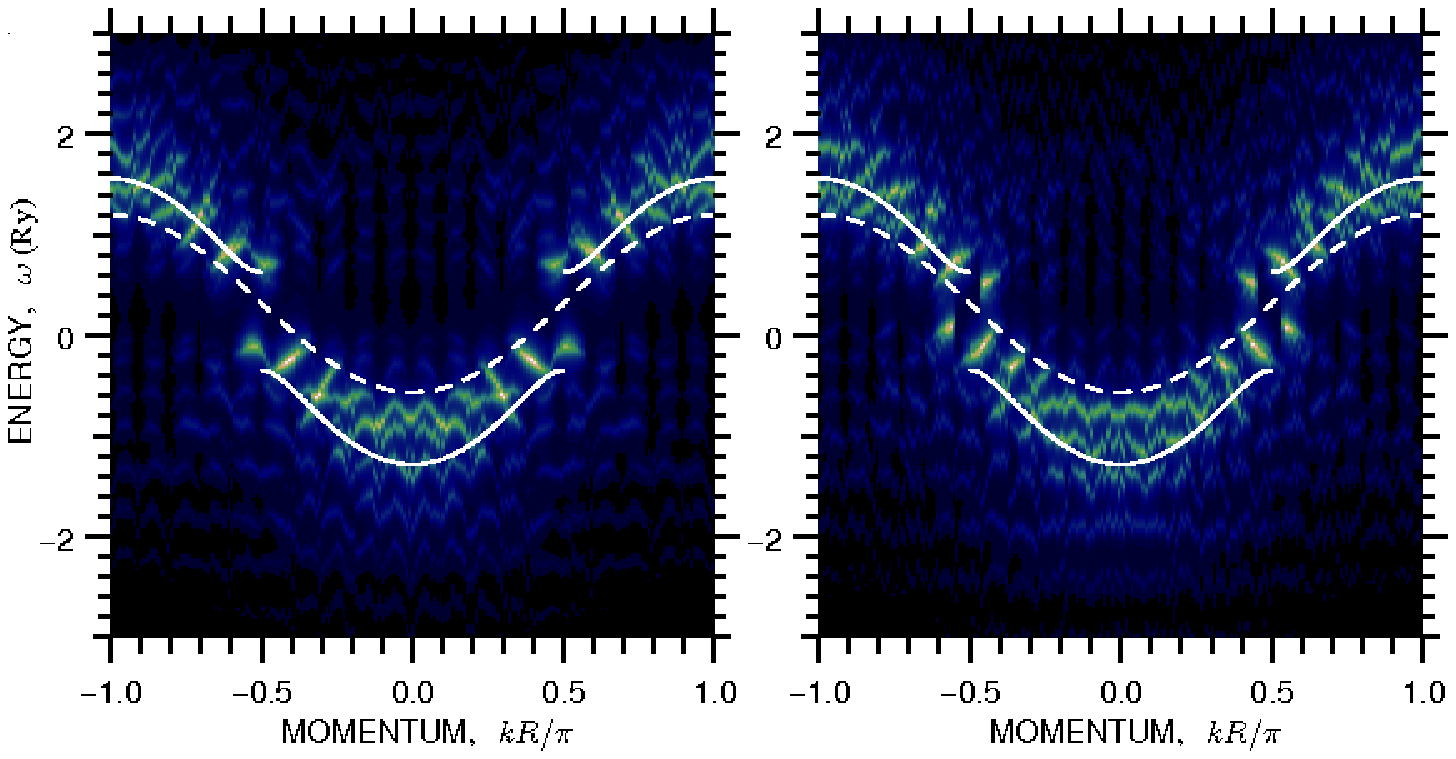}
} \caption{Spectral-density-peak positions for the nanochain of
$N=10$ (\emph{left} panel) and $N=11$ (\emph{right} panel) atoms
with generalized boundary conditions. The Hartree--Fock
(\emph{solid} line) and noninteracting system (\emph{dashed} line)
dispersion relations  are shown for comparison.} \label{ekcol}
\end{figure}
\subsection{Momentum distribution in the ground state: Fermi or Luttinger nano liquid localization?}
We now introduce the particle quasimomentum distribution $n_{{\bf k}
\sigma}$ in the ground state for the situation with one electron per
atom; this is displayed in Fig. \ref{Fig:MomentumDistribution} for
the number of atoms $N = 6 \div 14$. We observe a very universal
character of the curves provided the periodic boundary conditions
are taken for the chains of $N = 4n + 2$ atoms, whereas the
antiperiodic boundary conditions are taken for $N = 4n$ atoms, with
$n$ being a natural number. It is very tempting to regard the
distributions for $R / a_0 \leq 3.5$ as modified Fermi-Dirac
function characterizing the Fermi liquid even for such short chains
for which we have discrete momentum states. However, for $R/a_0
\rightarrow 4$ the distributions becomes continuous, i.e. without an
apparent Fermi ridge at Fermi momentum $k_F = \pi / (2R)$, in rough
agreement with the preliminary results for the case with the
Slater-type orbitals \cite{SpalekRycerz01}. So, a crossover with the
increasing interatomic distance from the Fermi liquid-like behavior
to the chain of atomic-like states is clearly seen. It should be
noted that the Luttinger liquid-like scaling fitting to the points
displayed in Fig. \ref{Fig:MomentumDistribution} is
equally convincing, at least for this half-filled band configuration, as discussed next.\\
\subsection{Single-particle spectral density function evolution}
The state of a nanochain viewed through the statistical distribution
of momentum states can be discussed as follows. The ideal Mott
localized state, as is seen in the bulk systems, is not possible for
finite system. This is simply because the length of the system is
finite and therefore, the probability of the electron tunneling from
an  atomic state on one chain end to the other end is {\em nonzero}.
The quantitative question is, whether it is of the order
$e^{-N\alpha R}$ or larger. Due to the fact that the overlap
integral is $\sim \frac{1}{3}(\alpha R)^2 e^{-\alpha R}$, the
probability  is enhanced by a nonzero hopping amplitude between more
distant neighbors. Having in mind this circumstance, it is at least
imaginable, that the distribution function displayed in Fig.
\ref{Fig:MomentumDistribution} is Fermi-Dirac like for $\alpha R
\sim 1$. One can thus try to formalize this observation further.
Namely, in Fig. \ref{nkscol} we plot $n_{{\bf k} \sigma}$ for a half
filled band system containing up to $N=14$ electrons. The continuous
lines represent the parabolic parametrization:
\begin{equation}
n_{{\bf k} \sigma} = \frac{1}{2} sgn (k - k_F) \left[ \alpha (k -
k_F)^2 + \beta |k - k_F|  + \gamma \right],
 \label{EParabolicParam}
\end{equation}
for both $k < k_F$ and $k > k_F$ (note that the Fermi point is not
occupied for $N$ even). This parametrization, in fact implies a
discontinuous distribution, allows us to interpret the distribution
discontinuity $\Delta n_{k_F} \equiv n_{k_F -0} - n_{k_F +0}$ in
terms effective mass enhancement at the Fermi point.
\begin{equation}
m^*_F/m_{BAND} = (\Delta n_{k_F})^{-1} \equiv Z^{-1},
\end{equation}
where $Z^{-1}$ is the usual Fermi-liquid enhancement factor. The
corresponding enhancement factor determined in that manner is
plotted in Fig. \ref{meffint} (squares), whereas the solid line
represents the finite-size scaling with interatomic distance,
$m^*_F/m_{BAND} = A | R-R_C|^{-\gamma}$, with $A \simeq 10.2$, the
localization thereshold is $R_C a_0 = 3.92$, and the critical
exponent has the approximate value $\gamma = 4/3$. This behavior
emulates a quantum critical behavior and should not be taken
literally. Nonetheless, the value of $R_C$ distinguishes {\em
qualitatively} between the nanometallic state $R<R_C$ (for which the
distribution (\ref{EParabolicParam}) has a jump at $k_F$) and the
Mott insulating (semiconducting) state
($R>R_C$, the distribution function is a continous function of $k$).\\
\indent However, the situation is not that simple. One can ask the
question whether this short one-dimensional system is not rather
resembling the Luttinger liquid-like behavior. This task has been
undertaken seriously and in Fig. \ref{llnnpabc} we plot the fitting
of the data displayed in Fig. \ref{Fig:MomentumDistribution} to the
corresponding dependence for the Tomanaga-Luttinger liquid, with the
logarithmic correction included, namely \cite{RycerzSpalek04}
\begin{equation}
\ln |n_{k_F} - n_{k \sigma}| = - \Theta \ln z + b \ln \ln z +c +
\theta(1/\ln z),
\end{equation}
where $z = \pi/|k_F - \pi/R|$,$\Theta$,$b$, and $c$ are constants.
The parameters depend on the distance $R$, as shown elsewhere
\cite{RycerzSpalek04}. This fitting provides also the localization
thereshold $R_C \simeq 2.60 a_B$, which is reduced by 50\% from the
value for the Fermi liquid interpretation of $n_{k\sigma}$. One
should also note that the present interpretation of the exact
solution displayed in Fig. \ref{Fig:MomentumDistribution} does not
admit a discontinuity at $k=k_F$. However, it is quite amazing that
both the Fermi- and Luttinger-liquid interpretations can provide
satisfying interpretation of $n_{k\sigma}$ to an equal degree.  This
means that there must be an underlying universal behavior of a new
type, incorporating Fermi- and Luttinger-liquid concepts, at least
semiquantitatively. Nonetheless, for the not too large $R$ the
Fermi-Dirac-like distribution fits better the $k$ dependence. The
situation for odd $N$ requires an explicit discussion of boundary
conditions and is provided below.\\
 \indent  To  characterize the spectrum of single-particle
excitations we use the definition of the spectral-density function
\begin{equation}
A({\bf k},\omega )= \sum_n \left| \langle \Phi_n^{N \pm 1} |
a^\dagger_{{\bf k} \sigma} | \Phi_0^N \rangle \right|^2 \delta \left[ \omega
- ( E_n^{N+1} - E_0^N ) \right], \label{E73}
\end{equation}
where upper(lower) sign corresponds to the energies with $\omega >
\mu$ ($\omega < \mu$), respectively, $| \Phi_n^N \rangle$ is the
n-th eigenstate of the system containing $N$ particles, $E_n^N$ is
the corresponding eigenenergy, and $<..>$ is the matrix element are
calculated within the Lanczos algorithm \cite{dago}. In Fig.
\ref{Fig:SpectralFunction} we present the panel of $A({\bf k},\omega
)$ for three distances $R$. For small $R$, a clear two-peak
structure appears at the Fermi momenta $k_F =\pm \pi / (2R)$ for $N
= 12$; an artificial broadening of the peaks appears because we use
the approximate expression for the $\delta (x)$ function: $\delta(x)
\approx ( 1/\pi ) \epsilon / (x^2 + \epsilon^2)$, with $\epsilon =
10^{-2}Ry$. The splitting is caused by the antiferromagnetic
correlations depicted in Fig. \ref{ssfig}a. In the range $R/a_0 \leq
2$ the quasiparticles are well defined, but an incoherent part grows
with the increasing $R$. For $R/a_0 \sim 3 \div 4$ the {\em Hubbard
subbands} are formed and evolve continuously into atomic levels
located respectively at $\omega = \epsilon_a$ and $\omega =
\epsilon_a + U$ for $R \rightarrow \infty$.  Those two limiting peak
positions correspond to the ground $H^0$ and excited $H^-$ states,
respectively. Combining the last results with the corresponding
discussion for the $H_N$ clusters one sees that {\em the Hubbard
subband structure represents a universal feature of nanoscopic
systems}. In the limit of larger interatomic separation this
structure is clearly distinguishable from the discrete level
structure coming from geometric quantization of
this confined system. \\
\indent To demonstrate the importance of {\em incoherent} part of
the spectrum we have calculated the density of states $\mathcal{N}(
\omega ) = \sum_{\bf k} A({\bf k},\omega )$, displayed in Fig.
\ref{Fig16} for $N = 10$ atoms, for the interatomic distances
specified. In the regime $R = 3 \div 5 a_0$ well resolved
quasiparticle peaks coalesce into a complicated random-like
(incoherent) spectrum, out of which clean atomic peaks emerge for
 larger $R$. Note that the intermediate regime
corresponds to the situation for which the bare bandwidth $W = 4
|t|$ of the single particle states fulfils the condition $W \sim U$,
i.e. the electronic system switches from the  weak- to the rather
strong-correlations regime, corresponding to the
delocalization-localization crossover of the Mott-Hubbard type
taking place. For one electron per atom the lower energy manifold is
filled with electrons, whereas the upper is empty. The presence of
the incoherent spectrum for $R \approx 3 \div 5 a_0$ seems to
represent also a universal feature, as it appears also for larger
values of $N$.
\subsection{Slater vs. Gaussian trial basis}
We now compare the results obtained within EDABI when either the
adjustable Slater or the STO-3G functions are used. Probably, most
interesting is to compare the results for the Hubbard model with
intersite Coulomb interactions taken in the Hartree-Fock
approximation (only then the results for the Slater 1s-like basis
are energetically stable. In Table \ref{Tab9} we list the results
for the Slater basis containing both $E_G$ and microscopic
parameters for a ring of $N = 10$ atoms. The same for the case of
STO-3G basis is shown in Table \ref{Tab10}. One sees that the
Gaussian basis provides a lower value of $E_G$. This is because in
the Slater case we have neglected 3- and 4- site interactions in the
atomic basis that represents a rather crude approximation. On the
contrary, the results for $E_G$ obtained for the STO-3G basis for $N
= 10$ atoms are very close (within $10^{-2} Ry$) to those obtained
for an exact solution for an infinite chain \cite{LiebWu} and the
values of the microscopic parameters within $10^{-3}$
\cite{GorlichKurzyk}. Because of the good accuracy of the results
for the Gaussian basis we have also provided in Table \ref{Tab11}
the results for the chain of $N = 10$ atoms when the long-range part
of the Coulomb interactions $\sim K_{ij}$ are included. These
results provide the range of variation of the microscopic parameters
when compared to either $H_2$ (cf. Table \ref{Tmicropar}) or $U_1$
value for $He$ atom (cf. Table \ref{TH2micro}).
\subsection{The role of boundary-conditions: parity effects}
\begin{figure}
\resizebox{0.75\columnwidth}{!}{%
  \includegraphics{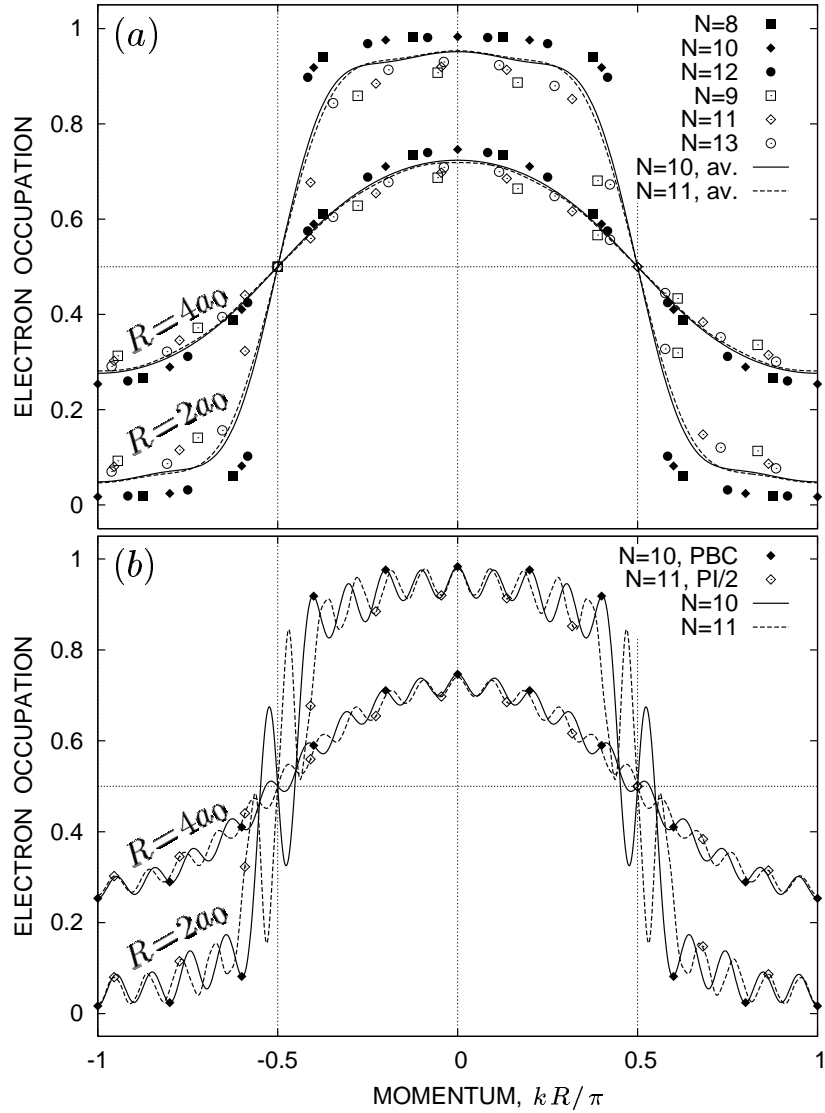}
} \caption{Electron momentum distribution for chains of $N=8\div 13$
atoms: $(a)$ datapoints for \emph{optimal} boundary conditions
(BC\textit{s}) and the sample $n(k)$ curves averaged over
BC\textit{s} (\emph{solid} lines); $(b)$ the original $n(k_q(\phi))$
functions used for the averaging.} \label{parinks}
\end{figure}

It is believed that system properties in the thermodynamic limit ($N
\rightarrow \infty$) are the same regardless of the boundary
conditions used. This claim has been tested on many model systems.
However, for finite clusters the boundary conditions are crucial.
For the chains studied here when $N = 4n + 2$ (where n is a natural
number) the periodic boundary conditions (PBC) are used, whereas for
$N = 4n$ the antiperiodic boundary conditions (ABC) lead to a
lower-energy state. Namely, the terminal-atomic annihilation
operators should be defined as follows
\begin{equation}
\left\{ \begin{array}{ll} a_{N+1,\sigma} \rightarrow a_{1 \sigma} &  \; for PBC, \\
a_{N+1,\sigma} \rightarrow -a_{1 \sigma} &  \; for ABC.
\end{array} \right.
\end{equation}
Therefore, the terminal hopping term involving the end atoms changes
sign for ABC, while the interaction terms $\sim n_{i \uparrow} n_{i
\downarrow}$, $n_i n_j$ and ${\bf S}_i \cdot {\bf S}_j$ remain
unaltered. Obviously, the periodic boundary conditions are proper
for any $N$, when we have a ring geometry. The situation becomes
more involved when $N$ is an odd number. Namely, we write down the
Hamiltonian (\ref{HamAdam}) in the form
\begin{equation}
H = \epsilon^{eff}_a \sum_{i \sigma} n_{i \sigma} + t \sum_{j
\sigma} e^{-i\Phi / N} a^{\dagger}_{j \sigma} a_{j+1 \sigma} + U
\sum_i n_{i \uparrow} n_{i \downarrow} + \sum_{i < j} K_{i j} \delta
n_i \delta n_j , \label{HamAdamPhi}
\end{equation}
where $\Phi$ is the fictious (dimensionless) flux through the ring.
One can show that the unitary transformation $c_{j\sigma}
\rightarrow e^{-i\Phi / N} c_{j\sigma}$ allows for accumulation of
complex phase factor in the terminal hopping term, which then takes
the form $t(e^{-i \Phi} a^\dagger_{N \sigma} a_{1 \sigma} + h.c.)$
and this can be regarded as generalized boundary condition $a_{N+1
\sigma} \rightarrow e^{i\Phi / N} a_{1\sigma}$. We do not
distinguish between the system with a fictious flux and the
generalized boundary conditions. The presence of the flux can be
regarded as an accumulation of the Berry phase during the motion of
individual electron in the milieu of all other electrons. With such
interpretation the boundary conditions apply also to linear chain.\\
\indent The electron momentum for nanochains with such boundary
conditions is displayed in Fig. \ref{parinks}. The discrete momenta,
corresponding to the solution of the single particle part of
(\ref{HamAdamPhi}) for a finite $N$, are given by
\begin{equation}
k_n = \frac{2\pi n - \Phi}{N}, 0 \leq n \leq N
\end{equation}
The optimal BCs, corresponding to the minimal ground-state energy
$E_G$, with  respect to $\Phi$, are realized for $\Phi=0$ when
$N=6,10,14, \ldots$ (periodic BCs), $\Phi = \pi$ when
$N=4,8,12,\ldots$ (antiperiodic BCs), and $\Phi = \pi/2, 3/2 \pi$
when $N$ is odd. A basic analysis of Eq. (\ref{HamAdamPhi}) shows,
that for the optimal BCs, the Fermi momentum value $k_F = \pi/(2R)$
is never reached for even $N$, whereas for odd $N$ it happens for a
single value of $n$. This circumstance has important implication for
the nanochain electronic structure, however, almost does not affect
its transport properties, as discussed below. The exact meaning of
the averaging procedure when drawing Fig. \ref{parinks}, is
elaborated elsewhere. Analogously, the effect of BCs on the
spin-spin correlation function $<{\bf S}_i \cdot {\bf S}_j>$ is also
touched upon there.\\
\indent An explicit form of the dispersion relation derived from the
spectral density functions with inclusion of the generalized
boundary conditions, is shown in Fig. \ref{ekcol}. We see again,
that the spin splitting is present. Also, one can see the difference
between either dispersion relation for noninteracting particles or
with that calculated in the Hartree-Fock approximation for the
Slater antiferromagnetic chain.

\section{Transport in nanosystems}
In this Section we complement the discussion of nanochains by providing
the values of Drude weight (charge stiffness) and optical gap. We also
show, how the EDABI method can be used to calculate the conductance of an
open system: a quantum dot attached to the leads.

\subsection{Drude weight for nanochains}

The real part of the optical conductivity at zero temperature is determined by
the corresponding real part of the linear response to the applied electric
field \cite{shamil}, and can be written as $\si(\om)=D\de(\om) +
\si_{\rm reg}(\om)$, where the regular part is
\begin{equation}
\label{sireg:def}
  \si_{\rm reg}(\om)=\frac{\pi}{N}\sum_{n\neq 0}\frac{\abs{\bra{\Psi_n}j_p
  \ket{\Psi_0}}^2}{E_n-E_0}\de\left(\om-(E_n-E_0)\right),
\end{equation}
whereas the Drude weight (the \emph{charge stiffness}) $D$ is defined by
\begin{equation}
\label{d:def}
  D=-\frac{\pi}{N}\bra{\Psi_0}T\ket{\Psi_0}-\frac{2\pi}{N}
    \sum_{n\neq 0}\frac{\abs{\bra{\Psi_n}j_p\ket{\Psi_0}}^2}{E_n-E_0},
\end{equation}
where the kinetic--energy term $T$ is the same as the second term in Eq.\
(\ref{HamAdamPhi}), and the diamagnetic current operator is defined as usual:
$j_p= it\sum_{j\si}(a_{j\si}^{\dagger}a_{j+1\si}-\mbox{h.c.})$.
Here the states $\ket{\Psi_n}$ in Eqs.\ (\ref{sireg:def}) and (\ref{d:def})
are the eigenstates of the Hamiltonian (\ref{HamAdamPhi}) corresponding to the
eigenenergies $E_n$, again with boundary conditions which minimize the
ground--state energy for a given system size $N$ (see \emph{previous}
Section).
We calculate matrix elements $\bra{\Psi_n}j_p\ket{\Psi_0}$ numerically
by adapting the method developed by Dagotto \cite{dago}, which is stable
for \emph{even} number of electrons only.
Namely, we diagonalize the Hamiltonian (\ref{HamAdamPhi}) in the Fock space
with the help of Davidson technique \cite{davidson}, and then repeat
the procedure starting from a specially prepared initial state
$j_p\ket{\Psi_0}$. Replacing the Lanczos scheme, utilized in Ref.\ \cite{dago},
by the Davidson one provides an excellent accuracy for either \emph{even} or
\emph{odd} number of electrons.
\begin{figure}
\resizebox{0.75\columnwidth}{!}{%
  \includegraphics{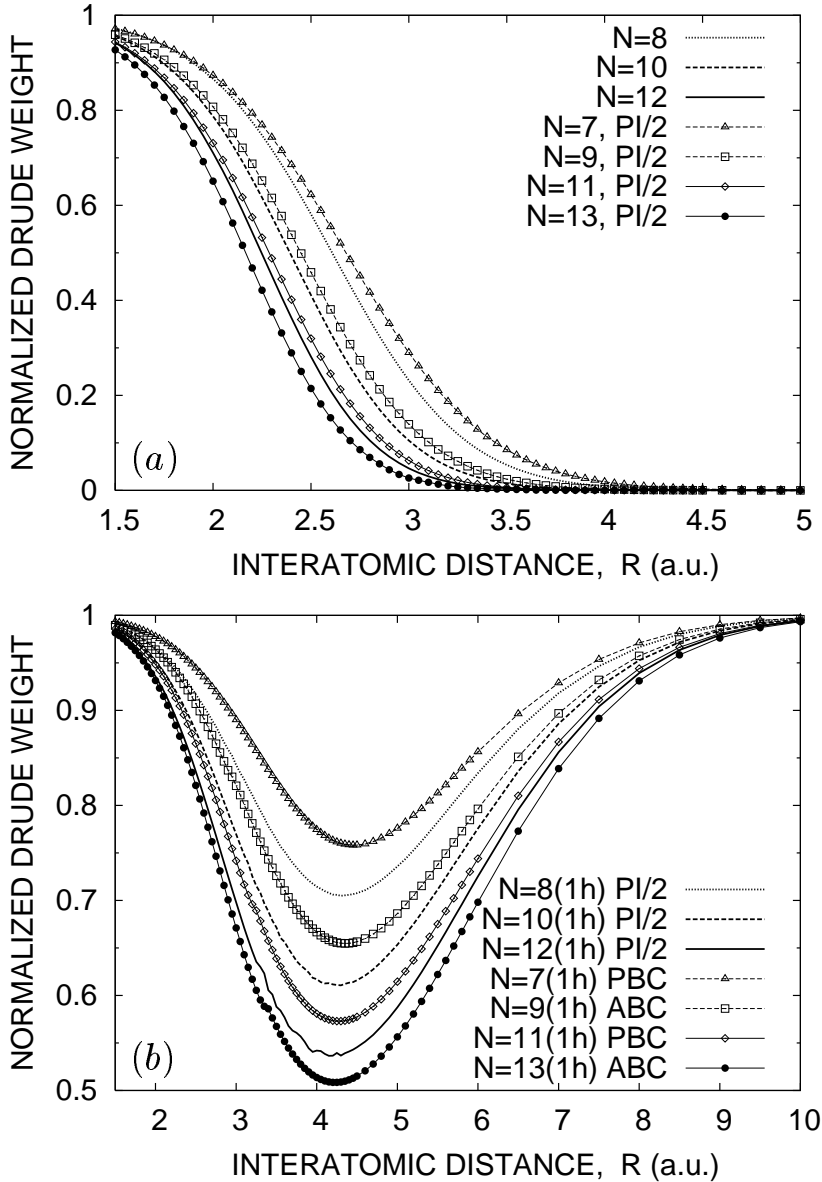}
}
\caption{Normalized Drude weight for nanochains in the \emph{half--filled}
case $(a)$, and for a system with a single hole $(b)$. The \emph{optimal}
boundary conditions are specified for each dataset.}
\label{sigd}
\end{figure}
\begin{figure}
\resizebox{0.75\columnwidth}{!}{%
  \includegraphics{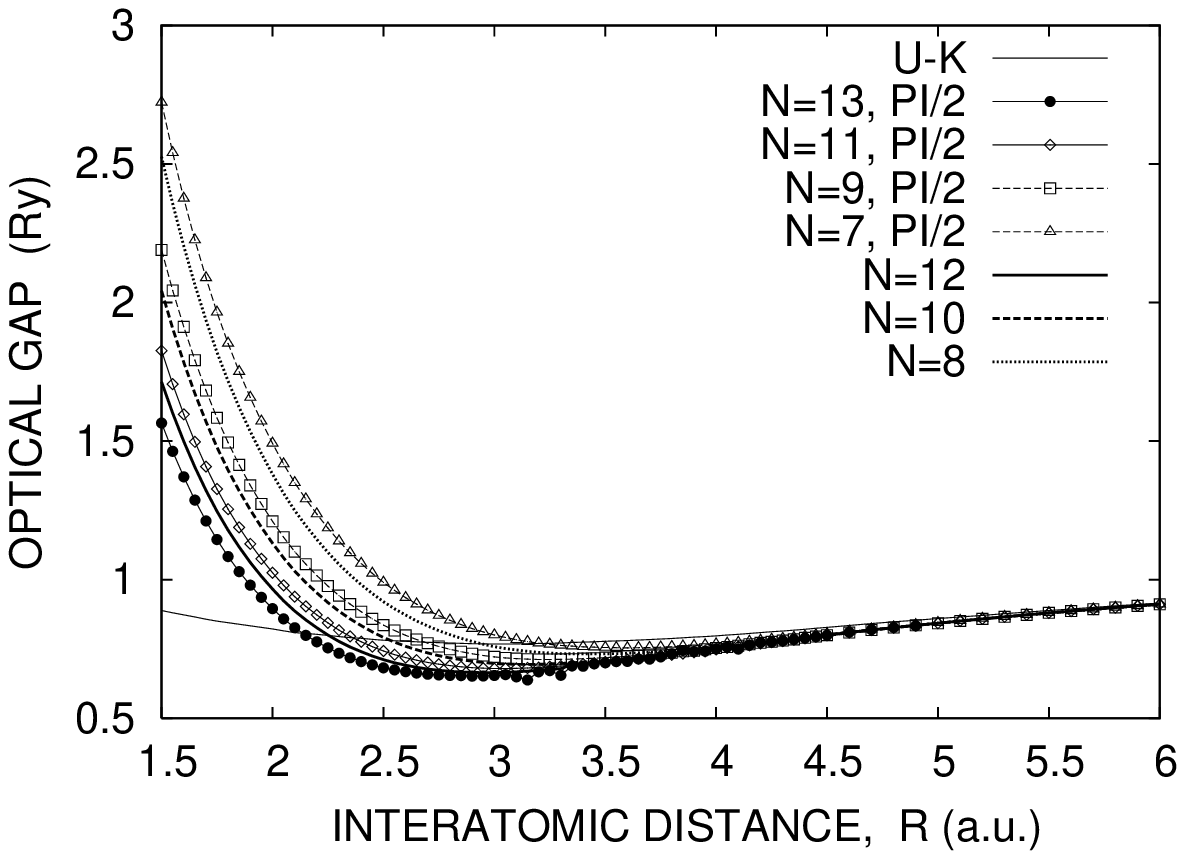}
} \caption{Optical gap for nanochains in the \emph{half--filled}
case.} \label{ogap}
\end{figure}
The evolution of Drude weight with $N$ and $R$ is shown in Fig.\
\ref{sigd}. In the half--filled case $N_e=N$ (\emph{cf.}\ Fig.\
\ref{sigd}a) Drude weight gradually decrease with $N$, as we have
shown for even $N$ \cite{ryspa}. The most interesting feature of
these results is, that the curves for odd $N$ fits smoothly between
those for even $N$, with very weak parity effect, totally
incomparable with that present in the charge gap and electron
momentum distribution \cite{paripss}. This observation can be
understood when we take into account, that the Drude weight defined
by Eq.\ \ref{d:def} is the \emph{integral} quantity, involving the
summation over all the excited states of the Hamiltonian Eq.\
(\ref{HamAdamPhi}), so it cannot be determined only by the
electronic structure near the Fermi points, particularly for a small
system.

The parity effect on Drude weight disappears for the system with
a single hole ($N_e=N-1$, \emph{cf.}\ Fig.\ \ref{sigd}b), in which
the magnetic frustration is absent.
For this case, the Drude weight evolution with $R$ is very interesting.
In the weak--correlation range ($R/a_0\lesssim 2$) the chain shows a
highly--conducting behavior for each $N$. Next, in the intermediate range
($R/a_0=4\div 5$) the Drude weight decrease rapidly with $N$, indicating an
insulating (Mott--Hubbard) state in the large--$N$ limit.
In the strongly--correlated range ($R/a_0\sim 10$) the Drude weight
approaches again its maximal value $D=1$.
Such a behavior can be explained, when we analyze the situation in two steps:
\emph{First,} for low values of $R$, the bandwidth--to--interaction ratio
is small, and the system with a single hole does not differ significantly
from a half--filled one. This is why in both cases Drude weight decreases
gradually with both $N$ and $R$, as the tunneling amplitude through the barrier
of a finite width.
\emph{Second,} for the largest values of $R$, the system can be described
by an effective $t-J$ model \cite{spatj} with a coupling constant
$J=4t^2/(U-K)\ll |t|$  (where $K\equiv K_{j,j+1}$ denotes the nearest
neighbor Coulomb repulsion), which corresponds to an asymptotically--free
hole motion.
Then, it become clear that in the intermediate range the Drude weight has
to suppressed, what can be interpreted in terms of a partially localized
spin--ordered state.
It would be very interesting to test experimentally this result,
possibly for a mesoscopic atomic ring.

The smooth behavior of the system transport properties with $N$ is illustrated
in Fig.\ \ref{ogap} on the example of other relevant quantity, the
\emph{optical gap} $\Delta E_\mathrm{opt}=E_i-E_0$ (where $E_i$ is the energy
lowest-lying excited level $\ketaa{\Psi_i}$, for which
$\braaa{\Psi_i}j_p\ketaa{\Psi_0}\neq 0$).
Again, adding a single hole totally removes the weak parity effect (not shown).
The large-$R$ behavior of $\Delta E_\mathrm{opt}$ could be explained by
the value for an insulating limit $\Delta E_\mathrm{opt}\approx U-K$,
whereas small $R$ we have a strong $N$ dependence similar to those observed
for $D$. The later shows, that the contribution of the state $\ketaa{\Psi_i}$
to the sum (\ref{d:def}) is dominating for such a \emph{partly-localized}
quantum nanoliquid \cite{ryspa}.

\subsection{H$_2$ molecule as a quantum dot}

We have recently shown, that EDABI method is useful to investigate
the zero--temperature conductance of diatomic molecule, modeled as a
correlated double quantum dot attached to noninteracting leads
\cite{RycerzSpalek06}. This became possible when utilizing the
Rejec--Ram\v{s}ak formulas, relating the linear--response
conductance to the ground--state energy dependence on magnetic flux
\cite{reram}. Here we complement the discussion with the stability
analysis showing, that large coupling between molecule and the leads
may provide the possibility for interatomic distance manipulation in
an experiment.

\begin{figure}
\resizebox{0.75\columnwidth}{!}{%
  \includegraphics{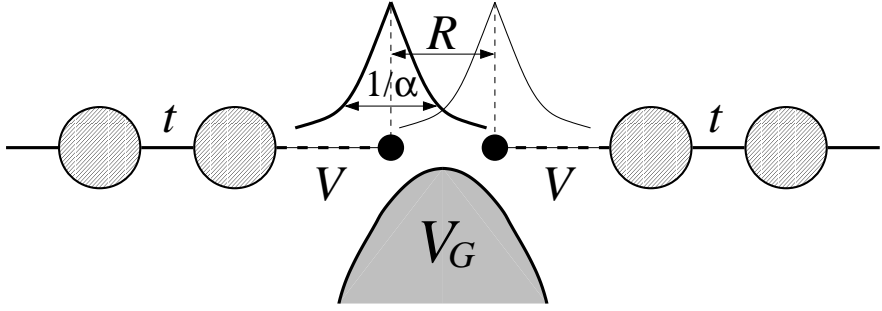}
}
\caption{Diatomic molecule as a double quantum dot.}
\label{qd2fig}
\end{figure}

We start with the Hamiltonian which is a generalization of the
Anderson impurity model \cite{ander}, and can be written as
\begin{equation}
\label{ham5}
  H = H_L+V_L+H_C+V_R+H_R,
\end{equation}
where $H_C$ models the central region,
$H_{L(R)}$ describes the left (right) lead, and $V_{L(R)}$ is the coupling
between the lead and the central region.
Both $H_{L(R)}$ and $V_{L(R)}$ terms have a tight--binding form, with the
hopping $t$ and the tunneling amplitude $V$
\begin{equation}
\label{hamlr}
  H_{L(R)}=t\sum_{j\neq L(R),\si}(a^{\dagger}_{j\si}a_{j+1,\si}+\mbox{h.c.}),
\end{equation}
\begin{equation}
\label{hamv}
  V_{L(R)}= V\sum_\si(a^{\dagger}_{L(R)\si}a_{1(2)\si}+\mbox{h.c.}),
\end{equation}
as depicted schematically in~Fig.\ \ref{qd2fig}.
The index $j=L(R)$ corresponds to the last atom of the left (right) lead.
The central--region term $H_C$ is a version of Hamiltonian (\ref{HamAdam})
for two atoms, and describes a double quantum dot with
electron--electron interaction
$$
  H_C = (\epsilon_a-eV_G)\sum_{j=1,2}n_i
  -t'\sum_{\sigma=\uparrow,\downarrow}(a_{1\sigma}^{\dagger}a_{2\sigma}
 +\mbox{h.c.})
$$
\vspace{-1.5em}
\begin{equation}
\label{hamc}
 + U\sum_{j=1,2}n_{i\uparrow}n_{i\downarrow} + Kn_1n_2 + (Ze)^2/R,
\end{equation}
where $\epsilon_a$ is atomic energy, $V_G$ is an external gate voltage,
$t'$ is the internal hopping integral, $U$ and $K$ represents the
intra-- and inter--site Coulomb interactions, respectively, and the last term
describes the Coulomb repulsion of the two ions at the distance $R$.
Here we put $Z=1$ and calculate all the parameters $\epsilon_a$, $t'$, $U$,
and $K$ as the Slater integrals \cite{slat} for $1s$--like hydrogenic orbitals
$\Psi_{1s}(\mathbf{r})=\sqrt{\alpha^3/\pi}\exp(-\alpha|\mathbf{r}|)$,
where $\alpha^{-1}$ is the orbital size (\emph{cf.}\ Fig.\ \ref{qd2fig}).
The parameter $\alpha$ is optimized to get a minimal ground state energy
for whole the system described by the Hamiltonian (\ref{ham5}). Thus,
following the idea of EDABI method \cite{ryspa}, we reduce the number
of physical parameters of the problem to just a three: the interatomic
distance $R$, the gate voltage $V_G$, and the lead--molecule coupling
$V$ (we put the lead hopping $t=1\ \mathrm{Ry} = 13.6\ \mathrm{eV}$ to work in
the wide--bandwidth limit).

\begin{figure}
\resizebox{0.75\columnwidth}{!}{%
  \includegraphics{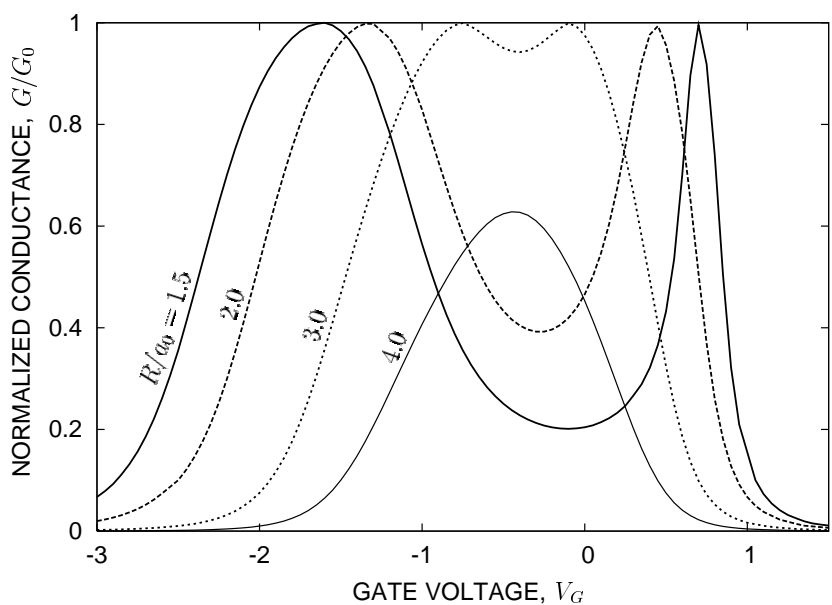}
}
\caption{
  Zero--temperature conductance of the system in Fig.\ \ref{qd2fig}
  as a function of the gate voltage $V_G$ and interatomic distance $R$.
  $1s$ orbital size $\alpha^{-1}$ is optimized variationally.
  The lead parameters are $t=1\mathrm{ Ry}$ and $V=0.5t$.
}
\label{gH2fig}
\end{figure}

\begin{figure}
\resizebox{\columnwidth}{!}{%
  \includegraphics{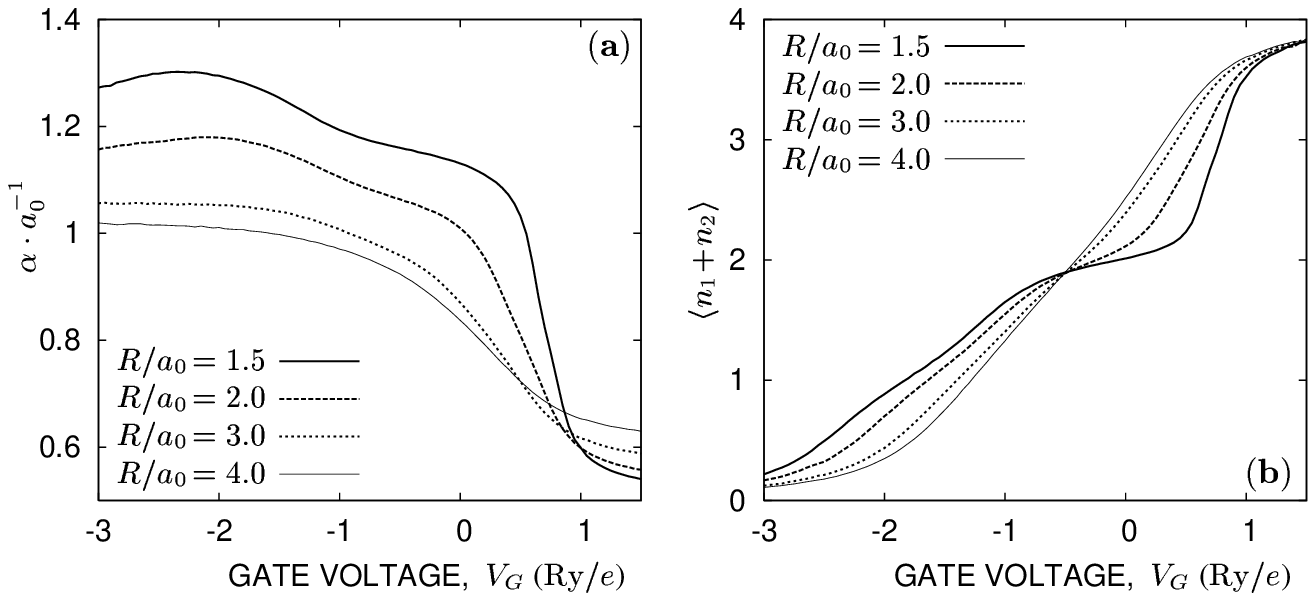}
}
\caption{
  The optimized inverse inverse orbital size $(a)$ and average central
  region occupancy $(b)$ for the diatomic molecule attached to noninteracting
  leads characterized by the parameters $t=1\mathrm{ Ry}$ and $V=0.5t$.
}
\label{AlpNjfig}
\end{figure}

We discuss now the molecule conductivity calculated from the Rejec--Ram\v{s}ak
two--point formula \cite{reram}:
\begin{equation}
  G=G_0\sin^2\pi[E(\pi)-E(0)]/2\Delta,
\end{equation}
where $G_0=2e^2/\hbar$ is the conductance quantum,
$\Delta=1/N\rho(\epsilon_F)$ is the average level spacing at Fermi energy,
determined by the density of states in an infinite lead $\rho(\epsilon_F)$,
$E(\pi)$ and $E(0)$ are the ground--state energies of the system with periodic
and antiperiodic boundary conditions, respectively.
$E(\phi)$ is calculated for $\phi=0,\pi$ within the Rejec--Ram\v{s}ak
variational method \cite{reram}, complemented by the inverse orbital size
$\alpha$ optimization, as mentioned above.
We use typically $N=10^2\div 10^3$ sites to reach the convergence.

In Fig.\ \ref{gH2fig} we show the conductivity for $V=0.5t$, and different
values of the interatomic distance $R$. The conductance spectrum evolves from
the situation well--separated peaks corresponding to the independent filling
of bonding and antibonding molecular orbitals ($R\lesssim 2a_0$, where
$a_0$ is the Bohr radius), to the single peak in the intermediate range
($R\approx 3a_0$), which decays when $t'\ll V$ for large $R$.

Probably, the most interesting feature of the conductance depicted
in Fig.\ \ref{gH2fig} is their strong asymmetry for small $R$.
Namely, the low--$V_G$ conductance peak, corresponding to the system
filling $\langle n_1\!+\!n_2\rangle\approx 1$ (one \emph{hole}) is
significantly wider than the high--$V_G$ peak for $\langle
n_1\!+\!n_2\rangle\approx 3$ (one \emph{extra electron}). Such a
particle--hole symmetry breaking is the novel feature, observed when
including the correlation--induced basis optimization, and absent in
the parametrized--model approach \cite{reram,bulk}. It is also a new
feature of a nanosystem, do not observed in mesoscopic double
quantum dots \cite{kouw}, where the particle--hole symmetry holds.

The relation between the observed asymmetry, basis renormalization,
and electron correlations can be clarified as follows.
\emph{First}, one can observe that the optimal values of the variational
parameter $\alpha$, provided in Fig.\ \ref{AlpNjfig}a, decrease dramatically
for high $V_G$, corresponding to the overdoped situation
$\langle n_1\!+\!n_2\rangle >2$ (\emph{cf.}\ Fig.\ \ref{AlpNjfig}b).
This is because the system minimize energy of double occupancies, which
is of the order $U\sim\alpha$ \cite{slat}.
\emph{Then}, we focus on an small $R$ limit, in which the well separation
of molecular orbitals allows one to approximate the expression for a single
impurity at zero temperature  \cite{meir}
\begin{equation}
\label{gsingle}
  G=G_0\sin^2(\pi\langle n_k\rangle/2),
\end{equation}
where $\langle n_k\rangle$ is the bonding ($k=0$) or antibonding ($k=\pi$)
orbital occupancy. Expanding around the maximum we have
$G(V_G)-G(V_G^{*,k})\approx -G_0(\chi_c\pi/2)^2(V_G-V_G^{*,k})$, where
$V_G^{*,k}$ ($k=0,\pi$) is the low/high voltage peak position and $\chi_c$
is the charge susceptibility, which may be approximated as
$\chi_c\approx\partial\langle n_1\!+\!n_2\rangle/\partial V_G$, since
the orbitals are filled separately. Values of the later derivative read
from Fig.\ \ref{AlpNjfig}b around the low and high voltage peak positions
($\langle n_1\!+\!n_2\rangle\approx1$ and $\approx 3$, respectively) provides
a clear asymmetry. Moreover, the expansion of $G(V_G)$ allows one to roughly
estimate the peak width as $\Delta V_G\approx\chi_c^{-1}\approx (U+K)/2
\sim\alpha$, what provides an indirect correspondence between the spectrum
asymmetry and basis renormalization.

\begin{table}
\caption{
  The binding energy $\Delta E$ and the bond length $R_{\min}$ for different
  coupling to the lead $V$. The conductance at the energy minimum is also
  provided.\label{stabtab}
}
\begin{center}
\begin{tabular}{rrrr}
\hline\hline
$V/t$ & $\mbox{ }\Delta E\mbox{\hspace{2ex}}$ & $\mbox{ }R_{\min}/a_0$
  & $\mbox{ }G_{\min}^{V_G=0}/G_0$ \\
\hline
0.0  & -0.296  & 1.43  & 0 \\
0.1  & -0.293  & 1.44  & 10$^{-4}$ \\
0.2  & -0.282  & 1.46  & 0.004 \\
0.3  & -0.180  & 1.52  & 0.023 \\
0.4  & -0.040  & 1.59  & 0.093 \\
0.5  & +0.074  & 1.93  & 0.422 \\
\hline\hline
\end{tabular}
\end{center}
\end{table}

The practical possibility of the conductance measurement involving atoms
manipulation (changing of $R$) were also explored in terms of the system
stability. Namely, the binding energies
$\Delta E\equiv E_{R_{\min}}-E_{\infty}$ (where $R_{\min}$ is the bond length),
listed in Table \ref{stabtab}, shows that the system become metastable around
$V=0.5t$ ($\Delta E>0$ indicates a \emph{local} energy minimum).
Therefore, the binding of the atoms to the lead become stronger than
interaction between the atoms, that may allow individual atoms manipulation.

We would like to stress here, that \emph{ab initio} analysis have
never been performed for the \emph{open} system with strong electron
correlation. Apart from simplifying the discussion (the model
parameters are calculated as a function of interatomic distance),
such an approach leads to  new physical effects, since the
particle--hole symmetry is broken, as exemplified on the example of
the conductivity calculations.


\section{Concluding remarks}
In this paper we have provided a detailed discussion of the method
based on the diagonalization in the Fock space, when combined with
with an ab initio adjustment of single-particle wave functions in
the interacting state of N-particle system \emph{(the EDABI
method)}. The method can be improved upon by systematically
increasing the number of wave functions in the basis $[w_i({\bf r})
]$ when defining the field operator for the system (thus enriching
the model). We have illustrated
our method with the discussion of several nanoscopic systems ranging from atoms to nanochains and quantum dots.\\
\indent The method is useful when the exact solution of a model is
available in the Fock space. Such situation happens also for one
dimensional atomic chain represented  by the Lieb-Wu solution of the
Hubbard model \cite{LiebWu}. In that situation, exact Wannier or
Bloch wave functions can be constructed. Other applications such as
solving the magnetic impurity  in a nonmetallic environment
 are also possible, although not carried out
explicitly as yet.\\
\indent The importance of our approach is, among others, in showing
explicitly that the concept of the statistical distribution of
particles with quasimomentum, as a good quantum number, is feasible
for a relatively small interatomic distances. This means, that the
corresponding N-electron states form a quantum nanoliquid for not
too-large inter-distance in nanowire. In connection with this, in
replying the question {\em how small a piece of metal can be}, we
can say that the nanoliquid exhibits metallic conductivity for $e V
\geq \Delta E (N,R)$, where $V$ is the voltage applied to the system
and $\Delta E (N,R)$ is the energy difference between the highest
occupied and the lowest unoccupied state, i.e. that the momentum is
a good quantum number then. For an intermediate interatomic distace,
we have a gradual transition to a Mott insulator, above which the
monoatomic  (quantum) nanowire is useless for electronic
applications.
\section{Acknowledgments}
The authors were supported by the Ministry of Science and Higher
Education, Grant No, 1P~03B~001~29. We also thank the fellowships of
the Polish Science Foundation (FNP) for a Senior Fellowship (J.S.)
and Foreign Postdoc Fellowship (A.R.) The discussions with Jan
Kurzyk, Maciek Ma\'ska, Robert Podsiad\l y and W\l odek W\'ojcik were very important.\\


\begin{thebibliography}{}
\bibitem{ref1}
cf. e.g. Y. Imry, {\em Introduction to Mesoscopic Physics} (Oxford
University Press, Oxford, 2002); {\em Quantum Dots: a Doorway to
Nanoscale Physics}, edited by W.D. Heiss (Springer Verlag, Berlin,
2005);L. Jacak, P. Hawrylak, and A. Wojs, {\em Quantum Dots}
(Springer-Verlag, Berlin, 1998).
\bibitem{SpalekPodsiadl00}
J. Spa\l ek, R. Podsiad\l y, W. W\'ojcik, and A. Rycerz, Phys. Rev.
B {\bf 61}, 15676 (2000); J. Spa\l ek, E.M. G\"orlich, A .Rycerz, R.
Zahorbe\'nski and W. W\'ojcik, in {\em Concepts in Electron
Correlation}, edited by A.C. Hewson and V. Zlati\'c (Kluwer Academic
Publishers, Dordrecht, 2003) pp. 257-268.
\bibitem{Slater}
J. C. Slater, {\em Quantum Theory of Molecules and Solidsw}
(McGraw-Hill, New York, 1963), vol. 1; P.Korbel, W. W\'ojcik, A.
Klejnberg, J. Spa\l ek, M. Acquarone, and M. Lavagna, Eur. Phys. B
{\bf 32}, 315 (2003).
\bibitem{RycerzSpalek06}
A. Rycerz and J. Spa\l ek, Physica B {\bf 378 - 380}, 935 (2006)
\bibitem{ThesisAR}
A. Rycerz, Ph. D. Thesis, Jagiellonian University, Krak\'ow, 2003;
for online version see {\tt http://th-www.if.uj.edu.pl/ztms}
\bibitem{ThesisEG}
E. G\"orlich, Ph. D. Thesis, Jagiellonian University, Krak\'ow,
2004; for online version see {\tt http://th-www.if.uj.edu.pl/ztms}
\bibitem{ThesisRZ}
R. Zahorbe\'nski, Ph. D. Thesis, Jagiellonian University, Krak\'ow,
2005; for online version see {\tt http://th-www.if.uj.edu.pl/ztms}
\bibitem{SpalekDattaHonig87}
J. Spa\l ek, A. Datta, and J.M. Honig, Phys. Rev. Lett. {\bf 59},
728 (1987); Phys. Rev. B {\bf 33}, 4891 (1986).
\bibitem{Schweber}
The relativistic and nonrelativistic field quantizations for an
arbitrary single-particle basis was compared first in: S.S.
Schweber, {\em An Introduction to Relativistic Quantum Field Theory}
(Row, Peterson and Co., Evanston, IL, 1961), ch. 6.
\bibitem{Feynman}
See e.g. R.P. Feynman, {\em Statistical Mechanics: A Set of
Lectures} (W.A. Benjamin, Menlo Park, CA, 1972) ch. 6.7. In the case
of fermions we use the spinor notation for the field operators, e.g.
$\hat \Psi^\dagger({\bf r}) \equiv \left( \hat \Psi^\dagger_\uparrow
({\bf r}),\hat \Psi^\dagger_\downarrow ({\bf r}) \right)$,
$a^\dagger_i \equiv \left( a^\dagger_{i\uparrow},
a^\dagger_{i\downarrow} \right)$, and $\Phi_i({\bf r}) \equiv \left(
\Phi_{i\uparrow}({\bf r}), \Phi_{i\downarrow}({\bf r}) \right)
\equiv \Phi_i({\bf r}) \left(\chi_{i\uparrow}, \chi_{i\downarrow}
\right).$
\bibitem{Noone}
One could obtain a complete solution of (\ref{Ham2Q}) directly by
determining the field operator from the Heisenberg equation of
motion for $\hat \Psi$ or for the Green function related to it.
However, this is usually not feasible if one has to go beyond the
Hartree-Fock approximation or the interaction part must be treated
nonperturbatively.
\bibitem{RycerzSpalek01}
A. Rycerz and J. Spa\l ek, Phys. Rev. B {\bf 63}, 073101 (2001);
{\em ibid}.
{\bf 65}, 035110 (2002).\\
For brief review see: J. Spa\l ek, A. Rycerz, and  W. W\'ojcik,
Acta Phys. Polonica B {\bf 32}, 3189 (2001).
\bibitem{SpalekRycerz01}
J. Spa\l ek and A. Rycerz, Phys. Rev. B {\bf 64},161105(R) (2001).
\bibitem{SpalekWojcik92}
The method presented  evolved from an approximate treatment of an
extended (one-band) system of correlated electrons discussed in: J.
Spa\l ek and W. W\'ojcik, Phys. Rev. B {\bf 45}, 3799 (1992); J.
Magn. Magn. Mat. {\bf 104 - 107}, 723 (1992). This method has also
been applied to correlated systems selecting the basis of single
Gaussian instead; see: M. Acquarone, J.R. Iglesias, M.A. Gusm\~ao,
C. Noce, and A. Romano, Phys. Rev. B {\bf 58}, 7626 (1998); cf. also
A. Fortunelli and A. Painelli, J. Chem. Phys. {\bf 106}, 8041
(1997).
\bibitem{Shavitt}
R. A. Shavitt, in {\em Methods of Electronic Structure Theory},
edited by H. Schaeffer (Plenum Press, New York, 1977), pp. 189-275.
\bibitem{Schroed}
The method is essentially the same as the variational principle
introduced originally in E. Schr\"odinger, Ann. Phys. {\bf 79}, 361
(1926); cf. E. Schr\"odinger, {\em Collected papers on wave
mechanics}, Chelsea Publ. Co., New York, 1978. The Schr\"odinger
equation is obtained by minimizing the expression for the system
energy, $<\Psi | H_1 | \Psi >$ under the condition $<\Psi | \Psi
>=1$. Here, our functional $E_G\{ \Phi_i ({\bf r}) \}$ is much more
complicated due to the many-particle nature of the system.
\bibitem{Robertson}
For a lucid introduction to the relation between Fock and
Hilbert-space representations of multiparticle states see e.g.: B.
Robertson, Am. J. Phys {\bf 41},678 (1973); c.f. also
\cite{Schweber}.
\bibitem{Mattis}
The parametrized models play a prominent role in the theory of
correlated fermionic and bosonic systems, for which exact
solutions are avaible in some cases, cf. {\em The Many-Body
Problem}, edited by D.C. Mattis (World Scientific, Singapore,
1993); V.E. Korepin and F.H.L. Essler, {\em Exactly Solvable
Models of Strongly Correlated Electrons} (World Scientific, River
Edge, NJ, 1994)..
\bibitem{Noone2}
Note that the parameters $t_{ij}$ and $V_{ijkl}$ contain the
functions $\{ w_i({\bf r})\}$ under the integral expressions
(\ref{Tint}) and (\ref{Vint}), so $E_G$ is indeed a functional of
$\{ w_i({\bf r})\}$ an $\{ \nabla w_i({\bf r})\}$.
\bibitem{Szabo}
A. Szabo and N.S. Ostlund, {\em Quantum Chemistry} (Dover, Mineola,
1996).
\bibitem{Noone3}
The notion of the {\em self-adjusted state} differs from the concept of {\em quasiparticle} state invoked by Landau in his theory of
Fermi liquids. Landau quasiparticle results from an {\em
adiabatic} switching of the interaction and therefore , the
quasiparticle states are in one-to-one correspondence with the
states of an ideal quantum gas forming the liquid. Here, the {\em
self-adjusted states} can be either localized or itinerant as the
interparticle interaction cannot be treated perturbationally.
The statistical distribution of correlated state can differ from the Fermi-Dirac distribution (cf. Ref. 8),
as quasimomentum may not be a good quantum number.
\bibitem{Noone3a}
In the context of the orbit-size relaxation one can talk about
wave-function renormalization. The statistical distribution is
determined by dynamics in the Fock space.
\bibitem{Noone4}
Note that the atomic orbitals 1s and 2s are not orthogonal to each
other for arbitrary values of their spatial extents $1/\alpha_i$.
The 2p orbitals are orthogonal to each other and to s orbitals,
since they contain a nontrivial angular dependence expressed via
spherical harmonics $Y^m_l (\theta,\phi)$.
\bibitem{Bethe}
See e.g. H.A. Bethe and E.E. Salpeter, {\em Quantum Mechanics of
One- and Two- Electron Atoms}(Academic Press, New York 1957), pp.
146-156 and references therein.
\bibitem{Hubbard}
J. Hubbard, Proc. Roy. Soc. (London) A {\bf 281}, 401 (1964).
\bibitem{RycerzSpalek06a}
A. Rycerz and J. Spa\l ek, phys. stat. solidi (b) {\bf 243}, 183
(2006).
\bibitem{RycerzSpalek04}
A. Rycerz and J. Spa\l ek, Eur. Phys. J. B {\bf 40}, 153 (2004).
\bibitem{GorlichKurzyk}
E.M. G\"orlich, J. Kurzyk, A. Rycerz, R. Zahorbe\'nski, R. Podsiad\l
y, W. W\'ojcik, and J. Spa\l ek, in {\em Molecular Nanowires and
Other Quantum Objects}, edited by A.S. Alexandrov, J. Demsar, and
I.K. Yanson (Kluwer, Dordrecht, 2004), p. 364.
\bibitem{Cit27}
J. Spa\l ek, A. Rycerz, E.M. G\"orlich, and R. Zahorbe\'nski, in
{\em Highlights of Condensed Matter Physics} (AIP Conf. Proc., New
York), edited by A. Avella et al. (AIP Conf. Proc. No. 695,
Melville, New York, 2003), pp.291-303.
\bibitem{LiebWu}
E. H. Lieb and F. Y. Wu, Phys. Rev. Lett. {\bf 20}, 1445 (1968).

\bibitem{shamil}
B.S.\ Shastry and B.\ Sutherland Phys.\ Rev.\ Lett.\ \textbf{65},
243 (1990); J.A.\ Millis and S.N.\ Coppersmith, Phys.\ Rev.\ B
\textbf{42}, 10807 (1990); D.\ G\'{o}ra, K.\ Ro\'{s}ciszewski, A.M.\
Ole\'{s}, J.\ Phys.\ Condensed Matter \textbf{10}, 4755 (1998).

\bibitem{dago}
E.\ Dagotto,
Rev.\ Mod.\ Phys.\ \textbf{66}, 763 (1994);
P.\ Prelov\v{s}ek and X.\ Zotos,
in \textit{Lectures on the Physics of Highly Correlated Electron Systems VI},
edited by F.\ Mancini,
AIP Conf.\ Proc.\ vol 629 (New York, 2002) p.\ 161.

\bibitem{davidson}
E.\ R.\ Davidson,
\textit{J.\ Comput.\ Phys.} \textbf{17}, 87 (1975).

\bibitem{ryspa}
A.\ Rycerz, J.\ Spa{\l}ek,
\textit{Eur.\ Phys.\ J.\ B} \textbf{40}, 153 (2004).

\bibitem{paripss}
A.\ Rycerz, J.\ Spa{\l}ek,
\textit{phys.\ stat.\ sol.\ (b)} \textbf{243}, 183 (2006).

\bibitem{spatj}
P.\ W.\ Anderson, \textit{Phys. Rev.} \textbf{115}, 2 (1959);
W.\ F.\ Brinkman and T.\ M.\ Rice,
\textit{Phys.\ Rev.\ B} \textbf{2}, 1324 (1970);
K.\ A.\ Chao, J.\ Spa{\l}ek, and A.\ M.\ Ole\'s, \textit{J.\ Phys.\ C}
\textbf{10}, L271 (1977).

\bibitem{reram}
T.\ Rejec, A.\ Ram\v{s}ak,
\textit{Phys.\ Rev.\ B} \textbf{68}, 035342 (2003);
\textit{ibid.}\ 033306.

\bibitem{ander}
P.\ W.\ Anderson, \textit{Phys.\ Rev.}\ \textbf{124}, 41 (1961).

\bibitem{slat}
J.\ C.\ Slater, \textit{Quantum Theory of Molecules and Solids},
McGraw--Kill (New York, 1963), Vol.\ 1, p.\ 50; J.\ Spa{\l}ek
\emph{et al.}, Ref.\ \cite{SpalekPodsiadl00}.

\bibitem{bulk}
T.\ Kostyrko, B.\ Bu{\l}ka,
\textit{Phys.\ Rev.\ B} \textbf{67}, 205331 (2003).

\bibitem{kouw}
W.\ G.\ van der Wiel \emph{et al.}, \textit{Rev.\ Mod.\ Phys.}\
\textbf{75}, 1 (2003).

\bibitem{meir}
Y.\ Meir, N.\ S.\ Wingreen,
\textit{Phys.\ Rev.\ Lett.} \textbf{68}, 2512 (1992);
H.\ M.\ Pastawski, \textit{Phys.\ Rev.\ B} \textbf{46}, 4053 (1992).
A.--P.\ Jauho, N.\ S.\ Wingreen, Y.\ Meir,
\emph{ibid.} \textbf{50}, 5528 (1994).

\end{thebibliography}
\end{document}